\documentclass[aps,prb,twocolumn,superscriptaddress]{revtex4-2}
\usepackage{graphicx}
\usepackage{textcomp}
\usepackage{gensymb}
\usepackage{xcolor}
\usepackage{xcolor}
\usepackage[export]{adjustbox}

\usepackage[normalem]{ulem}
\usepackage{amssymb, amsmath,bm,onlyamsmath}

\usepackage{siunitx}
\definecolor{codegreen}{rgb}{0,0.6,0}
\definecolor{codegray}{rgb}{0.5,0.5,0.5}
\definecolor{codepurple}{rgb}{0.58,0,0.82}
\definecolor{backcolour}{rgb}{0.95,0.95,0.92}

\newcommand{\Gb}{\sigma_b}
\newcommand{\Ge}{G_e}
\newcommand{\HBA}{HB10a}
\newcommand{\HBB}{HB70a}

\begin{document}

\title{Multi-probe analysis to separate edge currents from bulk currents in quantum spin Hall insulators and to analyze their temperature dependence}

\author{S.~Benlenqwanssa}
\thanks{These two authors contributed equally.} 
\affiliation{Institut d'Electronique et des Syst\`emes (IES), UMR 5214 CNRS-Universit\'e de Montpellier, Montpellier, France.}

\author{S.~S.~Krishtopenko}
\thanks{These two authors contributed equally.} 
\affiliation{Laboratoire Charles Coulomb (L2C), UMR 5221 CNRS-Universit\'e de Montpellier, Montpellier, France.}


\author{M.~Meyer}
\affiliation{Julius-Maximilians-Universit\"at W\"urzburg, Physikalisches Institut and W\"urzburg-Dresden Cluster of Excellence ct.qmat, Lehrstuhl f\"ur Technische Physik, Am Hubland, 97074 W\"urzburg, Deutschland.}

\author{B. Benhamou--Bui}
\affiliation{Laboratoire Charles Coulomb (L2C), UMR 5221 CNRS-Universit\'e de Montpellier, Montpellier, France.}

\author{L.~Bonnet}
\affiliation{Laboratoire Charles Coulomb (L2C), UMR 5221 CNRS-Universit\'e de Montpellier, Montpellier, France.}

\author{A.~Wolf}
\affiliation{Julius-Maximilians-Universit\"at W\"urzburg, Physikalisches Institut and W\"urzburg-Dresden Cluster of Excellence ct.qmat, Lehrstuhl f\"ur Technische Physik, Am Hubland, 97074 W\"urzburg, Deutschland.}



\author{C.~Bray}
\affiliation{Laboratoire Charles Coulomb (L2C), UMR 5221 CNRS-Universit\'e de Montpellier, Montpellier, France.}

\author{C.~Consejo}
\affiliation{Laboratoire Charles Coulomb (L2C), UMR 5221 CNRS-Universit\'e de Montpellier, Montpellier, France.}

\author{S.~Ruffenach}
\affiliation{Laboratoire Charles Coulomb (L2C), UMR 5221 CNRS-Universit\'e de Montpellier, Montpellier, France.}


\author{S.~Nanot}
\affiliation{Laboratoire Charles Coulomb (L2C), UMR 5221 CNRS-Universit\'e de Montpellier, Montpellier, France.}

\author{J.-B.~Rodriguez}
\affiliation{Institut d'Electronique et des Syst\`emes (IES), UMR 5214 CNRS-Universit\'e de Montpellier, Montpellier, France.}


\author{E.~Tourni\'e}
\affiliation{Institut d'Electronique et des Syst\`emes (IES), UMR 5214 CNRS-Universit\'e de Montpellier, Montpellier, France.}

\author{F.~Hartmann}
\affiliation{Julius-Maximilians-Universit\"at W\"urzburg, Physikalisches Institut and W\"urzburg-Dresden Cluster of Excellence ct.qmat, Lehrstuhl f\"ur Technische Physik, Am Hubland, 97074 W\"urzburg, Deutschland.}

\author{S.~H\"ofling}
\affiliation{Julius-Maximilians-Universit\"at W\"urzburg, Physikalisches Institut and W\"urzburg-Dresden Cluster of Excellence ct.qmat, Lehrstuhl f\"ur Technische Physik, Am Hubland, 97074 W\"urzburg, Deutschland.}

\author{F.~Teppe}
\affiliation{Laboratoire Charles Coulomb (L2C), UMR 5221 CNRS-Universit\'e de Montpellier, Montpellier, France.}

\author{B.~Jouault}
\email[]{benoit.jouault@umontpellier.fr}

\affiliation{Laboratoire Charles Coulomb (L2C), UMR 5221 CNRS-Universit\'e de Montpellier, Montpellier, France.}

\begin{abstract}
We present a multi-probe transport analysis that effectively separates bulk and edge currents in large Hall bar devices with standard geometries. Applied to transport measurements on all possible four-probe configurations of six-probe Hall bar devices made of inverted three-layer InAs/GaInSb quantum wells (QWs), our analysis not only reveals the presence of dissipative edge currents in the topological gap, but also allows the temperature dependence of bulk and edge conductivity to be evaluated separately.
The temperature dependence of the edge conductivity for Hall bar channels from 10 $\mu$m to 70~$\mu$m in the range of 1.5~K to 45~K is consistent with the theoretical expectation of weakly interacting helical edge electrons with backscattering due to localized magnetic moments of charge impurities. We argue that these charge impurities are naturally associated with intrinsic Ga-antisite defects, which act as double acceptors in InAs/Ga(In)Sb-based QWs.
\end{abstract}

\maketitle

\section{\label{Int} Introduction}
The resistivity of two-dimensional (2D) systems, like quantum wells (QWs) or monolayer materials is usually measured either by means of the van der Pauw method~\cite{1958Pauw}, or by using specially designed devices, such as Hall bars (HBs). These methods are well suited for trivial 2D systems that conduct only through the bulk, while they must be modified for topological materials with additional conduction at the edges, superimposed on the conventional bulk resistivity. Typical examples of such 2D topological materials are quantum spin Hall (QSH) insulators~\cite{2005Kane}. The one-dimensional (1D) edges of QSH materials are metallic, while their bulk is insulating.

In the HB devices of QSH materials, the edge current is usually determined by analyzing non-local resistances~\cite{2007Konig,2009Roth,2011Knez,2013Grabecki,2013Suzuki,2015Du,2017Mueller}, because the bulk contribution becomes exponentially small when the distance between the probes is large. However, the existing non-local measurements have not used the full potential of this method offered by the presence of all the contacts in HB devices. 
Indeed, conventional non-local transport measurements involve only certain contacts, which does not allow one to accurately determine the contribution of the residual bulk current that may coexist with the edge current due to hopping or thermal activation. 
Very often, to evaluate the residual bulk current, one has to perform additional measurements based on Corbino disks~\cite{2015Du,2016Nichele,2017Mueller} or devices of specific geometry~\cite{2016Couedo,2016Nguyen}.
{The purpose of this work is to demonstrate a new multi-probe analysis for the separation of edge and bulk current in HB devices made from QSH insulators. The essential point of the proposed method is that it is based on the transport measurements involving \emph{all 45 geometries} of conventional six-probe HB devices. The use of all possible geometries allows us to obtain maximum accuracy in the separation of bulk and edge currents.}

Despite the fact that the first experimental evidence of QSH insulator state was found in HgTe~QWs~\cite{2007Konig}, which made this 2D system canonical for studying transport through topological edge states, for our HB devices, we chose QSH insulators on the basis of III-V semiconductor QWs. The main reason for this choice is that III-V semiconductors have well-established technological process for manufacturing devices, while good ohmic contacts of well-reproduced geometry are still difficult to achieve in HgTe-based QSH devices. Incorrect determination of device geometry or poor contact ohmicity would greatly compromise the quantitative multi-probe analysis presented below.

The first 2D system based on III-V semiconductors, in which the QSH insulator state was predicted~\cite{2008Liu} and then experimentally observed~\cite{2011Knez} are inverted InAs/GaSb QW bilayers (2Ls). Due to the inverted bandgap typically lower than 4~meV~\cite{2024Meyer}, the observation of topological edge states in InAs/GaSb QW 2Ls is more challenging than in HgTe QWs. Particularly, the small bandgap results in high residual bulk conductivity and additional parasitic edge conductance, which persists even in the trivial regime~\cite{2016Nichele}, thus hindering a clear observation of the transport via topological edge states. However, these initial difficulties inherent in InAs/GaSb QW 2Ls were later overcome and the QSH effect was observed,
either by reducing the bandgap mobility~\cite{2015Du}, or by engineering strained InAs/GaInSb QW 2Ls~\cite{2017Dua} with higher bandgap values.  

Not long ago, some of us predicted that adding a third InAs layer to the InAs/Ga(In)Sb QW 2Ls would significantly increase the bandgap values in the QSH insulator state~\cite{2018Krishtopenkoa}. Subsequent magneto-optical measurements of three-layer (3L) InAs/GaSb QWs confirmed that the inverted bandgap was much higher than in the QW 2Ls~\cite{2018Krishtopenko}, and was independent of temperature, unlike the one in canonical HgTe QWs~\cite{2017Marcinkiewicz,2018Kadykov}. Later, numerous THz spectroscopy investigations and magnetotransport studies of large-scale QSH devices confirmed the whole phase diagram of topological states predicted for 3L InAs/GaSb QWs~\cite{NewRef1,NewRef2,2019Krishtopenkoa,2019Krishtopenko,2021Meyer,2022Schmid,2023Meyer}. Finally, we have recently been able to experimentally realize strained 3L InAs/GaInSb QWs with a large inverted bandgap up to 45 meV, making these structures promising candidates for high-temperature QSH insulators~\cite{2022Avogadri}. By combining local and non-local measurements of the HB devices, we detected the traces of edge conductivity at temperatures up to 40~K with equilibrium lengths of a few micrometers. Our observed edge conductivity was much lower than its predicted quantized value~\cite{2022Avogadri}, {which can be explained by the presence of backscattering in the edge channels of large HB devices made from QSH insulators~\cite{NewRef3,2013Grabecki,NewRef4,2013Vaeyrynen,2014Vaeyrynen,2016Vaeyrynen,2021Hsu}.}

{To perform any characterization of the backscattering mechanism, one has to carefully extract the edge contribution dependence on \emph{temperature} ($T$) or \emph{voltage drop} ($V$) in the edge channel from the experimental data~\cite{NewRef3,NewRef4,2013Vaeyrynen,2014Vaeyrynen,2016Vaeyrynen,2021Hsu}. In the large scale HB devices, the edge contribution into non-local transport always coexists with the residual bulk contribution~\cite{2015Du,2014Knez}. The residual bulk contribution, in turn, can also depend on temperature, which distorts the temperature dependence of non-local transport, making it different from the real dependence of the edge contribution. Currently, there are only a few works~\cite{2015Li,2014Spanton,NewRef7} devoted to studying the dependence of edge conductance on temperature or voltage drop in the channel of HB devices made from InAs/Ga(In)Sb-based QSH insulators.}

In particular, transport measurements of Li~\emph{et~al.}~\cite{2015Li} at $T<2$~K, performed in an HB device with a channel length of $1.2$~$\mu$m made of InAs/GaSb QW 2L, revealed a power-law dependence of the edge conductance on the voltage drop and temperature, which was interpreted as the observation of a helical Luttinger liquid at the HB edge.
In contrast, scanning superconducting quantum interference imaging of the edge current in 50-$\mu$m-long channels revealed temperature-independent edge resistances in the range up to $30$~K within the limits of experimental resolution~\cite{2014Spanton}. In both works~\cite{2015Li,2014Spanton}, the InAs/GaSb interface was additionally doped with silicon to reduce the residual bulk conductivity. 
Finally, the transport measurements at $T<2$~K of inverted InAs/Ga$_{0.68}$In$_{0.32}$Sb QW 2L of Li~\emph{et al.}~\cite{NewRef7} showed that the edge conductance in a 12-$\mu$m long HB channel increases with increasing $T$ or $V$. We note, however, that the interpretation of the results of these measurements did not take into account the residual bulk conductivity, assuming that it should be negligible in a 2L QW with a bandgap of about $20$~meV at $T<2$~K.

In this work, by applying the multi-probe analysis to the transport measurements of several HB devices made from large-gap 3L InAs/GaInSb QWs, we first check that the edge current in the inverted gap can be detected by a sign reversal in some four-probe resistances of the HBs. Such a sign reversal is a signature of strongly modified current flows that was used, for instance, for detecting the appearance of the hydrodynamic regime in graphene~\cite{2016Bandurin}. Then, by analyzing transport measurements in all possible four-probe configurations at different temperatures, we separately evaluate the temperature dependencies of the bulk and edge current contributions for two HB devices. We show that the temperature dependence of edge conductivity in both HB devices, as well as previous experimental results~\cite{2015Li,2014Spanton,NewRef7}, are well explained by the backscattering due to localized magnetic moments~\cite{2016Vaeyrynen}. We argue that the presence of these localized magnetic moments is naturally associated with intrinsic Ga-antisite defects, acting as double acceptors in InAs/Ga(In)Sb-based QWs.
\textcolor[rgb]{0.00,0.00,0.00}
{
\begin{figure}
\includegraphics[width=1.0\linewidth]{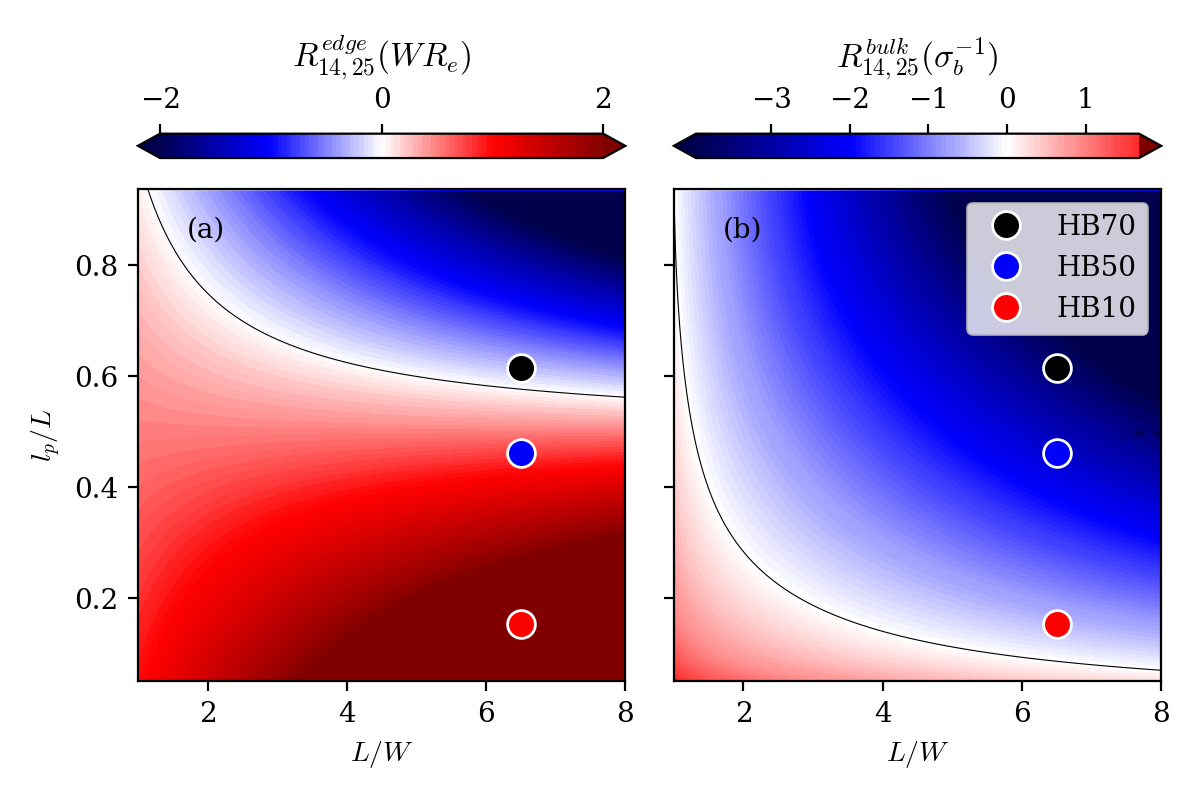}
\caption{\textcolor[rgb]{0.00,0.00,0.00}{Color maps of the four-probe resistances $R_{14,25}^\mathrm{edge}$ (panel a) and $R_{14,25}^\mathrm{bulk}$ (panel b), as a function of the parameters $L/W$ and $l_p/L$ defining the shape of the HB.  The black  lines are equipotential lines at zero resistance. The three solid dots represent the geometry of devices HB10 (red dot), HB50 (blue dot) and HB70 (black dot). The inset shows the HB geometry and contact labeling.}}
\label{fig:sign}
\end{figure}
}
\begin{figure*}
\includegraphics[width=1.0\linewidth]{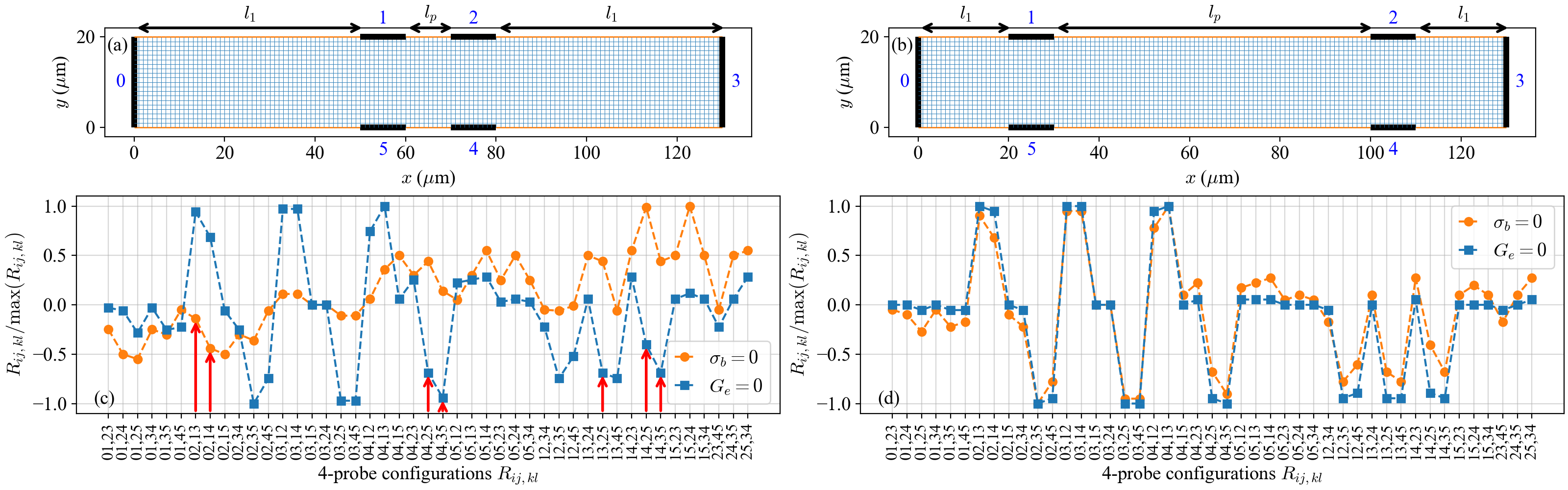}
\caption{(a,b) Geometry of (a) device HB10, and (b) device HB70, with a distance $l_p= \qty{10}{\micro\metre}$ and $l_p= \qty{70}{\micro\metre}$ between the middle probes respectively. The grid used for the numerical calculation is indicated in blue (bulk part) and orange (edge part) colors. The mesh size is $1$~$\mu$m. The ohmic contacts are shown with thick black lines. (c,d)  \textcolor[rgb]{0.00,0.00,0.00}{Numerically calculated four-probe resistances $R_{ij,kl}$ both for bulk conduction only ($G_e=0$) and edge conduction only ($\Gb{}=0$), for (c) device HB10 and (d) device HB70. Resistances have been normalized to their maximum value for each device and conduction type.}}
\label{fig:model_HB10_HB70}
\end{figure*}

\section{Detection of edge currents}
First, let us determine some geometries in which a sign reversal of the resistance is expected as the current flows from the bulk to the edges.
We start with a simple model of a HB, as shown in the inset of Fig.~\ref{fig:sign}.
The HB represented as a 2D rectangle of width $W$ and total length $L$,
has six infinitely small contacts labeled from 0 to 5,
and the distance between two adjacent side contacts is marked as $l_p$. By definition,
$R_{ij,kl}$ is $V_{kl}/I_{ij}$, where $V_{kl}$ is the voltage difference between contacts $k$ and $l$, and $I_{ij}$ is the current from contact $i$ to contact $j$.

If the current flows only at the edge of the HB and is dissipative,
we can introduce a one dimensional edge resistance per unit length, $R_e$.
This situation can occur in the QSH state when the HB dimensions are larger than the helical edge backscattering length induced by the presence of charge puddles~\cite{2007Abanin}.
It may also happen if parasitic edge currents are present\cite{2016Nichele}.
In these situations, the corresponding resistances $R_{ij,kl}$ can be calculated analytically.
For example, $R_{14,25} = (P/2-2 l_p) R_e /2 $, where $P$ is the perimeter of the HB. Note that $R_{14,25}$ can be either positive or negative. In the opposite case, if the HB is perfectly homogeneous with an isotropic conductivity $\sigma_b$, then $R_{ij,kl}$
can be calculated by a numerical Schwartz-Christoffel (SC) mapping~\cite{2021Driscoll} from the Hall rectangle onto an infinite half-plane, as originally proposed by van der Pauw\cite{1958Pauw}.

In the case where the HB hosts a QSH state, a transition between these two configurations can be observed by adjusting the carrier density using a gate.
From basic geometric considerations that arise from the SC mapping onto the half-plane, it can be shown that $R_{ij,kl}$
always has the same sign if the probes $k$ and $l$ are adjacent and not separated by one of the probes $i$ or $j$ ({\it{i.e.}}, the probes $i$,$j$,$k$,$l$ define one of the usual van der Pauw configurations).
Such a relationship does not hold if the probes $i,j,k$ and $l$ are interleaved.

Let us focus in particular on the interleaved configuration $R_{14,25}$.
We have calculated $R_{14,25}$ in the case of edge current only ($R_{14,25}^\mathrm{edge})$ and bulk current only ($R_{14,25}^\mathrm{bulk}$)
as a function of the dimensionless parameters $L/W$ and $l_p/L$ of the HB.
Figure~\ref{fig:sign} shows the mapping of $R_{14,25}^\mathrm{edge}$ and $R_{14,25}^\mathrm{bulk}$ as a function of $L/W$ and $l_p/L$.
It appears that the region  where $R_{14,25}^\mathrm{edge}$ and $R_{14,25}^\mathrm{bulk}$ have opposite signs is quite extended.
Moreover, for large $L/W$ and small $l_p/L$, $R_{14,25}^\mathrm{edge}$ can be quite large.
Therefore, to detect a change in current flow induced by a topological transition, these peculiar HB geometries are particularly well suited.

\section{Separation of the bulk and edge contributions}
We now develop a numerical model adapted for realistic HB geometries.
It takes into account the finite size of the contacts and includes
both edge and bulk conduction via the two parameters $G_e$ and $\sigma_b$.
We assume the existence of a unique electrochemical potential $\psi$ in the HB.
In the inner part of the HB, current conservation and Ohm's law yield the usual
Laplace equation $\Delta \psi = 0$. On the edges, the boundary condition
is imposed by current continuity:
\begin{equation}
\label{Eq:BC}
G_e \frac{\partial^2\psi}{\partial x^2}+
\sigma_b \frac{\partial\psi}{\partial y}=0,
\end{equation}
where $x$ ($y$) is the direction parallel (normal) to the edge and $G_e = 1/R_e$.
Note that $G_e$ is in \unit{\per\ohm\metre} and $\sigma_b$ is in \unit{\per\ohm}.
The validity of this model for QSH states depends on additional assumptions discussed in Appendix~\ref{appendix:QSH}.
We note that similar models have already been successfully applied to decouple edge and bulk conductance
for both 2D~\cite{2016Nguyen} and 3D~\cite{2016Durand} topological insulators.

 We assume an infinite  conductivity for the contacts.
For the source and the drain contacts $i$ and $j$ we apply Dirichlet boundary condition. This completes the set of boundary conditions, and $\psi$, as well as the associated current distribution, can be calculated numerically for a given device
and a given experiment.

For the numerical solution of the boundary value problem described by Eq.~(\ref{Eq:BC}), 
we used a finite difference scheme on a square grid.
%
We have chosen standard HB geometries with six contacts,
with a HB width $W=20$~$\mu$m, and
a HB total length $L=130$~$\mu$m.
Typical geometries are shown in Fig.~\ref{fig:model_HB10_HB70}(a,b).
The size of the four lateral contacts of infinite conductivity is fixed to $l_c=10$~$\mu$m.
We let the distance $l_p$ between the middle probes vary from device to device,
and label devices with the prefix HB followed by a number corresponding to $l_p$. Hence,
$l_p=\qtylist{10;30;50;70}{\micro\metre}$ for HB10, HB30, HB50 and HB70 respectively.
The six probes are labeled from $0$ to $5$, see Fig.~\ref{fig:model_HB10_HB70}(a,b).
In a device with six probes, we obtain 45 possible $R_{ij,kl}$ resistances, imposing $i<j$, $k<l$ and excluding
$R_{kl,ij}$ if $R_{ij,kl}$ is already present,
because the Onsager-Casimir relation imposes $R_{kl,ij}= R_{ij,kl}$.
In what follows, these 45 configurations are used to check the integrity of the HBs.
However, the number of configurations can be reduced to 12 by considering the HB symmetry and removing certain redundancies (for example, $R_{01,24}= R_{01,23}+R_{01,34}$).


Figure~\ref{fig:model_HB10_HB70}(a) shows the geometry of HB10, as well as the finite difference grid
used for the numerical calculation.
We used a mesh size $a=1$~$\mu$m~$\ll W,L$,
which ensures a good numerical estimate of $\psi$ in the inner part of the HB.
Figure~\ref{fig:model_HB10_HB70}(c) shows the 45 resistances $R_{ij,kl}$
calculated for HB10
for two extreme cases:
i) bulk conduction only ($G_e=0$) and ii) edge conduction only ($\sigma_b=0$).
For clarity, in these two cases the resistances have been normalized with the highest value of the 45 resistances for bulk and edge conduction:
${0.75/\sigma_b}$~$\Omega$ and $45/G_e~\unit{\ohm\per\micro\metre}$, respectively.
Interestingly, the $R_{ij,kl}$ signatures
for these two cases are very different.
First, $R_{15,24}$, often nicknamed the ``non-local'' resistance,
greatly exceeds $R_{03,12}$ in case of edge conduction.
Second and as expected, the previously studied  configuration $R_{14,25}$  changes its sign
when the conduction evolves from bulk to edge.
Moreover, a similar sign reversal is also observed for the configurations $R_{02,13}$, $R_{02,14}$, $R_{04,25}$, $R_{04,35}$,
$R_{13,25}$, and $R_{14,35}$.
These configurations are indicated by vertical red arrows in Fig.~\ref{fig:model_HB10_HB70}(c).

Similarly, Figs.~\ref{fig:model_HB10_HB70}(b,d) show the geometry of HB70 and the associated  $R_{ij,kl}$
signature for bulk and edge conduction. Again, the resistances have been normalized with the highest resistance value for bulk and  edge conduction:
${3.5/\sigma_b}~\unit{\ohm}$ and ${37/G_e}~\unit{\ohm\per\micro\metre}$, respectively.
As expected, the non-local resistance $R_{15,24}$
is much larger for edge conduction than for bulk conduction. However, when compared to
Fig.~\ref{fig:model_HB10_HB70}(c), the differences between bulk and edge conduction appear less obvious.
In particular, there is no more sign reversal when the current flow moves from edge to bulk.
The same analysis has been performed for HB50. For this geometry, a sign reversal is detected only for $R_{14,25}$.
The three different geometries: HB10, HB50 and HB70 are reported
in Fig.~\ref{fig:sign} as solid colored circles.
The positions of these HBs in Fig.~\ref{fig:sign} confirm that a sign reversal
between $R_{14,25}^\mathrm{edge}$ and $R_{14,25}^\mathrm{bulk}$ is indeed expected for HB10 and HB50, while in contrast,
no sign reversal should be expected for the HB70 geometry.

\begin{figure}
\includegraphics[width=0.97\linewidth]{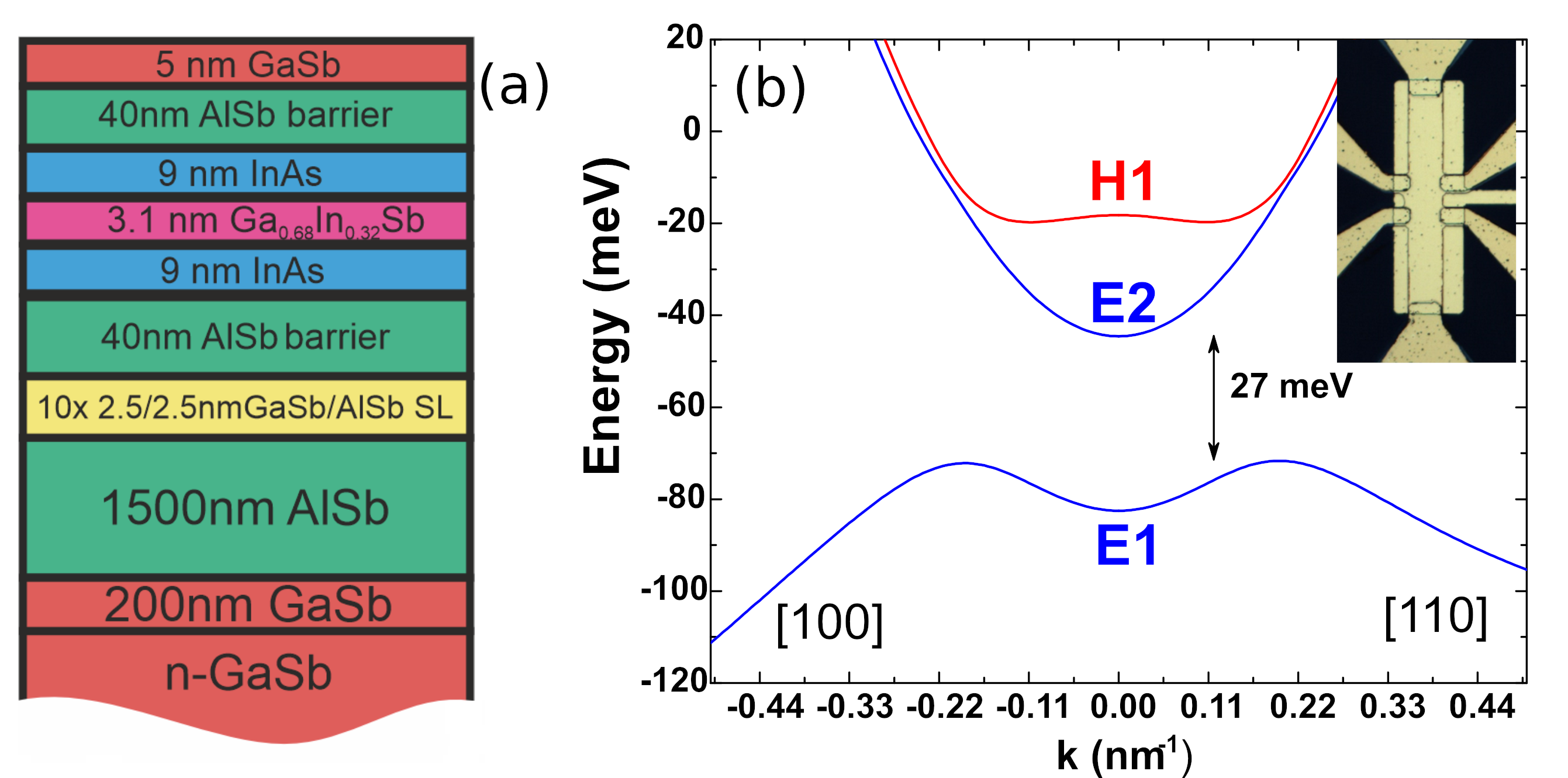}
\caption{Growth scheme (a) and band structure (b) of the 3L InAs/Ga$_{0.68}$In$_{0.32}$Sb QW used for fabrication of HB devices. The blue and red curves represent the energy-dispersion of the electron-like (\emph{E}1, \emph{E}2) and hole-like (H1) subbands, respectively. The positive and negative values of quasimomentum $k$ correspond to the [110] and [100] crystallographic orientations. The inset shows an optical image of device \HBA{}.}
\label{fig:structure}
\end{figure}

%
\begin{figure*}
\includegraphics[width=0.97\linewidth]{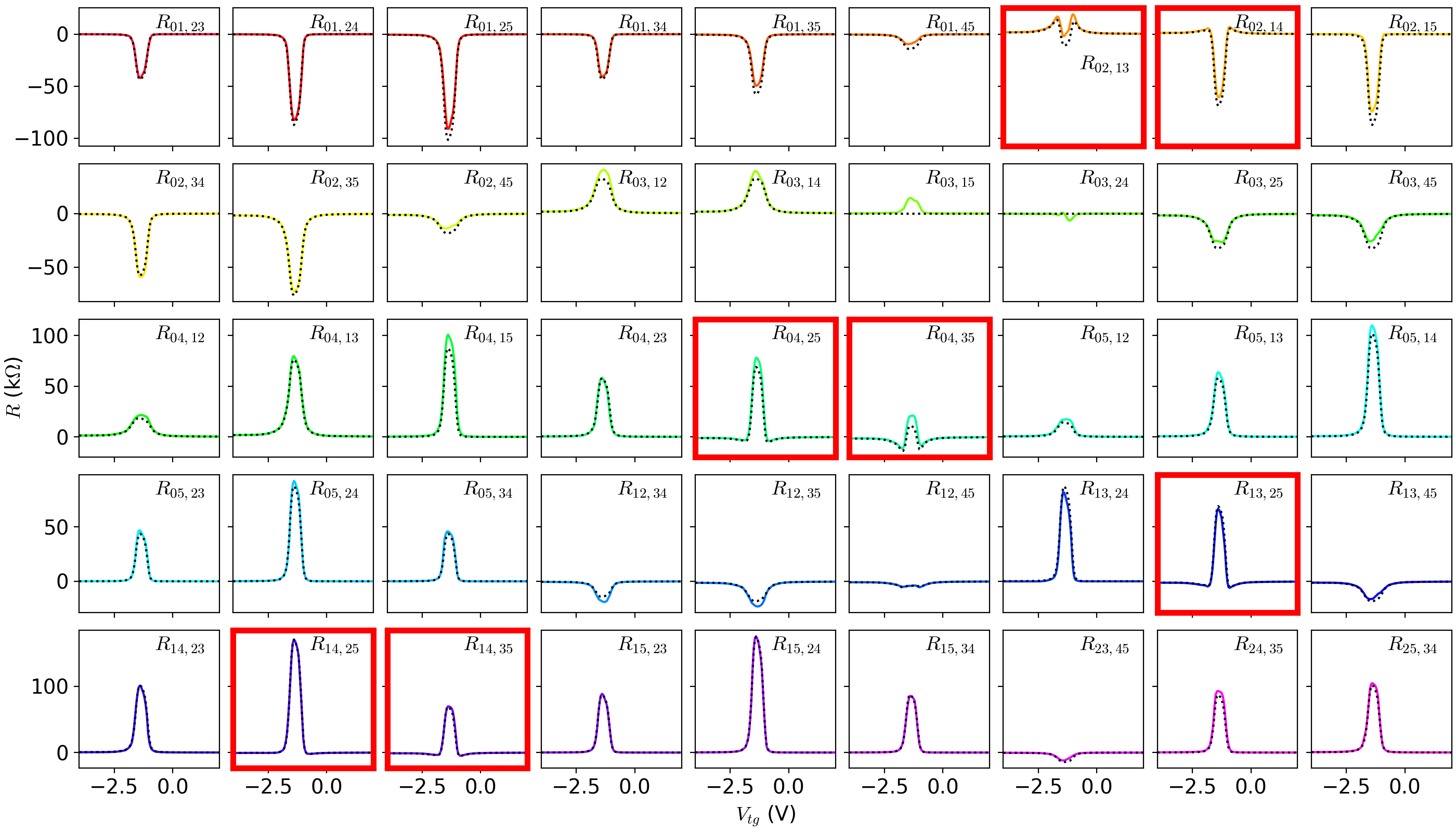}
\caption{\textcolor[rgb]{0.00,0.00,0.00}{(a) Evolution of the 45 $R_{ij,kl}$ resistances as a function of the top-gate voltage $V_\mathrm{tg}$ for device \HBA{}
at T= \qty{1.7}{\kelvin}. 
The colored lines are the experimentally measured four-probe resistances. 
The black dotted lines are the best nonlinear least-squares fit of the 45 resistances with the two-parameter model based on the edge conductance $G_e$ and the bulk conductivity $\sigma_b$. The fit is performed for each gate voltage value $V_\mathrm{tg}$.
The seven configurations in which a sign inversion of the resistance is expected, as shown in Fig.~\ref{fig:model_HB10_HB70}, and demonstrated experimentally, are highlighted with red boxes.}}
\label{fig:model_HB10_45conf}
\end{figure*}

\begin{figure}
\includegraphics[width=0.97\linewidth]{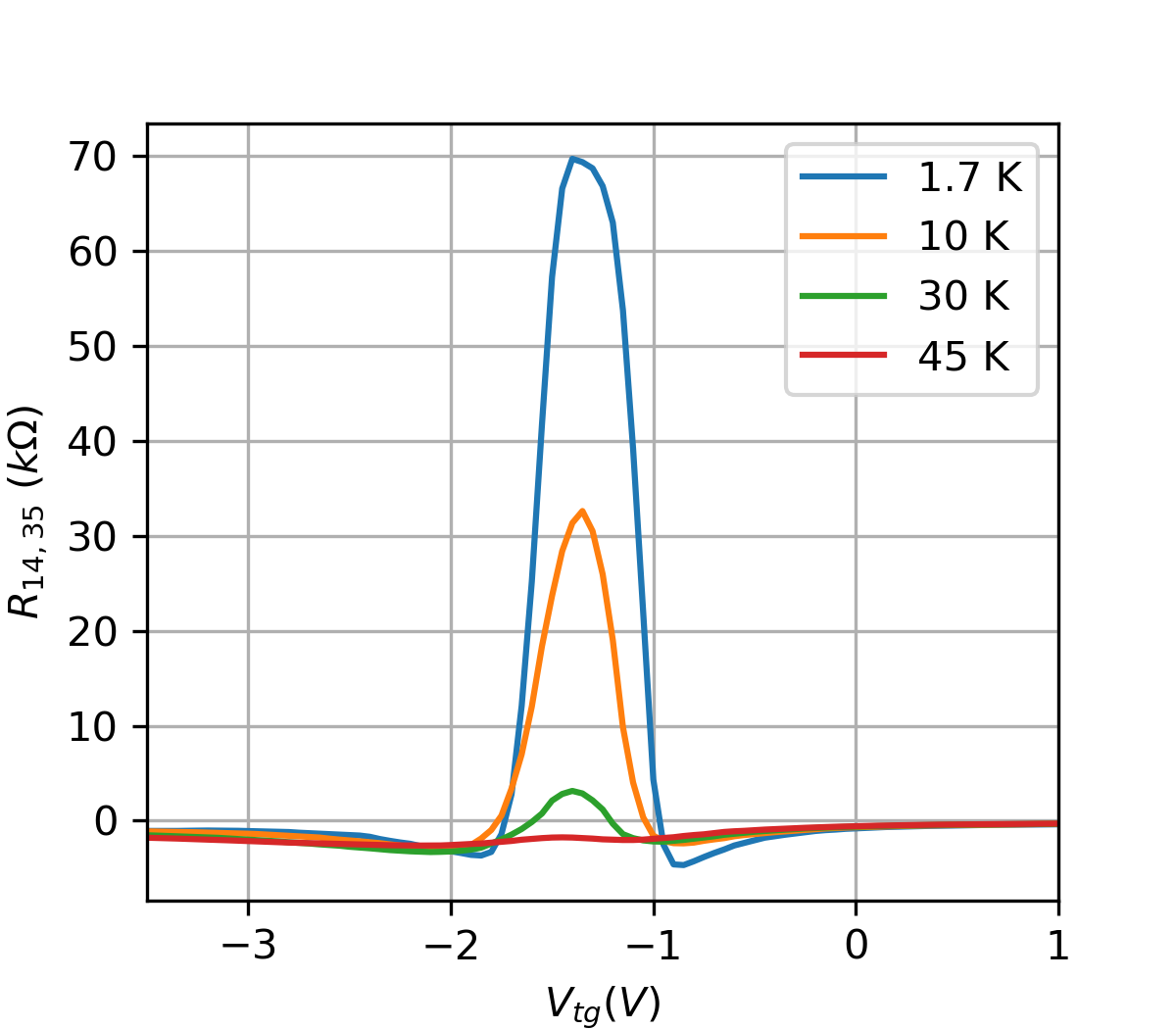}
\caption{$R_{14,35}$ as a function of $V_{tg}$ for \HBA{} at four different temperatures: $1.7$~K, $10$~K, $30$~K and $45$~K.}
\label{fig:R1435Tdep}
\end{figure}

\begin{figure*}
\includegraphics[width=1.0\linewidth]{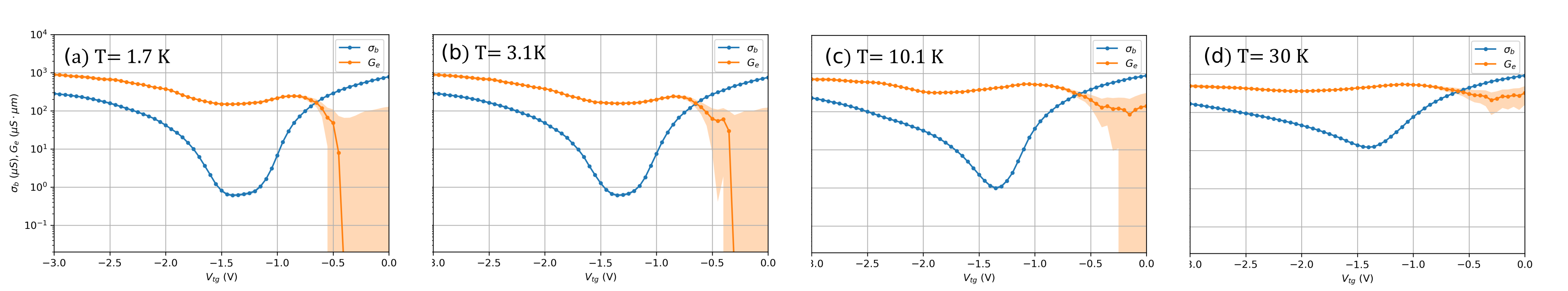}
\caption{Evolution of $G_e$ (orange dots) and $\sigma_b$ (blue dots) as a function of top-gate voltage $V_\mathrm{tg}$ at several temperature values for \HBA{}. The error intervals are indicated as colored areas.}
\label{fig:model_allHB_GeGb}
\end{figure*}

The structure used for the fabrication of the HB devices was grown by molecular beam epitaxy (MBE) on a (001) \emph{n}-type doped GaSb substrate.
On the substrate, a 700~nm GaSb layer is grown, followed by a 1500~nm AlSb buffer,
then a $10\times2.5$nm$/2.5$nm AlSb/GaSb superlattice (SL).
After the SL, a symmetrical 3L InAs/Ga$_{0.68}$In$_{0.32}$Sb QW is grown between two 40~nm AlSb barriers. The 3L QW consists of a 3.1~nm thick Ga$_{0.68}$In$_{0.32}$Sb layer surrounded by two 9~nm wide InAs layers. Finally, the top AlSb barrier is followed by a 5~nm thick GaSb cap layer to prevent oxidation of the AlSb barrier in air (see Fig.~\ref{fig:structure}(a)).

Band structure calculations based on the eight-band $\mathbf{k}\cdot\mathbf{p}$ Hamiltonian were performed by expanding the eight-component envelope wave functions in the basis set of plane waves and by numerically solving the eigenvalue problem. In the Hamiltonian, which directly accounts the interaction between $\Gamma_6$, $\Gamma_8$ and $\Gamma_7$ bands of bulk materials, we also include the terms describing the strain effect due to the mismatch of lattice constants in the buffer, QW layers and AlSb barriers. Details of the calculations and the form of the Hamiltonian can be found in Ref.~\onlinecite{2016Krishtopenko}. The bulk material parameters and valence band offsets used in the calculations are taken from Ref.~\cite{2001Vurgaftman}.
\textcolor[rgb]{0.00,0.00,0.00}{These calculations show that the grown 3L InAs/Ga$_{0.68}$In$_{0.32}$Sb QW has an inverted band structure, when the electron-like \emph{E}1 subband lies below the hole-like \emph{H}1 level~\cite{2006Bernevig}) with a band gap of $27$~meV (see Fig.~\ref{fig:structure}(b)).
When the Fermi energy is in the inverted gap, QSH edge conduction should appear. On the contrary, when the Fermi energy is in the valence or conduction band, bulk conduction comes into play. The position of the Fermi energy can be modulated by a gate voltage.}

More than ten HBs have been realized and measured in detail. The HBs have a metal top-gate and the geometry discussed above. In the following, we focus mainly on two HBs, named \HBA{} and \HBB{}, whose geometries HB10 and HB70 are shown in Fig.~2(a,b). As can be seen, each HB device has, in addition to the two edge channels of length $l_p$ between the middle contacts, four edge channels with a length
$l_1= (L- l_p - 2 l_c)/2$,
{\it i.e.}, $l_1= \qty{50}{\micro\metre}$ for \HBA{} and $l_1= \qty{20}{\micro\metre}$ for \HBB{}.

 The experiments were performed at low temperatures down to $T=1.5$~K.
All measurements were made with high impedance 1~T$\Omega$ preamplifiers, using standard lock-in techniques and a current of $10$~nA at a frequency close to 10~Hz. Measurements in four-probe configurations were automated using a mechanical switch system. During the measurement, the top-gate voltage is continuously swept from negative to positive and back to negative values over a period of several minutes. A small hysteresis occurs and the amplitude of the peaks depends on the direction of the gate voltage sweep~\cite{2023Meyer}.  Unless otherwise stated, the data presented below were taken for a negative gate sweep.

Figure~\ref{fig:model_HB10_45conf} shows the 45 four-probe configurations measured on \HBA{}
at $T=1.7$~K, as a function of the top-gate voltage $V_\mathrm{tg}$. The solid colored lines
correspond to the experimental data.  A prominent peak is observed for all configurations, and attributed to the inverted gap. On the left (right) side of the peak, the Fermi energy lies in the valence (conduction) band and bulk conduction  occurs. On the two sides, the charge of the carriers was confirmed by Hall measurements.  At the peak maximum, edge conduction is expected to take place.
As some details are difficult to see in Fig.~\ref{fig:model_HB10_45conf}, some important curves have been enlarged and are shown Fig.~\ref{fig:zoom_reversal}, Appendix~\ref{appendix:ZOOM}.

Remarkably, the sign reversal predicted by the model for the HB10 geometry at the transition from of bulk to edge conduction is observed  in the evolution of $R_{02,14}$, $R_{04,25}$, $R_{04,35}$, $R_{13,25}$, and $R_{14,35}$ as a function of the gate voltage.
Only for $R_{02,13}$ and $R_{14,25}$ the expected change of sign is not obvious in Fig.~\ref{fig:model_HB10_45conf}.
However, a close look on the data reveals that a sign reversal also takes place for these two configurations.
For $R_{14,25}$, the central positive peak attributed to edge conduction is so intense (more than $170$~k$\Omega$) that it masks the negative resistance on its aisles (which peaks at $-2$~k$\Omega$). Moreover,  for some of the $R_{ij,kl}$ peaks, a double peak structure can also be observed, see \emph{e.g.} $R_{02,45}$, $R_{03, 25}$ and many other resistances. Such a double-peak structure indicates that between the two peaks dramatic modifications of the current flow take place.

The sign reversal can be used to determine the highest temperature at which the edge conduction dominates. Figure~\ref{fig:R1435Tdep} shows  $R_{14,35}$ for \HBA{} at four different temperatures. The sign reversal is seen up to $T=30$~K, which demonstrates the presence of the edge currents in \HBA{} at least up to this temperature, without further analysis.

%

For each $V_\mathrm{tg}$, the 45 resistances were fitted together with the parameters $G_e$ and $\sigma_b$, as described by the model above.  The fit involves the resolution of a nonlinear least-squares problem, which we solved numerically. 
The details of the procedure are given in Appendix~\ref{appendix:NLLSF}. 
The dashed black lines in Fig.~\ref{fig:model_HB10_45conf} are the best obtained fits, with a standard deviation which does not exceed $7$~k$\Omega$ even at the peak maxima.
%
By comparison, the maxima of the resistance peaks can largely exceed $100$~k$\Omega$,
see \emph{ e.g.} $R_{14,25}$. Qualitatively, the sign reversal of the resistances and the double peak structures are well reproduced.
The largest discrepancy is obtained for the resistance peak of $R_{03,15}$, 
which is a transverse (Hall) resistance, sensitive to any residual asymmetry.

The fitting parameters $G_e$ and $\sigma_b$ are shown in Fig.~\ref{fig:model_allHB_GeGb}(a) for \HBA{},
as a function of $V_\mathrm{tg}$ and at $T\simeq1.8$~K. The error bars are also indicated as colored areas.
Some remarkable features can be spotted. First, edge conduction largely dominates at the peak center,
where bulk conduction becomes very small, with a resistivity greater than 1~M$\Omega$.
%
%
In the bandgap region, at $V_\mathrm{tg}\simeq-1.5$~V, $G_e=200$~$\mu$S$\cdot\mu$m. In the hypothesis of diffusive helical edge states, one expects the edge conductance $G_\mathrm{edge}$ to be defined as $G_\mathrm{edge}=(e^2/h)\lambda/l$, where $\lambda$ is the backscattering length and $l$ the length of the edge. 
Note that $G_{e}$ presented in Fig.~\ref{fig:model_allHB_GeGb} is defined as $G_{e}=G_\mathrm{edge}l$. Thus, the backscattering length is around $5$~$\mu$m, which is in good agreement with typical values in the literature~\cite{2015Du,2016Couedo,2017Dua,NewRef7,2022Avogadri}.

As $G_e$ decreases dramatically above $V_\mathrm{tg}=-0.5$~V, there is no edge conduction in the conduction band. This can be evidenced even if the error bar is quite large. By contrast, edge conduction in the valence band region is revealed. 
\textcolor[rgb]{0.00,0.00,0.00}{The persistence of such an edge conduction is confirmed by a direct and close analysis of the non-local resistances like $R_{15,23}$, which remains anomalously large on the left side of the peak: $R_{15,23}/R_{03,12} \simeq 10\%$ in the valence bands, while we would expect 2--3 \% for the bulk conduction only.}

The nonzero edge contribution in the valence-band region is not surprising -- some theoretical studies~\cite{2018Krishtopenkob} indeed predict the coexistence of edge and bulk states in complex valence band, described beyond the Bernevig-Hughes-Zhang model~\cite{2006Bernevig}. Alternatively, the edge conduction may be attributed to additional parasitic edge currents. 

\textcolor[rgb]{0.00,0.00,0.00}{
The fitting parameters $G_e$ and $\sigma_b$ have been extracted at different temperatures, as shown in Figs.~\ref{fig:model_allHB_GeGb}(a-d). The edge conductance in the bandgap persists up to \qty{30}{\kelvin} or more, with a weak temperature dependence. By contrast the bulk conductance in the gap increases.
Six HBs at different temperatures were investigated with this technique. All devices showed a very weak temperature dependence of the edge conductance below $T=\qty{10}{\kelvin}$. 
At the lowest temperature of $T=\qty{1.7}{\kelvin}$,
$G_{e}$ varied from \qty{50}{\micro\siemens\micro\metre} to
\qty{250}{\micro\siemens\micro\metre} from device to device,
and the bulk conduction varied from \qty{1E-2}{\micro\siemens\micro\metre} 
up to $\qty{1}{\micro\siemens\micro\metre}$. It turned out that $G_e$ and $\sigma_b$ are very sensitive
to the device preparation, and the gate voltage sweeps performed.}

\textcolor[rgb]{0.00,0.00,0.00}{To elucidate the temperature dependence of the edge conduction, thorough measurements were performed on \HBA{} and \HBB{} with a small temperature step, from \qty{5}{\kelvin} up to \qty{80}{\kelvin}, always using the same voltage sweeps.}
\begin{figure}
\includegraphics[width=1.0\linewidth]{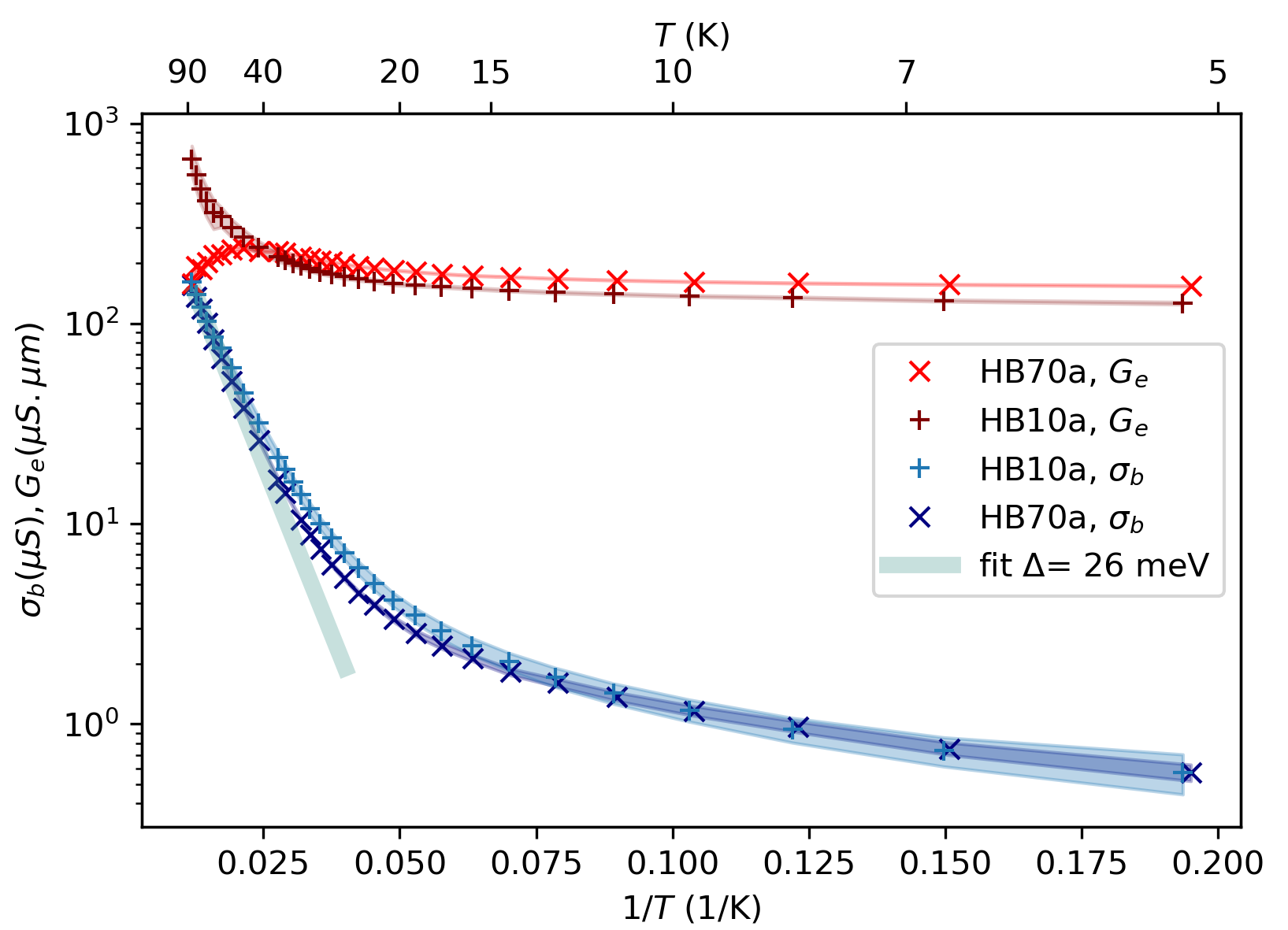}
\caption{Bandgap bulk conductivity $\sigma_b$ as a function of temperature for \HBA{} (blue symbols $+$) and \HBB{} (navy blue symbols $\times$). The thick cyan line represents the fit obtained by $\exp(-\Delta/2k_BT)$ with $\Delta=26\pm 1$~meV.  Edge conductances $G_e$ are also shown as brown symbols $+$ and red symbols $\times$ for \HBA{} and \HBB{} respectively. Colored areas indicate error intervals.}
\label{fig:trends}
\end{figure}
Figure~\ref{fig:trends} shows the temperature dependence of $\sigma_b$ extracted from these measurements.
The $\sigma_b$ and $G_e$ values have been taken at the gate voltages corresponding to the minima of $\sigma_b(V_\mathrm{tg})$.
At high temperatures, above $T= \qty{20}{\kelvin}$, $\sigma_b$ has an activated behavior and can be fitted by $\exp(-\Delta/2k_BT)$, where $\Delta$ is the energy gap. The fit on the range $\qty{40}{\kelvin} < T < \qty{90}{\kelvin}$ yields $\Delta=26\pm 1$~meV, which is in very good agreement with the theoretical bandgap value provided in Fig.~\ref{fig:structure}(b). 
On the contrary, the temperature dependence of $\sigma_b$ is weak at $T<20$~K, which might indicate the onset of hopping transport. The logarithm of $\sigma_b$ can be fitted with a power law of the kind $T^{1/\alpha}$, with $\alpha<4$. Unfortunately, the limited temperature range available for our measurements does not allow us to determine $\alpha$ with good accuracy. Hence, the underlying hopping mechanism cannot not be identified. Finally, $\sigma_b$ for both \HBA{} and \HBB{} almost coincide over the whole temperature range 
$\qty{5}{\kelvin} < T < \qty{80}{\kelvin}$. 
The fact that $\sigma_b$ depends only weakly on the device and its geometry suggests that the multi-probe model is adequate. 

Figure~\ref{fig:trends} also shows the temperature dependence of $G_e$ in the middle of the bandgap.
Note that $G_e$ was evaluated at the gate voltage where $\sigma_b$ is minimal.
The edge conductivities for both devices are quite similar at low temperatures, with a weak metallic behavior.
As seen, $G_e$ tends to saturate below 30 K at values around $\qty{200}{\micro\siemens\micro\metre}$. 
The extrapolation of $G_\mathrm{edge}$ to $T=\qty{0}{\kelvin}$ does not reach the quantized value 
({\it e.g.}, 
$l/G_e= \qty{50}{\kilo\ohm}$ ($\qty{350}{\kilo\ohm}$) with $l = \qty{10}{\micro\meter}$ ($\qty{70}{\micro\meter}$)).
Above 50 K, the edge conductances of both devices differ. In device \HBB{}, the edge conductance disappears
and merges with the bulk conductivity. On the contrary, an extra edge conductance persists in HB10a up to \qty{90}{\kelvin}.
This difference in behavior is unexpected and may be due to additional differences between the two HBs not accounted for by the model, as current leaks through the bottom gate.
In the following, we provide a more detailed analysis of the temperature dependence of the edge conductance below \qty{50}{\kelvin} only.


\section{Possible origins of the edge backscattering}
As mentioned in the introduction, the temperature and voltage drop dependence of the edge conductance may lead to the identification of the backscattering mechanism arising in the helical edge channels of large HBs~\cite{NewRef3,NewRef4,2013Vaeyrynen,2014Vaeyrynen,2016Vaeyrynen,2021Hsu}. In a very general case, the edge conductance depends dramatically on whether the 1D system is strongly correlated or not~\cite{2016Vaeyrynen,2021Hsu,NewRef8,NewRef9}. The parameter that characterizes the sign and the strength of the many-particle interaction in a 1D system is the Luttinger parameter $K$: $K<1$ for repulsion, $K>1$ for attraction, and $K=1$ for the non-interacting case. Besides, the value $K=1/4$ is a natural crossover between the Fermi liquid regime with relatively weak Coulomb interaction ($K>1/4$) and the strongly correlated case of 1D helical Luttinger liquid ($K<1/4$).

Despite numerous theoretical models, existing experimental results in various QSH materials~\cite{2013Grabecki,2014Knez,2014Spanton,NewRef11,NewRef12,NewRef13} are still interpreted based on the general concept of charge puddles~\cite{2013Vaeyrynen,2014Vaeyrynen} that trap edge carriers and within which the spin flips, allowing backscattering in the helical edge channels. 
In the following, let us first consider the most likely causes of backscattering in the edge channel of QSH insulators based specifically on semiconductor heterostructures, like HgTe QWs and InAs/Ga(In)Sb QWs.
Very recently, Dietl~\cite{NewRef5,NewRef6} noticed the critical role of localized spins of native acceptors with resonant levels in the bandgap on the properties of HgTe-based QSH insulators. Experimental studies of interband photoconductivity~\cite{NewRef14} and relaxation times of photo-excited carriers~\cite{NewRef15,NewRef16} demonstrated that these native acceptor states in HgTe QWs are associated with Hg vacancies, acting as double acceptors. In turn, the arising of double acceptors in InAs/Ga(In)Sb-based QWs is naturally associated with intrinsic Ga-antisite defects in Ga(In)Sb layers possessing the lowest formation energy~\cite{NewRef17,NewRef18,NewRef19,2017Segercrantz}. The presence of these defects in 3L InAs/GaSb QWs was previously revealed by the observation of its characteristic resonant line with an energy of about 35~meV in the terahertz photoluminescence spectra~\cite{2018Krishtopenko,2019Krishtopenko}. From these considerations, we can expect HgTe- and InAs/GaSb-based QWs to have similar backscattering mechanisms, based on the localized spin of the native acceptors. Interestingly, the conclusion on the similarity of backscattering mechanisms was previously made by Spanton~\emph{et~al.}~\cite{2014Spanton} by comparing experimental images of the edge current in InAs/GaSb 2L QWs and HgTe QWs~\cite{NewRef20}. Thus, as the main backscattering mechanism, we will further consider the  interaction of helical edge electrons with the localized spin of the impurities randomly distributed along the edge.

In what follows, to describe our experimental results on the temperature dependence of the edge conductance, we will use a model by V\"{a}yrynen~\emph{et al.}~\cite{2016Vaeyrynen} that takes into account the backscattering of edge electrons due to a ensemble of localized magnetic moments of random orientation.
Considering a channel hosting many diluted impurities near the edge, the conductance of a single edge $G_\mathrm{edge}$ of length $L$ in the high temperature regime $T>T^{*}$ can be written as:
\begin{equation}
\label{Eq:Thx1}
G_\mathrm{edge}\approx\dfrac{1}{N}\left(\dfrac{e^2}{h}\right)^2\left[\dfrac{\partial\langle\delta{I}\rangle}{\partial{V}}\right]^{-1},
\end{equation}
where $N$ is the number of impurities, $V$ is the voltage drop along the channel, and $\langle\delta{I}\rangle$ is the single-impurity backscattering current. For many impurities, one expects $N\propto{L}$ and hence resistive behavior $G\propto{L^{-1}}$.
The critical temperature $T^{*}$ which limits the use of Eq.~(\ref{Eq:Thx1}) is determined by the bulk bandgap $\Delta$, the Luttinger parameter $K$, and the effective constant $J_0$ of the \emph{isotropic part} of the exchange interaction between edge electrons and the localized impurity spin:
\begin{equation}
\label{Eq:Thx2}
T^{*}=\dfrac{\Delta}{k_B}\left(\dfrac{1}{\sqrt{2}}\dfrac{\rho J_0}{1-K}\right)^{1/(1-K)},
\end{equation}
where we assume $\rho J_0\ll 1-K$ (here, $\rho$ is the electron density of states per spin per unit edge length)~\cite{2014Vaeyrynen}.

The temperature and voltage-drop dependence of the backscattering current in Eq.~(\ref{Eq:Thx1}) can be represented as a product of several terms:
\begin{eqnarray}
\label{Eq:Thx3}
\langle\delta{I}\rangle=\delta{G}_{0}\left[\dfrac{\Delta}{2\pi{k_B}T}\right]^{2-2K}
\dfrac{B(K,K)f\left(x\right)V}{\left[1+A(K)\left(\dfrac{eV}{2k_B T}\right)^{2}\right]^{1-K}},~~\nonumber\\
f(x)=\frac{\beta+x^2}{1+x^2},~~~~~x=\dfrac{eV}{2k_{B}T}\dfrac{2}{\pi\zeta},~~~~~~~~~~~~~~
\end{eqnarray}
where $\delta{G}_{0}$ is independent of $T$ and $V$, $B(K,K)$ is the Euler beta function, and $A(K)=\pi^{-2}\Gamma(K)^{\frac{2}{1-K}}$.
\textcolor[rgb]{0.00,0.00,0.00}{Note that the model presented above does not imply any restrictions on the values of the Luttinger parameter $K$.}

Parameters $\beta$ and $\zeta$ in Eq.~(\ref{Eq:Thx3}) both origin from \emph{isotropic} and \emph{anisotropic parts} of the exchange interaction between the localized spins and the edge electrons~\cite{2016Vaeyrynen}. In general, $\beta$ and $\zeta$ are functions of temperature, which depend strongly on the spatial distribution of impurities along the QW growth direction and on the crystal symmetry of the 2D system~\cite{NewRef5,NewRef6}. Theoretical estimates for the simplest case~\cite{2016Vaeyrynen} predict that $\beta$ is well approximated by a constant of order 1 ($0.67<\beta(T)<0.83$) and $\zeta$ is small for $K\simeq 1$ ($2\rho J_0<\zeta(T)<3 (1-K)$). In the following we assume that both $\beta$ and $\zeta$ are independent of temperature.

As can be seen from Eq.~(\ref{Eq:Thx3}), the dependence of the backscattering current on the temperature and voltage drop has two well-separated crossover scales associated with the ratio $eV/k_{B}T$. The smaller scale, $eV/k_{B}T\sim\zeta$, is attributed to the dynamics of the localized spin in the presence of the effective field generated by the electron spin imbalance due to a finite voltage drop $V$~\cite{2016Vaeyrynen}. The crossover at the higher scale, $eV/k_{B}T\sim{1}$, occurs between the linear and weakly nonlinear regimes of the voltage dependencies of $\langle\delta{I}\rangle$ in Eq.~(\ref{Eq:Thx3}). Far away from this crossover, the edge conductance can be written in a simple form:
\begin{equation}
\label{Eq:Thx4}
G_\mathrm{edge}\sim\dfrac{1}{N}\left[\max\left(k_{B}T,eV\right)\right]^{2-2K}.
\end{equation}

\begin{figure}
\includegraphics[width=1.0\linewidth]{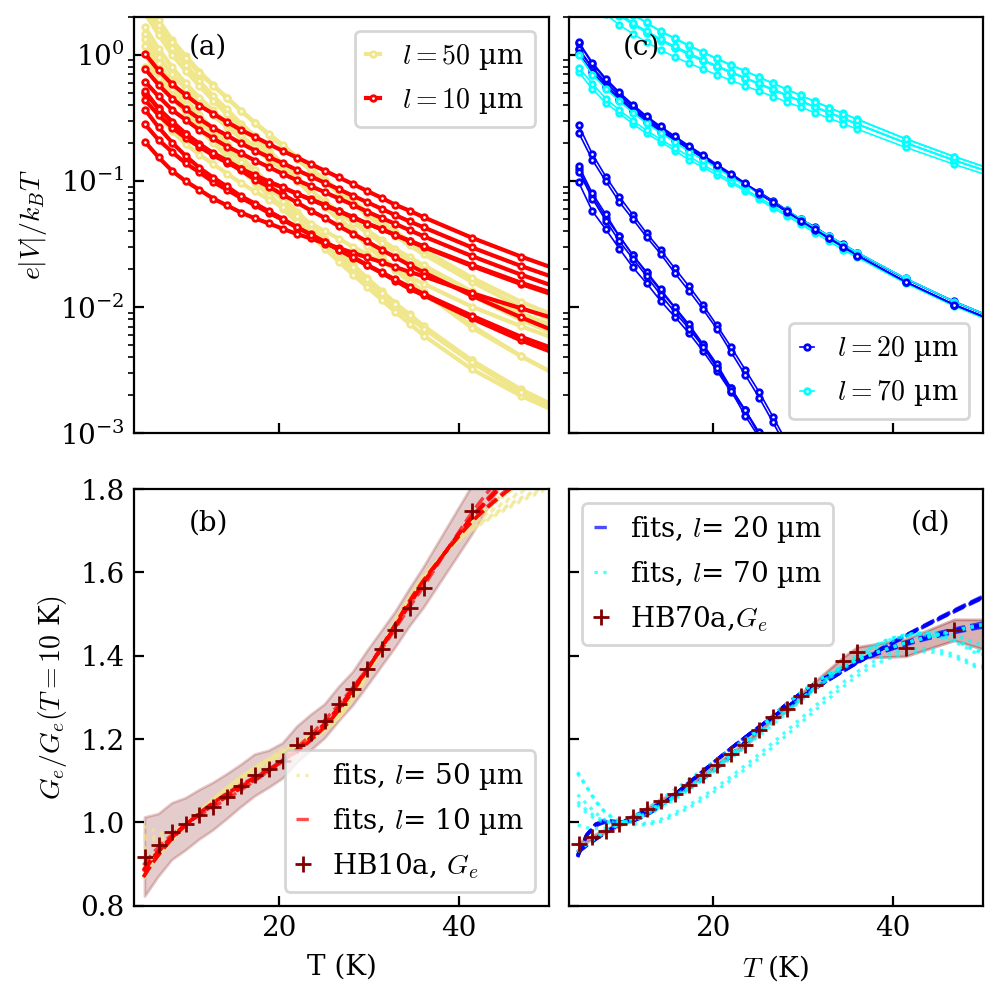}
\caption{\textcolor[rgb]{0.00,0.00,0.00}{(a) Solid lines: all measured voltage drops $e |V_{ij,kl}|/k_B T$ measured for \HBA{} between two adjacent probes, as a function of $T$. The voltages probing edges of length \qty{10}{\micro\metre} (\qty{50}{\micro\metre}) are highlighted in red (dark yellow). (b) Brown $+$ symbols: $G_e(T)$ for \HBA{}, normalized to its value at $10$~K. The shaded area is the error interval (1$\sigma$). The dashed red (dotted dark yellow) lines correspond to the fits based on Eqs.~(\ref{Eq:Thx1}) and~(\ref{Eq:Thx3})) of the voltages probing the \qty{10}{\micro\metre}-length (\qty{50}{\micro\metre}-length) edges shown in panel (a), evidencing the convergence of all the fits on $G_e(T)$. (c) The same as panel (a) but for \HBB{}. The voltages probing edges of length \qty{20}{\micro\metre} (\qty{70}{\micro\metre}) are highlighted in blue (cyan). (d) The brown $+$ symbols are experimental values of $G_e(T)$ for \HBB{}, normalized to its value at $10$~K. The dashed blue (dotted cyan) lines represent the fits of the voltages probing the \qty{20}{\micro\metre}-length (\qty{70}{\micro\metre}-length) edges shown in panel (c).}}
\label{fig:edge}
\end{figure}

We argue in the appendices that all currently available data in literature on the dependence of edge conductance on temperature and voltage-drop in InAs/Ga(In)Sb QW 2Ls~\cite{2015Li,2014Spanton,NewRef7} are consistent with the theoretical expectation of weakly interacting helical edge electrons with the backscattering due to localized magnetic moments of charge impurities. In the following, we discuss the temperature evolution of the edge conductivity in \HBA{} and \HBB{} devices made from 3L InAs/Ga$_{0.68}$In$_{0.32}$Sb QW (see Fig.~\ref{fig:structure}). 
The symbols in Fig.~\ref{fig:edge}(b,d) represent the experimental values of $G_e(T)$ for \HBA{} and \HBB{}, obtained by the multi-probe analysis of the transport data as a function of temperature.
The data are the same as the ones previously presented in Fig.~\ref{fig:trends},
taken on a limited temperature range $T < \qty{50}{\kelvin}$ 
to ensure negligible residual parasitic contributions. 
Here, $G_e$ is normalized to its value at \qty{10}{\kelvin} for clarity.
%
%
As the edge conductance $G_\mathrm{edge}$ is defined as $G_\mathrm{edge}=G_{e}/l$, this normalized form allows us to analyze the data in Fig.~\ref{fig:edge}(b,d) in terms of temperature dependence of the edge conductance. 
As seen from Figs.~\ref{fig:edge}(b,d), the edge conductance in \HBA{} and \HBB{} increases with temperature  up to \qty{50}{\kelvin}.

\textcolor[rgb]{0.00,0.00,0.00}
{Figure~\ref{fig:edge}(a,c) shows all the measured experimental ratios $e|V_{ij,kl}|/k_{B}T$ as a function of $T$ for adjacent probes $k$ and $l$. Here, we define $V_{ij,kl} = R_{ij,kl} I$, and  $I= \qty{10}{\nano\ampere}$. For each voltage $V_{ij,kl}$ and its temperature dependence, we evaluate $G_\mathrm{edge}$ from Eqs.~(\ref{Eq:Thx1}) and (\ref{Eq:Thx3}). For \HBA{}, the ratio $e|V_{ij,kl}|/k_BT$ stays in the range $5\times10^{-3}$ to $1$ over the whole temperature range when $V_{ij,kl}$ probes one of the edges of the length \qty{10}{\micro\metre}. These voltages are shown in Fig.~\ref{fig:edge}(a) as red lines. If we fit separately these specific voltages using the model based on Eqs.~(\ref{Eq:Thx1}) and (\ref{Eq:Thx3}), all these fits converge to the $G_e(T)$ curve, as shown in Fig.~\ref{fig:edge}(b) by the set of red lines. Moreover, the variance of the three fitting parameters $K$, $\beta$ and $\zeta$ over the different fits is rather small. The fits yield $K=0.89\pm 0.01$, $\beta=0.73\pm 0.03$, $\zeta=0.02\pm 0.01$. Repeating the same procedure for the edge channels of 50~$\mu$m long of \HBA{} device, we get $K=0.87\pm0.01$, $\beta=0.80\pm0.02$, and $\zeta=0.13\pm0.06$.}

\textcolor[rgb]{0.00,0.00,0.00}
{The same analysis was also performed for \HBB{} device, as shown in Fig.~\ref{fig:edge}(c,d). If we choose the voltages $V_{ij,kl}$ for the \qty{20}{\micro\metre}-length edges that do not exceed $k_{B}T/e$, the fits again all converge to the experimental curve $G_e(T)$, as shown in Fig.~\ref{fig:edge}(d). These fits yield $K=0.88\pm 0.04$, $\beta=1.1\pm 0.2$, and $\zeta=0.04\pm 0.02$. On the contrary, by using the voltage drops corresponding to the \qty{70}{\micro\metre} edges, the fits become poorer, as clear from Fig.~\ref{fig:edge}(b,d). This may indicate the emergence of an additional backscattering mechanism that is absent in the shorter edge channels. Clarification of the nature of this mechanism requires additional experimental research beyond the scope of this work. Nevertheless the fits for the 70~$\mu$m-channels still give values of $K$ corresponding to weak electron correlation. The fitting parameters for the different edge lengths \qtylist{10;20;50;70}{\micro\metre} are summarized in Table~\ref{tab:parameters}.}

\textcolor[rgb]{0.00,0.00,0.00}
{By construction, the multi-probe analysis assumes that $G_e$ is independent of voltage.
Consequently, our multiprobe analysis is eligible only in the regimes, where $G_e$ depends only on temperature but not on $eV$: (i) $\zeta<eV/k_BT<1$ if $\beta$ differs significantly from unity and (ii) $eV/k_BT<1$ for $\beta\sim1$. One can directly verify these limitations by means of Eqs.~(\ref{Eq:Thx1}),~(\ref{Eq:Thx3}) (see also Fig.~1 in Ref.~\onlinecite{2016Vaeyrynen}). As can be seen from Tab.~\ref{tab:parameters}, the obtained $\beta$ values are of the order of unity for all HB devices, thus justifying the use of the numerical model underlying the  multi-probe analysis (see Section~\ref{appendix:NLLSF}).}

\begin{table}
\caption{\label{tab:parameters} 
\textcolor[rgb]{0.00,0.00,0.00}
{A set of fitting parameters obtained for different edge channel lengths of \HBA{} and \HBB{} devices or extracted from the analysis of experimental data available in the literature~\cite{2015Li,2014Spanton,NewRef7} (see Appendix~\ref{appendix:LiteratuData}).}}
\begin{ruledtabular}
\begin{tabular}{c|c|c|c|c}
HB device & $l$~($\mu$m) & $K$ & $\beta$ & $\zeta\times{10}$ \\
\hline
\HBA{} & 10 & $0.89 \pm 0.01$ & $0.73 \pm 0.03$ & $0.2  \pm 0.1$   \\
 & 50 & $0.87 \pm 0.01$ & $0.80  \pm 0.02$ & $0.13 \pm 0.06$ \\
\hline
\HBB{} & 20 & $0.88 \pm 0.04$ & $1.1  \pm 0.2$  & $0.4 \pm 0.2 $  \\
 & 70 & $0.85 \pm 0.06$ & $1.2  \pm 0.3$  & $0.48 \pm 0.05$  \\
\hline
Ref.~\onlinecite{2015Li}  & 1.2 & 0.82\footnote[1]{Analysis of the experimental data at high voltage drops ($eV{\gg}k_{B}T$) on the basis of Eq.~(\ref{Eq:Thx4}).} & -- & -- \\
  & 1.2 & 0.84\footnote[2]{Analysis of the experimental data at high temperatures ($eV{\ll}k_{B}T$) on the basis of Eq.~(\ref{Eq:Thx4}).} & -- & -- \\
\hline
Ref.~\onlinecite{2014Spanton}  & 50 & $[0.77,0.97]$ & $[0,5]$ & $[0,0.3]$ \\
\hline
Ref.~\onlinecite{NewRef7}  & 12 & $0.986$ & $1.35$ & $0.06$
\end{tabular}
\end{ruledtabular}
\end{table}

Before concluding, let us discuss the strong difference in the parameters $\zeta$ and $\beta$ obtained from the analysis of experimental results for the HB devices made from QW 2L~\cite{NewRef7} (see Fig.~\ref{fig:edgeRRD} in Appendix F)  and 3L QW based on InAs/Ga$_{0.68}$In$_{0.32}$Sb, provided in Fig.~\ref{fig:edge}. The main difference between 3L QWs and QW 2Ls is the absence of strong structural inversion asymmetry (SIA), which allows not only to obtain a larger inverted bandgap~\cite{2018Krishtopenkoa}, but also to draw a definite conclusion about the edge state polarization. Indeed, if we neglect small effects associated with the absence of an inversion center in the unit cell of the bulk materials, then in the absence of SIA, the spin polarization of the edge states is oriented perpendicular to the QW plane. The latter is clear from the absence of the terms in low-energy Hamiltonian that mix different basic spin states of the QW at the $\Gamma$ point of the Brillouin zone~\cite{2018Krishtopenkoa,2006Bernevig}. The presence of SIA in QW 2Ls causes momentum-dependent terms to mix basis states with different spins~\cite{2008Liu}, making the orientation of the edge state polarization dependent of the momentum along the edge. Since both parameters $\beta$ and $\zeta$ arise from the exchange interaction between localized spins and edge electrons~\cite{2016Vaeyrynen}, they must depend on the orientation of the spin polarization of the edge electrons. Therefore, one can indeed expect their strong difference in InAs/GaInSb-based QW 2Ls and 3L QWs.

\section{Conclusion}
We have shown numerically and experimentally that by placing the lateral probes of a HB in close proximity to each other, one can detect edge currents as a sign reversal of the four probe resistances. We have also developed a numerical model to evaluate the relative contribution of edge and bulk currents in large-scale HB devices, taking into account most of the accessible four-probe configurations. 
This model, which can also be applied to other structures such as simple van der Pauw devices, fits remarkably well our experimental data obtained on HB structures based on inverted 3L InAs/GaInSb QWs designed to host quantum spin Hall insulator state.
In particular, we used the developed model 
(i) to show that bulk conduction is negligibly small in the best HB devices,
suggesting that edge quantization should be achieved in devices of smaller sizes; and
(ii) to separately evaluate the temperature dependencies of the bulk and edge current contributions for two HB devices with \qtylist{10;20;50;70}{\micro\metre}-long channels.
We have also shown that the temperature dependence of edge conductivity in both HB devices, as well as previous experimental results obtained in InAs/Ga(In)Sb QW 2Ls~\cite{2015Li,2014Spanton,NewRef7}, are well consistent with the theoretical expectation of weakly interacting helical edge electrons with the backscattering due to localized magnetic moments of charge impurities~\cite{2016Vaeyrynen}. We argue that these charge impurities are naturally associated with intrinsic Ga antisite defects, the presence of which in inverted InAs/Ga(In)Sb QWs was previously confirmed by means of terahertz photoluminescence~\cite{2018Krishtopenko,2019Krishtopenko}.

\begin{acknowledgments}
This work was supported by the Elite Network of Bavaria within the graduate program ``Topological Insulators'' and by the Occitanie region through the programs ``Terahertz Occitanie Platform'' and ``Quantum Technologies Key Challenge'' (TARFEP project). We also acknowledge financial support from the French Agence Nationale pour la Recherche through ``Demegras'' (ANR-19-GRF1-0006-03), ``HYBAT'' (ANR-21-ESRE-0026)
 and ``Cantor'' (ANR-23-CE24-0022) projects, 
and the DFG within project HO 5194/19-1 and through the W\"{u}rzburg-Dresden Cluster of Excellence on Complexity and Topology in Quantum Matter -- ct.qmat (EXC 2147, project-id 390858490). We thank Bertrand Mongellaz for his constant technical support in cryogenics.
\end{acknowledgments}



\appendix
\section{Validity of the Model for QSH states}
\label{appendix:QSH}

Let us assume that  the edges of the device host the QSH edge state.
A complete model of QSH states in the diffusive regime~\cite{2007Abanin,2019Gusev,2019Tagliacozzo} implies the use
of two phenomenological constants $\gamma$ and $g$, which
represent edge to edge and bulk to edge inverse scattering length, respectively, and the definition
of four local chemical potentials:
$\Psi_i$, the potentials in the bulk, and $\varphi_i$, the potentials of the helical states at one edge.
Here,  $i=1,2$ labels the states with different projections of the spins.

In this work, we largely simplify this model. We define a unique electrochemical potential $\varphi$ at the edge,
which is related to the potentials of 'virtual contacts' all along the edge, in which the counter-propagating edge states equilibrate.
The distance between the virtual contacts defines the edge to edge backscattering length.
Assuming fast spin relaxation in the bulk, $\Psi_1 \sim \Psi_2$ and we get the equation~\cite{2007Abanin}:
\begin{equation}
G_e \frac{\partial^2\varphi}{\partial x^2}=
g \left( \psi - \varphi \right),
\label{eq:edge_BC}
\end{equation}
where $x$ is the direction parallel the edge.
%
In the inner part of the bar, current conservation and Ohm's law yield the usual
Laplace equation $\Delta \Psi = 0$. The boundary condition
is imposed by current continuity:
\begin{equation}
\sigma_b~(\mathbf{n} \cdot \nabla \psi) = g \left( \psi - \varphi \right),
\label{eq:bulk_BC}
\end{equation}
where $\sigma_b$ is the bulk conductivity,
and $\mathbf{n}$ is the normal to the edge.
As clear, Eqs.~\ref{eq:edge_BC} and \ref{eq:bulk_BC} is the set of partial differential equations to be solved to determine $\psi$ and $\varphi$.  However, in the limit of high coupling:  $g \gg  G_e/L, \sigma_b/L^2$ where $L$ is the typical size of the device, we get $\varphi = \psi$ on the edge. The boundary condition for the bulk then simplifies as given in the main text.

\begin{figure*}
\includegraphics[width=1.0\linewidth]{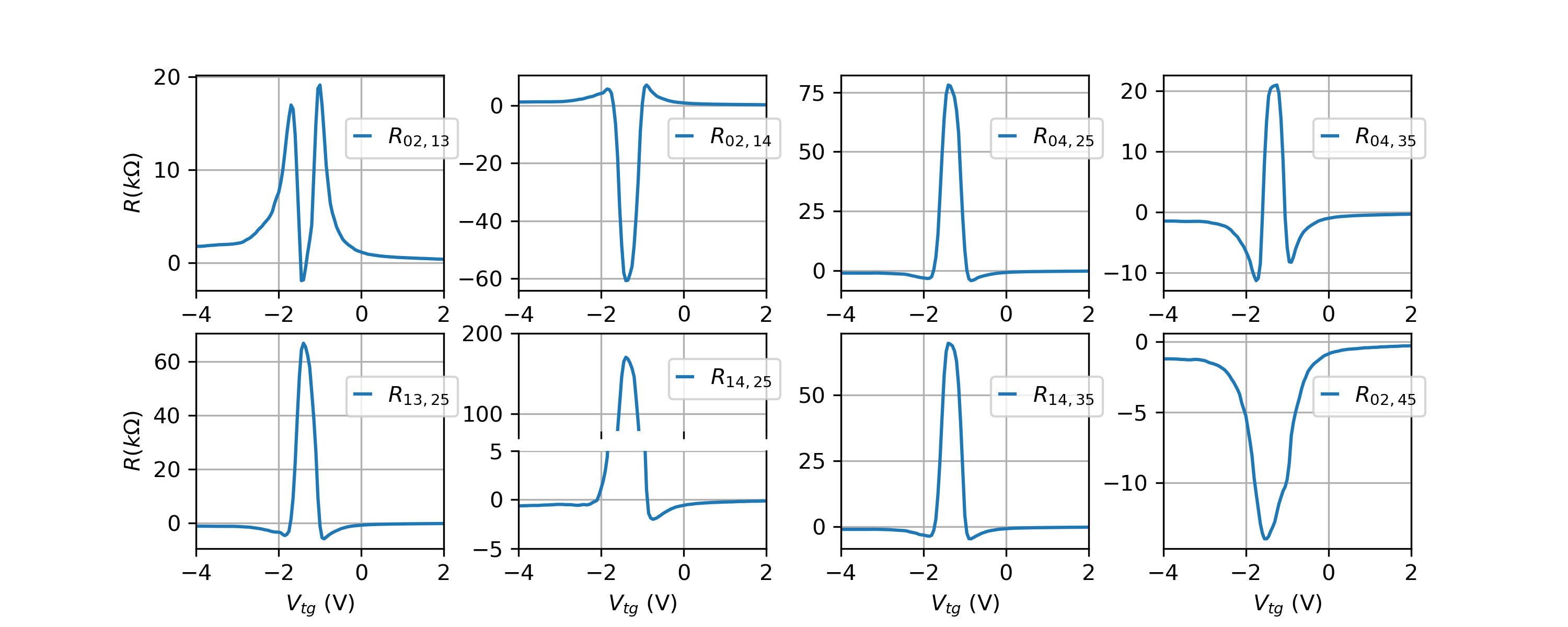}
\caption{\textcolor[rgb]{0.00,0.00,0.00}{The seven four-probe resistances measured at $T= \qty{1.7}{\kelvin}$ in \HBA{}, for which a sign inversion is expected and measured, and the resistance $R_{02,45}$ in which a double peak structure is evidenced. All data is the same as in Fig.~4 in the main text.}}
\label{fig:zoom_reversal}
\end{figure*}

\textcolor[rgb]{0.00,0.00,0.00}
{
\section{Enlargement of the HB10a resistances with a sign reversal}
\label{appendix:ZOOM}
Some of the fine features of the resistance peaks are difficult to spot in Fig. 4. 
Therefore, in Fig.~\ref{fig:zoom_reversal} we provide enlarged views of the most important four-probe resistances, including the seven resistances for which sign inversion is expected and measured, and also $R_{02,45}$, where a double peak structure is  visible.
}

\section{The nonlinear least-squares fit}
\label{appendix:NLLSF}
\textcolor[rgb]{0.00,0.00,0.00}{
The numerical procedure can be described as follows. The Hall bar is discretized onto a finite mesh by a finite difference method. For this study we used a mesh size a=\qty{1}{\micro\meter}. The two important parameters are the edge conductance and the bulk conductivity $G_e$ and $\sigma_b$.
The mesh gives a matrix $G$. The elements of the matrix $G$ are the conductance elements $G_{pq}$ between the nodes $p$ and $q$ which belong to the finite difference mesh.
From $G$, the equation to solve is the well-known formula for the multi-terminal conductors:
\begin{equation}
I_p=\sum_q  G_{pq}(V_p-V_q).
\end{equation}                                           
Here, $I_p$ is the current flowing from the external circuit into the node $p$, $V_q$ is the voltage at the node $q$. The $I_p$ are all zeros, except at the nodes corresponding to the source and drain. The voltages can then be determined by inversion of the $G$ matrix. }

\textcolor[rgb]{0.00,0.00,0.00}{
Once all the experimental four-probe resistances have been measured, this numerical model can be used to fit the data with a nonlinear least-squares method. For each top-gate voltage $V_{tg}$ , we minimize the chi-square $\chi^2$:
\begin{equation}
\chi^2=
\sum_{(ij,kl) \in \mathrm{configs}}
\frac{
\left[
R_{ij,kl}^\mathrm{exp} - R_{ij,kl}^\mathrm{model} (G_e,\sigma_b)\right] ^2
}
{\sigma_{ij,kl}^2},   
\end{equation}
where the sum is performed over the $N=45$ chosen configurations, $R_{ij,kl}^\mathrm{exp}$ are the experimental four-probe resistances, $R_{ij,kl}^\mathrm{model}\left(G_e,\sigma_b\right)$ the calculated four-probe resistances for the device geometry, $\sigma_{ij,kl\ }^2$ are the variances of the experimental resistance values.}

\textcolor[rgb]{0.00,0.00,0.00}{
For simplicity, we assume that all these variances are equal for a given gate voltage: $\sigma_{ij,kl\ }^2=\sigma^2$. The standard deviation of a unique resistance measurement with a lock-in technique is very small.  A larger error of a few percents comes from the necessity to perform a voltage sweep for each measured four-probe resistance. Hence, at each voltage sweep, the electrostatic environment of the 2D channel is slightly modified. For simplicity, we assume that the standard deviation $\sigma$ is equal to 5\% of the maximal resistance measured, for a given gate voltage. 
Then, the Hessian matrix is calculated:
\begin{multline}
H\left(G_e,\ \sigma_b\right)
=
\frac{1}{\sigma^2}\times\ \sum_{ij,kl\ \in\ configs}  \\ \left[
\begin{matrix}
\left(\frac{\partial R_{ij,kl}^{model}}{\partial G_e}\right)^2 & 
\frac{\partial R_{ij,kl}^{model}}{\partial G_e}\times\ \ \frac{\partial R_{ij,kl}^{model}}{\partial\sigma_b}\\
\frac{\partial R_{ij,kl}^{model}}{\partial G_e}\times \frac{\partial R_{ij,kl}^{model}}{\partial\sigma_b}&
\left(\frac{\partial R_{ij,kl}^{model}}{\partial\sigma_b}\right)^2\\
\end{matrix}
\right],
\end{multline}
and the covariance matrix is defined as the inverse of the Hessian, $C=H^{-1}$. The variances of the parameters are given by:  $\sigma_{G_e}^2=C_{G_e,G_e}$,  and  $\sigma_{\sigma_b}^2=C_{\sigma_b,\sigma_b}$.
Typical results are reproduced below in Fig.~\ref{fig:chi2}, where we also indicate 
the standard deviation of the data, and the reduced $\chi_\nu^2$ defined as $\chi_\nu^2=\chi^2/\nu$, where $\nu=45-2$ are the degrees of freedom.
This quantity $\chi_\nu^2$  is often used as a measure of the suitability of the chosen model for fitting to the given data, {\it i.e.}, the condition $\chi_\nu^2 \simeq 1$ is taken to indicate that the fit is meaningful.
\begin{figure}
\includegraphics[width=1.0\linewidth]{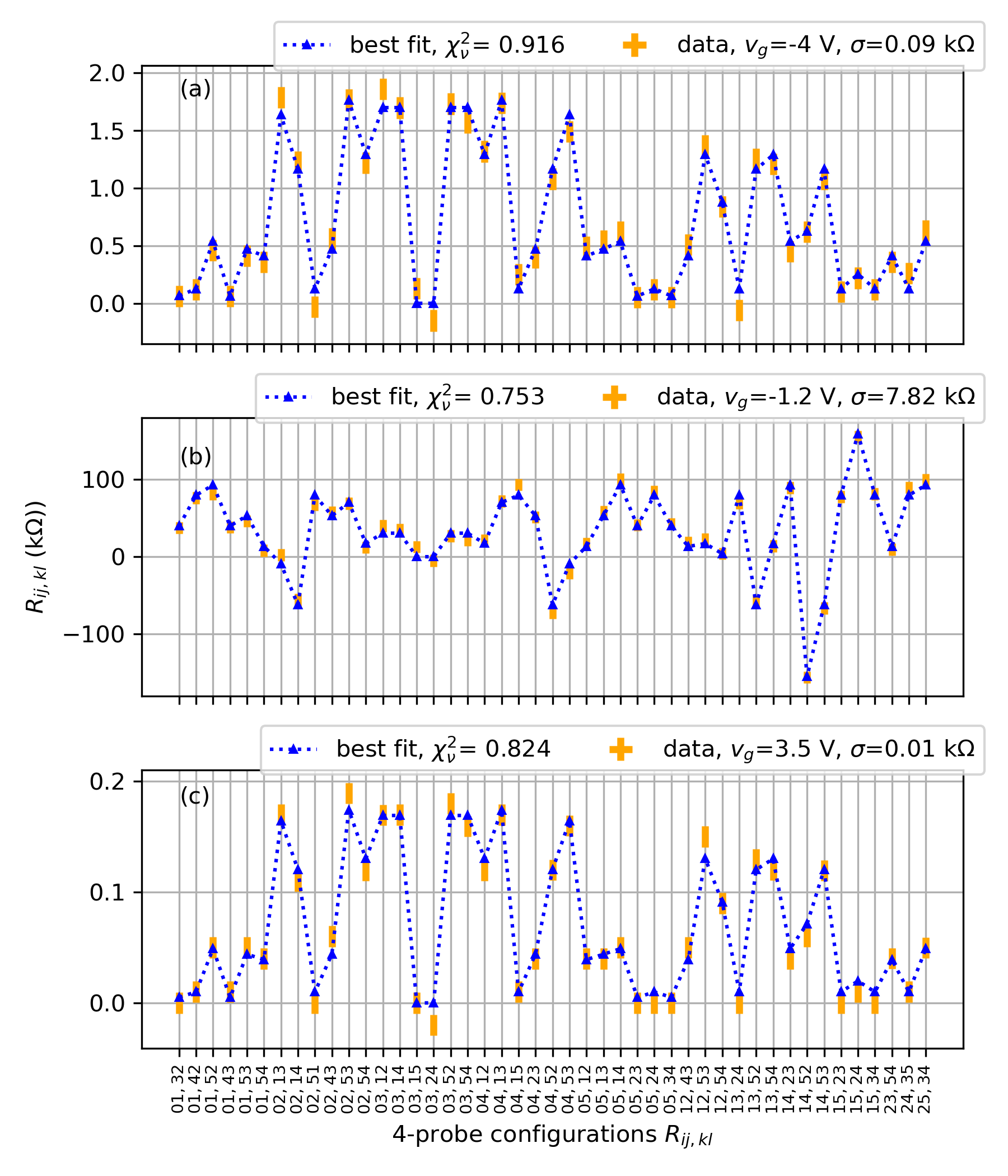}
\caption{\textcolor[rgb]{0.00,0.00,0.00}{(a) Four-probe resistances from \HBA{} at $T = \qty{1.7}{\kelvin}$ (the same data as in Fig.~4 in the main text) at three different gate voltages: (a) $V_{tg}= \qty{-4}{\volt}$ (valence band), (b) $V_{tg}= \qty{-1.2}{\volt}$ (peak resistance) and (c) $V_{tg}= \qty{3.5}{\volt}$ (conduction band). The experimental error (1$\sigma$) is indicated by orange bars. Best fits are shown with blue triangles. The signs of the four-probe resistances have been adjusted so that all resistances are positive for bulk conduction. The sign reversal is then better evidenced in panel (b).}}
\label{fig:chi2}
\end{figure}
}

\section{Decoupling edge and bulk contributions}
\begin{figure}[b]
\includegraphics[width=1.0\linewidth]{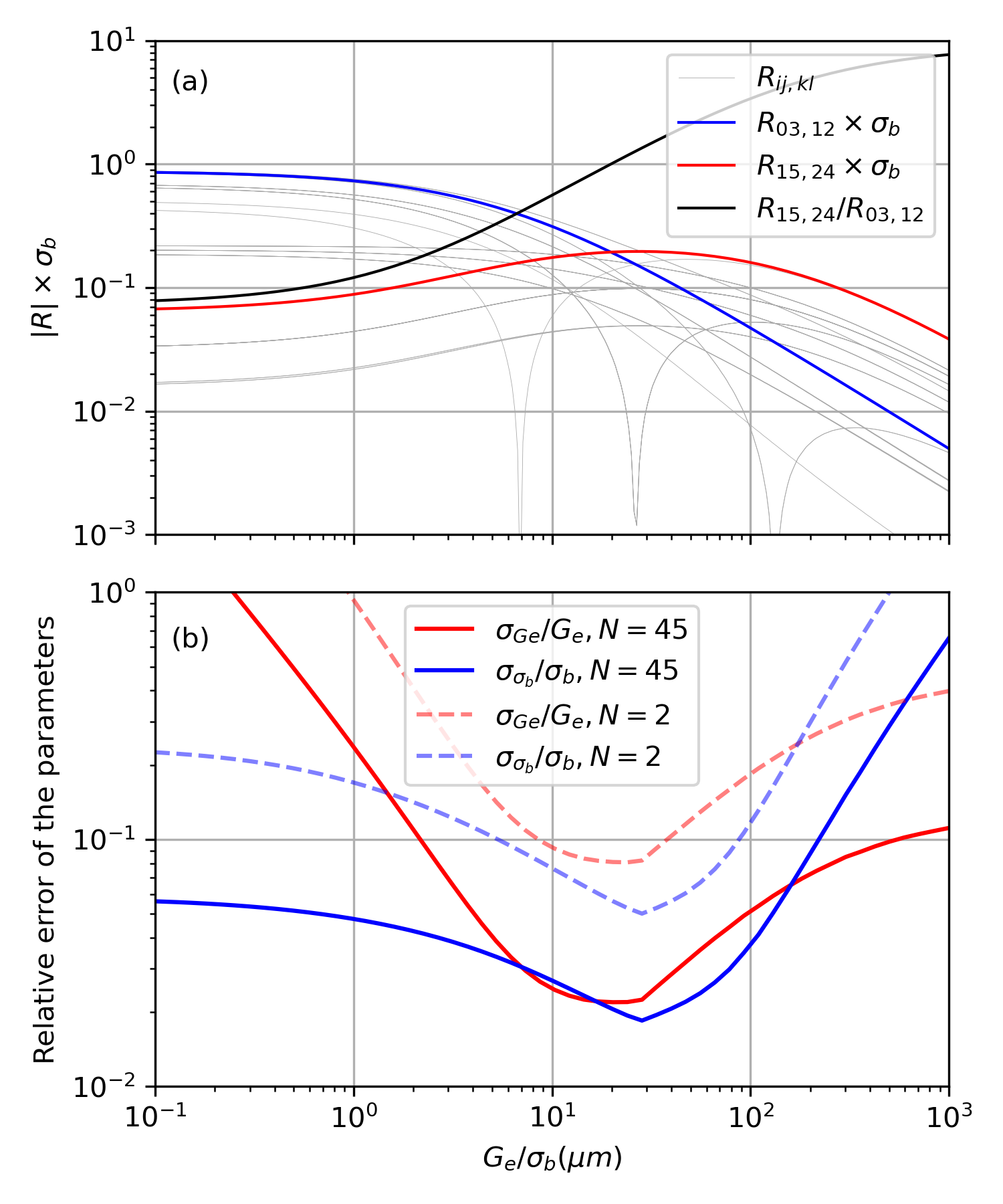}
\caption{\textcolor[rgb]{0.00,0.00,0.00}{(a) Evolution of the four-probes resistances $|R_{ij,kl}|\times\sigma_b$ as a function of $G_e/\sigma_b$ for device HB10. $R_{03,12} \times \Gb{}$ is indicated by a blue line, $R_{15,24}\times\sigma_b$ by a red line. The other resistances are not labeled and reported as thin gray lines. The ratio $R_{15,24}/R_{03,12}$ is also reported as a black gray line. (b) Relative error $\sigma_{Ge}/G_e$ (red line) and $\sigma_{\sigma b} /\sigma_b$ (blue line) for the nonlinear least-squares fit with i) only $N=2$ resistances $R_{15,24}$ and $R_{03,12}$ (dashed lines) and ii) all the $N=45$ resistances (solid lines).}}
\label{fig_model_HB10_GeGb}
\end{figure}

\begin{figure}
\includegraphics[width=1.0\linewidth]{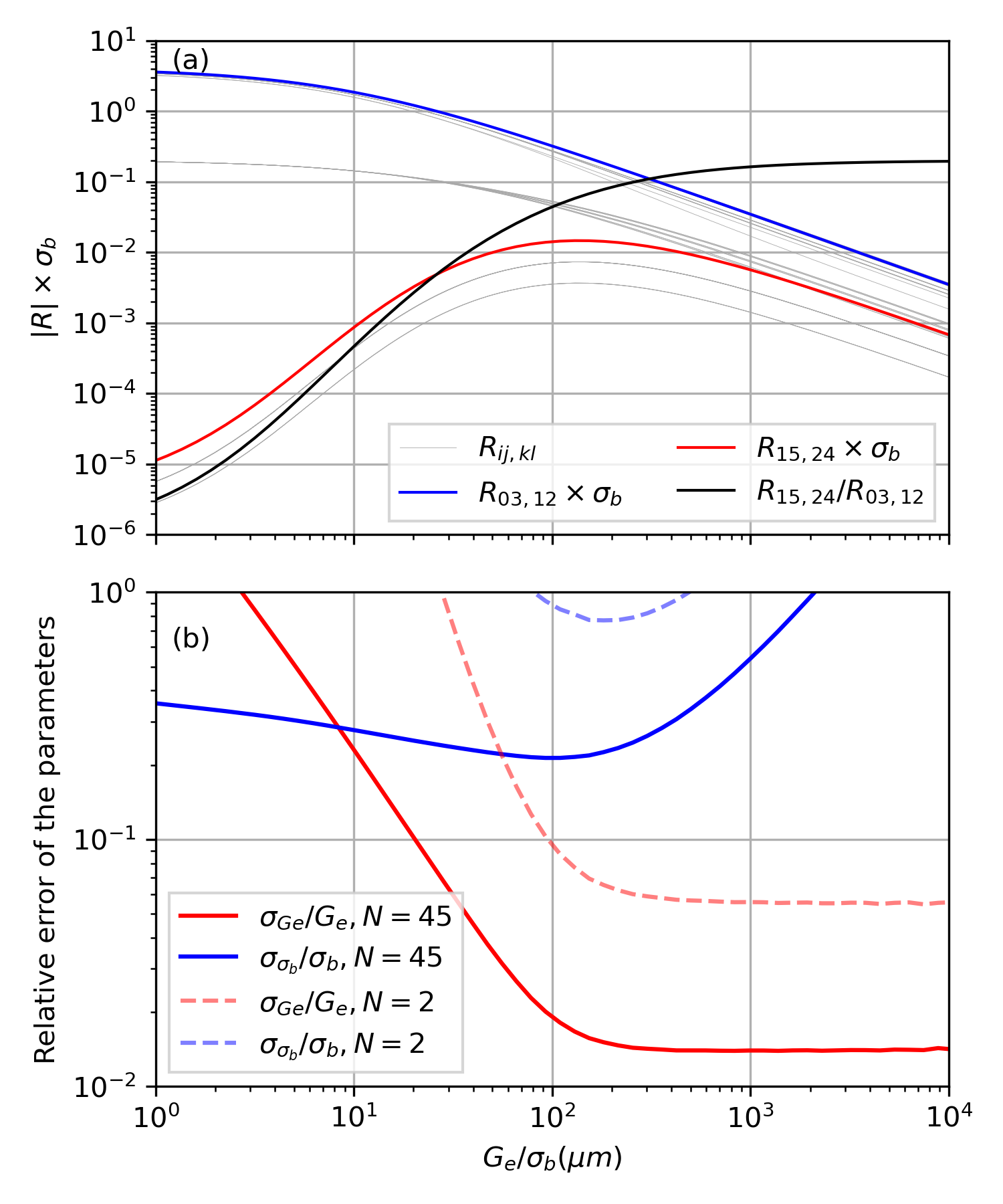}
\caption{\textcolor[rgb]{0.00,0.00,0.00}{(a) Evolution of the four-probes resistances $|R_{ij,kl}|\times\sigma_b$ as a function of $G_e/\sigma_b$ for device HB70. $R_{03,12}\times\sigma_b$ is indicated by a blue line, $R_{15,24} \times \sigma_b$ by a red line. The other resistances are not labeled and reported as thin gray lines. The ratio $R_{15,24}/R_{03,12}$ is also reported as a thick black line. (b) Relative error $\sigma_{Ge}/G_e$ (red line) and $\sigma_{\sigma b} /\sigma_b$ (blue line) for the nonlinear least-squares fit with i) only $N=2$ resistances $R_{15,24}$ and $R_{03,12}$ (dashed lines) and ii) all the $N=45$ resistances (solid lines).}}
\label{fig_model_HB70_GeGb}
\end{figure}

Let us illustrate here what are the conditions to determine the two components $G_e$ and $\sigma_b$ with reasonable error bars.
Fig.~\ref{fig_model_HB10_GeGb} shows the evolution of the absolute value of the different resistances $R_{ij,kl}$, normalized by the bulk conductance $\Gb{}$, as a function of $\Ge{}/\Gb{}$ for HB10. The two configurations
$R_{13,12}$ and $R_{15,24}$ are indicated by colored thick lines. The other non-null configurations
are indicated as thin gray lines, and the ratio $R_{15,24}/R_{03,12}$ is also reported as a thick black line. This ratio varies of two decades over a rather large $\Ge{}/\Gb{}$ interval, from $1\cdot10^{-1}$~$\mu$m to $1\cdot10^{3}$~$\mu$m. The situation is similar for HB70, with the main difference that $R_{15,24}/R_{03,12}$ varies
on more than six decades in this case, as shown in Fig.~\ref{fig_model_HB70_GeGb}.
We conclude that for both geometries, the analysis of $R_{15,24}$ and $R_{03,12}$
allows for a good estimate of both $G_e$ and $\Gb{}$ on several decades of the ratio $\Ge{}/\Gb{}$.
Only when
$G_e/\Gb{}<1\cdot10^{-1}$~$\mu$m ($G_e/\Gb{}>1\cdot10^{3}$~$\mu$m), $G_e$ ($\Gb{}$) cannot be evaluated precisely.

\textcolor[rgb]{0.00,0.00,0.00}{From this analysis it is clear that both $G_e$ and $\Gb{}$ can be determined from only $R_{13,12}$ and $R_{15,24}$ but this way has some limitations: 
i) it is not possible to determine
if the model ({\it i.e., the fitting function}) under consideration does faithfully represent the data, because
the best test for this is a low value of $\chi_\nu^2$. The latter however is is not defined with only two points and two parameters ($\nu=0$);
ii) the determination of both $\sigma_b$ and $G_e$ is much less precise with only $N=2$ parameters than with $N=45$. Generally speaking,
the variance of the parameters varies as $1/N$.
This last point is clearly evidenced in Fig.~\ref{fig_model_HB10_GeGb}(b) and Fig.~\ref{fig_model_HB70_GeGb}(b), where the relative standard deviation is shown for $G_e$ and $\sigma_b$, for $N=2$ and $N=45$, and for both HB10 and HB70 geometries.
The experimental standard deviation $\sigma$ was set at 5\% of the maximum resistances measured.
It is seen that the error is significantly reduced only when all the resistances are taken under consideration.
}

\textcolor[rgb]{0.00,0.00,0.00}{
\section{On the device uniformity}
{As seen from the data set for HB10a at $T=1.7$~K presented in Fig.~\ref{fig:model_HB10_45conf} in the main text, the fit at the maximum value of the resistivity peak is not perfect for some resistances like $R_{15,24}$.}
Another fit can be attempted in which, for simplicity, the low bulk conductance is neglected but the 
{\emph{six}} edge conductances are taken as 
{\emph{six}} independent fit parameters (instead of a single parameter 
{used in the main text}). 
{As a result, w}e get the following scattering lengths (assuming $\lambda=\left(\frac{h}{e^2}\right)G_e$): 
$\lambda_{01}= \qty{8.1}{\micro\metre}$, 
$\lambda_{12}= \qty{3.2}{\micro\metre}$,
$\lambda_{23}= \qty{6.4}{\micro\metre}$,
$\lambda_{34}= \qty{6.3}{\micro\metre}$, 
$\lambda_{45}= \qty{6.1}{\micro\metre}$, and 
$\lambda_{50}= \qty{7.1}{\micro\metre}$.
This gives $\lambda= \qty{6(2)}{\micro\metre}$, in agreement with $\lambda= \qty{5}{\micro\metre}$ given in the main text.
}
\section{Fitting data available in literature}
\label{appendix:LiteratuData}
Let us demonstrate that the model explains well the experimental results on the temperature and voltage-drop dependence of edge conductance reported previously~\cite{2015Li,2014Spanton,NewRef7}. Indeed, as shown by V\"{a}yrynen~\emph{et al.}~\cite{2016Vaeyrynen}, Eq.~(\ref{Eq:Thx4}) provides an alternative explanation of the results by Li~\emph{et~al.}~\cite{2015Li}, interpreted previously as the observation of 1D helical Luttinger Liquid with $K\simeq{1/4}$ in a HB bar of length \qty{1.2}{\micro\metre} made from inverted InAs/GaSb QW 2L. Indeed, matching the observed dependence $G_\mathrm{edge}\propto{V}^{0.37}$ in the regime $eV>k_{B}T$ with Eq.~(\ref{Eq:Thx4}) leads to $K=0.82$, which shows that in the presence of many impurities, even moderately weak interactions yields to the power law voltage dependence of the edge conductance. Analyzing in a similar way the temperature dependence $G_\mathrm{edge}\propto{T}^{0.32}$, when $eV<k_{B}T$, one can get $K=0.84$. Unfortunately, in short channels with $l <\lambda$ (where $\lambda$ is the backscattering length), it is not possible to distinguish a strongly correlated regime ($K\simeq{0.25}$) with a small amount of impurities from a weak-interaction regime ($K=0.82$) with a large $N$ based on the experimental data of Ref.~\cite{2015Li}.

\begin{figure}
\includegraphics[width=1.0\linewidth]{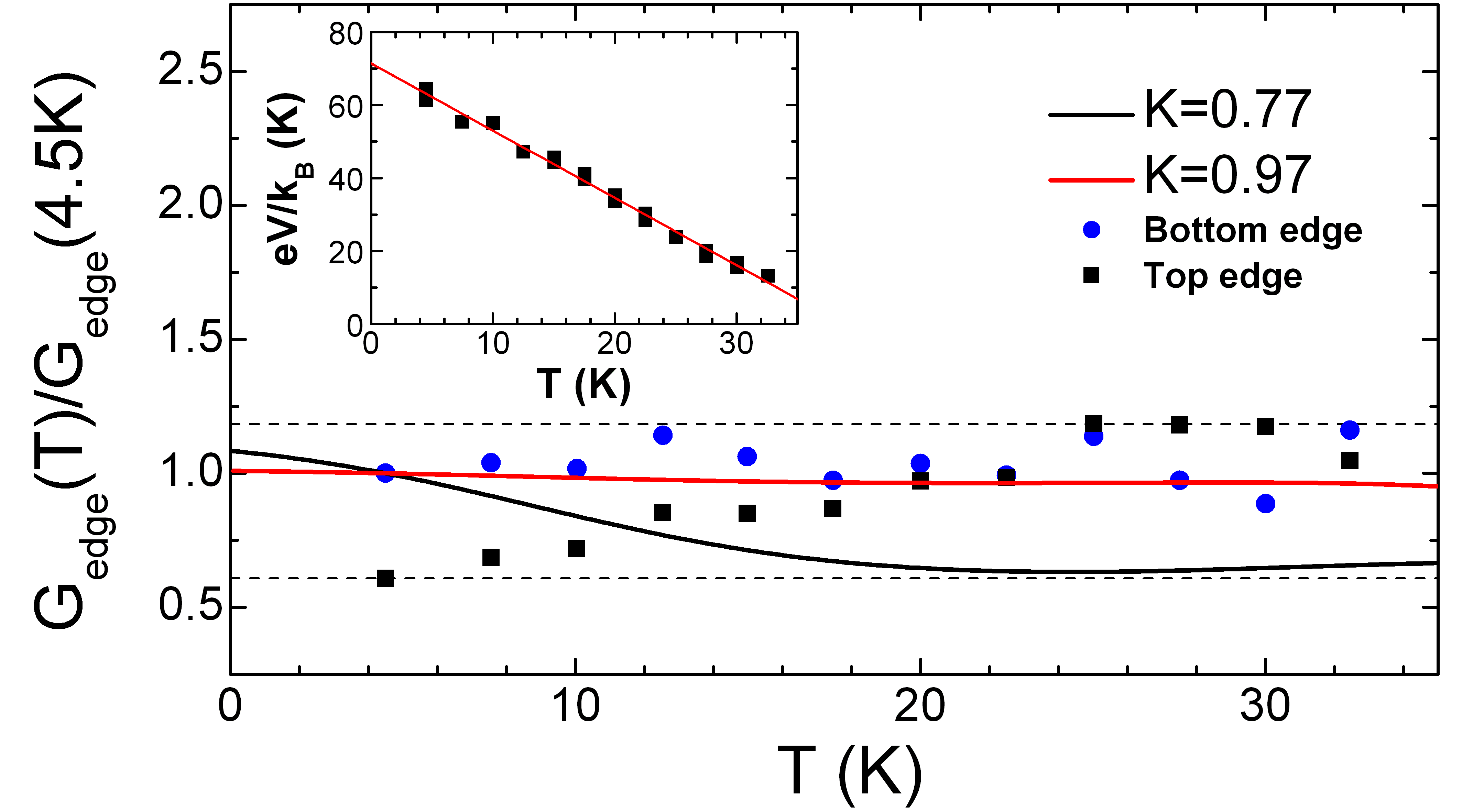}
\caption{Temperature dependence of the edge conductance in 50-$\mu$m-long channels of inverted InAs/GaSb QW 2L, normalized to the value at $4.5$~K (calculated from the data of Fig.~3 in Ref.~\cite{2014Spanton}). Symbols correspond to the experimental data, while black and red curves represent the fits at $K=0.77$ and $K=0.97$, respectively. Other parameters were chosen as $\Delta=3$~meV, $\zeta=0.03$ and $\beta=0.75$. The insets show the temperature dependence of the voltage drop in the edge channel, calculated from the data in Fig.~2(f) in Ref.~\cite{2014Spanton}. The red line in the inset represents the linear fit.}
\label{fig:Spanton2014}
\end{figure}

Alternatively, experimental results by Spanton~\emph{et~al.}~\cite{2014Spanton}, obtained by superconducting quantum interference device (SQUID) microscopy, have revealed that in 50-$\mu$m-long HB channels also made from inverted InAs/GaSb QW 2L, the edge resistance is independent of temperature in the range of up to $30$~K within the limits of experimental resolution. The latter, however, was quite mediocre and led to a wide spread of experimental values of edge resistance. The symbols in Fig.~\ref{fig:Spanton2014} represent experimental edge conductance values, obtained from the data of Fig.~3 in Ref.~\cite{2014Spanton}. The horizontal dashed curves that define the scattering range of experimental values of $G_\mathrm{edge}$ actually correspond to the limits of experimental resolution.

Using a current value of 10~nA and the resistance values in the bandgap region (see Fig.~2(f) in Ref.~\cite{2014Spanton}), one can estimate the voltage drop in the edge channels as a function of temperature. 
The results of this estimate are summarized in the inset of Fig.~\ref{fig:Spanton2014}.
The voltage drop has a linear dependence on temperature
and $eV \gtrapprox k_B T$.
Therefore, Eq.~(\ref{Eq:Thx4}) can explain the temperature independent behavior of $G_\mathrm{edge}$ obtained by Spanton~\emph{et~al.}~\cite{2014Spanton}.
A more accurate analysis based on Eqs.~(\ref{Eq:Thx1}) and (\ref{Eq:Thx3}) shows that in the scattering range of experimental values, $G_\mathrm{edge}$ can be well fitted by any sets of parameters $\zeta$, $\beta$ and $K$ from the intervals: $\zeta\in[0,0.03]$, $\beta\in[0,5]$ and $K\in[0.77,0.97]$ (see Fig.~\ref{fig:Spanton2014}). Estimating $T^{*}$ in Eq.~(\ref{Eq:Thx2}) based on these values gives $T^{*}<9.5$~mK, which fully ensures the fulfillment of the criterion $T>T^{*}$. Importantly, the range of $K\in[0.77,0.97]$ includes the values obtained above by analyzing the results by Li~\emph{et~al.}~\cite{2015Li}.

\begin{figure}
\includegraphics[width=1.0\linewidth]{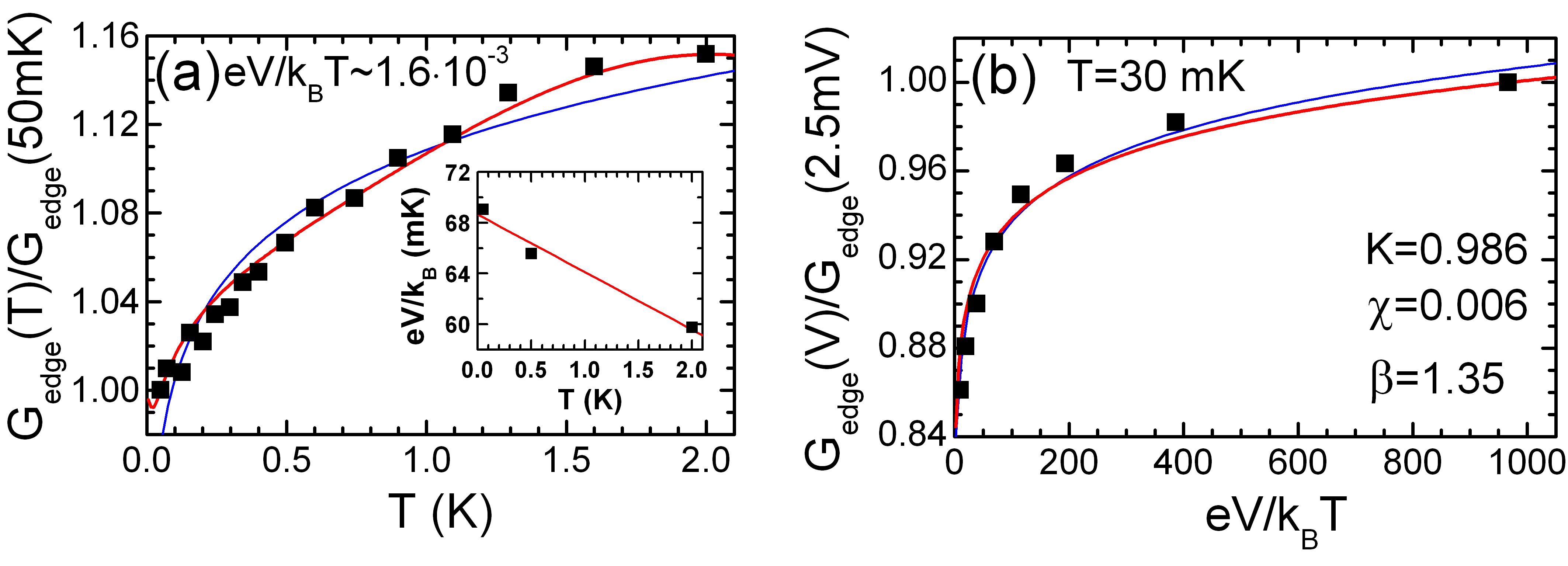}
\caption{(a) Temperature dependence of $G_\mathrm{edge}$ in a 12-\unit{\micro\metre}-long HB made from inverted InAs/Ga$_{0.68}$In$_{0.32}$Sb QW 2L, normalized to the value at $50$~mK. The symbols represent the data from the inset to Fig.~2(b) in Ref.~\cite{NewRef7}. The insets show the temperature dependence of the voltage drop, extracted from the longitudinal resistance $R_{xx}$ in the bandgap region (see Fig.~2(b) in Ref.~\cite{NewRef7}), and its linear fit. (b)~Voltage-drop dependence of $G_\mathrm{edge}$ at $T=30$~mK extracted from the zero-magnetic field values in Fig.~3(c) in Ref.~\cite{NewRef7}. The red curves in both panels correspond to the fit based on Eqs.~(\ref{Eq:Thx1}) and (\ref{Eq:Thx3}) with the \emph{same set} of parameters: $\Delta=20$~meV, $\zeta=0.006$, $\beta=1.35$ and $K=0.986$. The blue curves represents the power law fits with Eq.~(\ref{Eq:Thx4}) resulting in (a) $K=0.979$ and (b) $K=0.984$.}
\label{fig:edgeRRD}
\end{figure}

Since the bandgap in inverted InAs/GaSb QW 2Ls is very small, in order to reduce the residual bulk conductivity in the HB devices, the InAs/GaSb interface is usually doped by silicon~\cite{2015Li,2014Spanton}. As the bandgap increases, for example in InAs/GaInSb QW 2Ls, it is expected that the contribution of residual bulk conductivity should also decrease. Figure~\ref{fig:edgeRRD} shows temperature and drop-voltage dependence of $G_\mathrm{edge}$, extracted from the transport measurements by Li~\emph{et al.}~\cite{NewRef7} of a
12-\unit{\micro\metre}-long HB made from inverted InAs/Ga$_{0.68}$In$_{0.32}$Sb QW 2L. Note that it is not possible to estimate the contribution of residual bulk conductivity based on the available data~\cite{NewRef7}. One can only assume that it is small in a 2L QW with a bandgap of about $20$~meV at $T<2$~K.

As clear from Fig.~\ref{fig:edgeRRD}(a), the temperature dependence of the edge conductance corresponds to a regime where $eV{\ll}k_{B}T$. The latter can be understood by analyzing the dependence of the longitudinal resistance on temperature presented in Fig.~2(b) in Ref.~\cite{NewRef7}, taking into account the $0.1$~nA current used in the measurements (see also the inset to Fig.~\ref{fig:edgeRRD}(a)). A power law fit of the temperature dependence with Eq.~(\ref{Eq:Thx4}) yields $K=0.979$, which indicates the weakness of the electron-electron interaction in the helical edge channel. On the contrary, the voltage-drop dependence of the edge conductance at $T=30$~mK in Fig.~\ref{fig:edgeRRD}(b) represents a regime where $eV{\gg}k_{B}T$. By approximating the experimental $G_\mathrm{edge}$ values by a power-law dependence on the voltage drop given by Eq.~(\ref{Eq:Thx4}), we get a close value of $K=0.984$. A more accurate analysis shows that the experimental dependencies of $G_\mathrm{edge}$ on temperature and voltage drop, measured in the same HB device, can be well fitted by the theoretical dependencies based on Eqs.~(\ref{Eq:Thx1}) and (\ref{Eq:Thx3}) with the \emph{same set} of parameters: $\Delta=20$~meV, $\zeta=0.006$, $\beta=1.35$ and $K=0.986$. Estimates based on these values give $T^{*}<1$~$\mu$K, which certainly ensures that condition $T>T^{*}$ is satisfied throughout the entire temperature range used in the measurements. Finally, we note that the temperature dependence of $G_\mathrm{edge}$ in Fig.~\ref{fig:edgeRRD}(a) is also attributed to the proximity to the small-scale crossover $eV/k_{B}T\zeta=0.77$ ignored by Eq.~(\ref{Eq:Thx4}). This explains the better fitting based on Eqs.~(\ref{Eq:Thx1}) and (\ref{Eq:Thx3}).

%


\begin{thebibliography}{62}%
\makeatletter
\providecommand \@ifxundefined [1]{%
 \@ifx{#1\undefined}
}%
\providecommand \@ifnum [1]{%
 \ifnum #1\expandafter \@firstoftwo
 \else \expandafter \@secondoftwo
 \fi
}%
\providecommand \@ifx [1]{%
 \ifx #1\expandafter \@firstoftwo
 \else \expandafter \@secondoftwo
 \fi
}%
\providecommand \natexlab [1]{#1}%
\providecommand \enquote  [1]{``#1''}%
\providecommand \bibnamefont  [1]{#1}%
\providecommand \bibfnamefont [1]{#1}%
\providecommand \citenamefont [1]{#1}%
\providecommand \href@noop [0]{\@secondoftwo}%
\providecommand \href [0]{\begingroup \@sanitize@url \@href}%
\providecommand \@href[1]{\@@startlink{#1}\@@href}%
\providecommand \@@href[1]{\endgroup#1\@@endlink}%
\providecommand \@sanitize@url [0]{\catcode `\\12\catcode `\$12\catcode
  `\&12\catcode `\#12\catcode `\^12\catcode `\_12\catcode `\%12\relax}%
\providecommand \@@startlink[1]{}%
\providecommand \@@endlink[0]{}%
\providecommand \url  [0]{\begingroup\@sanitize@url \@url }%
\providecommand \@url [1]{\endgroup\@href {#1}{\urlprefix }}%
\providecommand \urlprefix  [0]{URL }%
\providecommand \Eprint [0]{\href }%
\providecommand \doibase [0]{https://doi.org/}%
\providecommand \selectlanguage [0]{\@gobble}%
\providecommand \bibinfo  [0]{\@secondoftwo}%
\providecommand \bibfield  [0]{\@secondoftwo}%
\providecommand \translation [1]{[#1]}%
\providecommand \BibitemOpen [0]{}%
\providecommand \bibitemStop [0]{}%
\providecommand \bibitemNoStop [0]{.\EOS\space}%
\providecommand \EOS [0]{\spacefactor3000\relax}%
\providecommand \BibitemShut  [1]{\csname bibitem#1\endcsname}%
\let\auto@bib@innerbib\@empty
\bibitem [{\citenamefont {van~der Pauw}(1958)}]{1958Pauw}%
  \BibitemOpen
  \bibfield  {author} {\bibinfo {author} {\bibfnamefont {L.~J.}\ \bibnamefont
  {van~der Pauw}},\ }\bibfield  {title} {\bibinfo {title} {A method of
  measuring specific resistivity and {H}all effect of discs of arbitrary
  shape},\ }\href@noop {} {\bibfield  {journal} {\bibinfo  {journal} {Philips
  Res. Repts}\ }\textbf {\bibinfo {volume} {13}},\ \bibinfo {pages} {1}
  (\bibinfo {year} {1958})}\BibitemShut {NoStop}%
\bibitem [{\citenamefont {Kane}\ and\ \citenamefont {Mele}(2005)}]{2005Kane}%
  \BibitemOpen
  \bibfield  {author} {\bibinfo {author} {\bibfnamefont {C.~L.}\ \bibnamefont
  {Kane}}\ and\ \bibinfo {author} {\bibfnamefont {E.~J.}\ \bibnamefont
  {Mele}},\ }\bibfield  {title} {\bibinfo {title} {Quantum spin {H}all effect
  in graphene},\ }\href {https://doi.org/10.1103/physrevlett.95.226801}
  {\bibfield  {journal} {\bibinfo  {journal} {Phys. Rev. Lett.}\ }\textbf
  {\bibinfo {volume} {95}},\ \bibinfo {pages} {226801} (\bibinfo {year}
  {2005})}\BibitemShut {NoStop}%
\bibitem [{\citenamefont {K{\"o}nig}\ \emph {et~al.}(2007)\citenamefont
  {K{\"o}nig}, \citenamefont {Wiedmann}, \citenamefont {Brune}, \citenamefont
  {Roth}, \citenamefont {Buhmann}, \citenamefont {Molenkamp}, \citenamefont
  {Qi},\ and\ \citenamefont {Zhang}}]{2007Konig}%
  \BibitemOpen
  \bibfield  {author} {\bibinfo {author} {\bibfnamefont {M.}~\bibnamefont
  {K{\"o}nig}}, \bibinfo {author} {\bibfnamefont {S.}~\bibnamefont {Wiedmann}},
  \bibinfo {author} {\bibfnamefont {C.}~\bibnamefont {Brune}}, \bibinfo
  {author} {\bibfnamefont {A.}~\bibnamefont {Roth}}, \bibinfo {author}
  {\bibfnamefont {H.}~\bibnamefont {Buhmann}}, \bibinfo {author} {\bibfnamefont
  {L.~W.}\ \bibnamefont {Molenkamp}}, \bibinfo {author} {\bibfnamefont {X.-L.}\
  \bibnamefont {Qi}},\ and\ \bibinfo {author} {\bibfnamefont {S.-C.}\
  \bibnamefont {Zhang}},\ }\bibfield  {title} {\bibinfo {title} {Quantum spin
  {H}all insulator state in {HgTe} quantum wells},\ }\href
  {https://doi.org/10.1126/science.1148047} {\bibfield  {journal} {\bibinfo
  {journal} {Science}\ }\textbf {\bibinfo {volume} {318}},\ \bibinfo {pages}
  {766} (\bibinfo {year} {2007})}\BibitemShut {NoStop}%
\bibitem [{\citenamefont {Roth}\ \emph {et~al.}(2009)\citenamefont {Roth},
  \citenamefont {Brüne}, \citenamefont {Buhmann}, \citenamefont {Molenkamp},
  \citenamefont {Maciejko}, \citenamefont {Qi},\ and\ \citenamefont
  {Zhang}}]{2009Roth}%
  \BibitemOpen
  \bibfield  {author} {\bibinfo {author} {\bibfnamefont {A.}~\bibnamefont
  {Roth}}, \bibinfo {author} {\bibfnamefont {C.}~\bibnamefont {Brüne}},
  \bibinfo {author} {\bibfnamefont {H.}~\bibnamefont {Buhmann}}, \bibinfo
  {author} {\bibfnamefont {L.~W.}\ \bibnamefont {Molenkamp}}, \bibinfo {author}
  {\bibfnamefont {J.}~\bibnamefont {Maciejko}}, \bibinfo {author}
  {\bibfnamefont {X.-L.}\ \bibnamefont {Qi}},\ and\ \bibinfo {author}
  {\bibfnamefont {S.-C.}\ \bibnamefont {Zhang}},\ }\bibfield  {title} {\bibinfo
  {title} {Nonlocal transport in the quantum spin {H}all state},\ }\href
  {https://doi.org/10.1126/science.1174736} {\bibfield  {journal} {\bibinfo
  {journal} {Science}\ }\textbf {\bibinfo {volume} {325}},\ \bibinfo {pages}
  {294} (\bibinfo {year} {2009})}\BibitemShut {NoStop}%
\bibitem [{\citenamefont {Knez}\ \emph {et~al.}(2011)\citenamefont {Knez},
  \citenamefont {Du},\ and\ \citenamefont {Sullivan}}]{2011Knez}%
  \BibitemOpen
  \bibfield  {author} {\bibinfo {author} {\bibfnamefont {I.}~\bibnamefont
  {Knez}}, \bibinfo {author} {\bibfnamefont {R.-R.}\ \bibnamefont {Du}},\ and\
  \bibinfo {author} {\bibfnamefont {G.}~\bibnamefont {Sullivan}},\ }\bibfield
  {title} {\bibinfo {title} {Evidence for helical edge modes in inverted
  {InAs}/{GaSb} quantum wells},\ }\href
  {https://doi.org/10.1103/physrevlett.107.136603} {\bibfield  {journal}
  {\bibinfo  {journal} {Phys. Rev. Lett.}\ }\textbf {\bibinfo {volume} {107}},\
  \bibinfo {pages} {136603} (\bibinfo {year} {2011})}\BibitemShut {NoStop}%
\bibitem [{\citenamefont {Grabecki}\ \emph {et~al.}(2013)\citenamefont
  {Grabecki}, \citenamefont {Wr\'obel}, \citenamefont {Czapkiewicz},
  \citenamefont {Cywi\ifmmode~\acute{n}\else \'{n}\fi{}ski}, \citenamefont
  {Giera\l{}towska}, \citenamefont {Guziewicz}, \citenamefont {Zholudev},
  \citenamefont {Gavrilenko}, \citenamefont {Mikhailov}, \citenamefont
  {Dvoretski}, \citenamefont {Teppe}, \citenamefont {Knap},\ and\ \citenamefont
  {Dietl}}]{2013Grabecki}%
  \BibitemOpen
  \bibfield  {author} {\bibinfo {author} {\bibfnamefont {G.}~\bibnamefont
  {Grabecki}}, \bibinfo {author} {\bibfnamefont {J.}~\bibnamefont {Wr\'obel}},
  \bibinfo {author} {\bibfnamefont {M.}~\bibnamefont {Czapkiewicz}}, \bibinfo
  {author} {\bibfnamefont {L.}~\bibnamefont {Cywi\ifmmode~\acute{n}\else
  \'{n}\fi{}ski}}, \bibinfo {author} {\bibfnamefont {S.}~\bibnamefont
  {Giera\l{}towska}}, \bibinfo {author} {\bibfnamefont {E.}~\bibnamefont
  {Guziewicz}}, \bibinfo {author} {\bibfnamefont {M.}~\bibnamefont {Zholudev}},
  \bibinfo {author} {\bibfnamefont {V.}~\bibnamefont {Gavrilenko}}, \bibinfo
  {author} {\bibfnamefont {N.~N.}\ \bibnamefont {Mikhailov}}, \bibinfo {author}
  {\bibfnamefont {S.~A.}\ \bibnamefont {Dvoretski}}, \bibinfo {author}
  {\bibfnamefont {F.}~\bibnamefont {Teppe}}, \bibinfo {author} {\bibfnamefont
  {W.}~\bibnamefont {Knap}},\ and\ \bibinfo {author} {\bibfnamefont
  {T.}~\bibnamefont {Dietl}},\ }\bibfield  {title} {\bibinfo {title} {Nonlocal
  resistance and its fluctuations in microstructures of band-inverted
  {HgTe/(Hg,Cd)Te} quantum wells},\ }\href
  {https://doi.org/10.1103/physrevb.88.165309} {\bibfield  {journal} {\bibinfo
  {journal} {Phys. Rev. B}\ }\textbf {\bibinfo {volume} {88}},\ \bibinfo
  {pages} {165309} (\bibinfo {year} {2013})}\BibitemShut {NoStop}%
\bibitem [{\citenamefont {Suzuki}\ \emph {et~al.}(2013)\citenamefont {Suzuki},
  \citenamefont {Harada}, \citenamefont {Onomitsu},\ and\ \citenamefont
  {Muraki}}]{2013Suzuki}%
  \BibitemOpen
  \bibfield  {author} {\bibinfo {author} {\bibfnamefont {K.}~\bibnamefont
  {Suzuki}}, \bibinfo {author} {\bibfnamefont {Y.}~\bibnamefont {Harada}},
  \bibinfo {author} {\bibfnamefont {K.}~\bibnamefont {Onomitsu}},\ and\
  \bibinfo {author} {\bibfnamefont {K.}~\bibnamefont {Muraki}},\ }\bibfield
  {title} {\bibinfo {title} {Edge channel transport in the {InAs/GaSb}
  topological insulating phase},\ }\href
  {https://doi.org/10.1103/physrevb.87.235311} {\bibfield  {journal} {\bibinfo
  {journal} {Phys. Rev. B}\ }\textbf {\bibinfo {volume} {87}},\ \bibinfo
  {pages} {235311} (\bibinfo {year} {2013})}\BibitemShut {NoStop}%
\bibitem [{\citenamefont {Du}\ \emph {et~al.}(2015)\citenamefont {Du},
  \citenamefont {Knez}, \citenamefont {Sullivan},\ and\ \citenamefont
  {Du}}]{2015Du}%
  \BibitemOpen
  \bibfield  {author} {\bibinfo {author} {\bibfnamefont {L.}~\bibnamefont
  {Du}}, \bibinfo {author} {\bibfnamefont {I.}~\bibnamefont {Knez}}, \bibinfo
  {author} {\bibfnamefont {G.}~\bibnamefont {Sullivan}},\ and\ \bibinfo
  {author} {\bibfnamefont {R.-R.}\ \bibnamefont {Du}},\ }\bibfield  {title}
  {\bibinfo {title} {Robust helical edge transport in gated {InAs}/{GaSb}
  bilayers},\ }\href {https://doi.org/10.1103/physrevlett.114.096802}
  {\bibfield  {journal} {\bibinfo  {journal} {Phys. Rev. Lett.}\ }\textbf
  {\bibinfo {volume} {114}},\ \bibinfo {pages} {096802} (\bibinfo {year}
  {2015})}\BibitemShut {NoStop}%
\bibitem [{\citenamefont {Mueller}\ \emph {et~al.}(2017)\citenamefont
  {Mueller}, \citenamefont {Mittag}, \citenamefont {Tschirky}, \citenamefont
  {Charpentier}, \citenamefont {Wegscheider}, \citenamefont {Ensslin},\ and\
  \citenamefont {Ihn}}]{2017Mueller}%
  \BibitemOpen
  \bibfield  {author} {\bibinfo {author} {\bibfnamefont {S.}~\bibnamefont
  {Mueller}}, \bibinfo {author} {\bibfnamefont {C.}~\bibnamefont {Mittag}},
  \bibinfo {author} {\bibfnamefont {T.}~\bibnamefont {Tschirky}}, \bibinfo
  {author} {\bibfnamefont {C.}~\bibnamefont {Charpentier}}, \bibinfo {author}
  {\bibfnamefont {W.}~\bibnamefont {Wegscheider}}, \bibinfo {author}
  {\bibfnamefont {K.}~\bibnamefont {Ensslin}},\ and\ \bibinfo {author}
  {\bibfnamefont {T.}~\bibnamefont {Ihn}},\ }\bibfield  {title} {\bibinfo
  {title} {Edge transport in {InAs} and {InAs/GaSb} quantum wells},\ }\href
  {https://doi.org/10.1103/physrevb.96.075406} {\bibfield  {journal} {\bibinfo
  {journal} {Phys. Rev. B}\ }\textbf {\bibinfo {volume} {96}},\ \bibinfo
  {pages} {075406} (\bibinfo {year} {2017})}\BibitemShut {NoStop}%
\bibitem [{\citenamefont {Nichele}\ \emph {et~al.}(2016)\citenamefont
  {Nichele}, \citenamefont {Suominen}, \citenamefont {Kjaergaard},
  \citenamefont {Marcus}, \citenamefont {Sajadi}, \citenamefont {Folk},
  \citenamefont {Qu}, \citenamefont {Beukman}, \citenamefont {de~Vries},
  \citenamefont {van Veen}, \citenamefont {Nadj-Perge}, \citenamefont
  {Kouwenhoven}, \citenamefont {Nguyen}, \citenamefont {Kiselev}, \citenamefont
  {Yi}, \citenamefont {Sokolich}, \citenamefont {Manfra}, \citenamefont
  {Spanton},\ and\ \citenamefont {Moler}}]{2016Nichele}%
  \BibitemOpen
  \bibfield  {author} {\bibinfo {author} {\bibfnamefont {F.}~\bibnamefont
  {Nichele}}, \bibinfo {author} {\bibfnamefont {H.~J.}\ \bibnamefont
  {Suominen}}, \bibinfo {author} {\bibfnamefont {M.}~\bibnamefont
  {Kjaergaard}}, \bibinfo {author} {\bibfnamefont {C.~M.}\ \bibnamefont
  {Marcus}}, \bibinfo {author} {\bibfnamefont {E.}~\bibnamefont {Sajadi}},
  \bibinfo {author} {\bibfnamefont {J.~A.}\ \bibnamefont {Folk}}, \bibinfo
  {author} {\bibfnamefont {F.}~\bibnamefont {Qu}}, \bibinfo {author}
  {\bibfnamefont {A.~J.~A.}\ \bibnamefont {Beukman}}, \bibinfo {author}
  {\bibfnamefont {F.~K.}\ \bibnamefont {de~Vries}}, \bibinfo {author}
  {\bibfnamefont {J.}~\bibnamefont {van Veen}}, \bibinfo {author}
  {\bibfnamefont {S.}~\bibnamefont {Nadj-Perge}}, \bibinfo {author}
  {\bibfnamefont {L.~P.}\ \bibnamefont {Kouwenhoven}}, \bibinfo {author}
  {\bibfnamefont {B.-M.}\ \bibnamefont {Nguyen}}, \bibinfo {author}
  {\bibfnamefont {A.~A.}\ \bibnamefont {Kiselev}}, \bibinfo {author}
  {\bibfnamefont {W.}~\bibnamefont {Yi}}, \bibinfo {author} {\bibfnamefont
  {M.}~\bibnamefont {Sokolich}}, \bibinfo {author} {\bibfnamefont {M.~J.}\
  \bibnamefont {Manfra}}, \bibinfo {author} {\bibfnamefont {E.~M.}\
  \bibnamefont {Spanton}},\ and\ \bibinfo {author} {\bibfnamefont {K.~A.}\
  \bibnamefont {Moler}},\ }\bibfield  {title} {\bibinfo {title} {Edge transport
  in the trivial phase of {InAs}/{GaSb}},\ }\href
  {https://doi.org/10.1088/1367-2630/18/8/083005} {\bibfield  {journal}
  {\bibinfo  {journal} {New Journal of Physics}\ }\textbf {\bibinfo {volume}
  {18}},\ \bibinfo {pages} {083005} (\bibinfo {year} {2016})}\BibitemShut
  {NoStop}%
\bibitem [{\citenamefont {Cou{\"e}do}\ \emph {et~al.}(2016)\citenamefont
  {Cou{\"e}do}, \citenamefont {Irie}, \citenamefont {Suzuki}, \citenamefont
  {Onomitsu},\ and\ \citenamefont {Muraki}}]{2016Couedo}%
  \BibitemOpen
  \bibfield  {author} {\bibinfo {author} {\bibfnamefont {F.}~\bibnamefont
  {Cou{\"e}do}}, \bibinfo {author} {\bibfnamefont {H.}~\bibnamefont {Irie}},
  \bibinfo {author} {\bibfnamefont {K.}~\bibnamefont {Suzuki}}, \bibinfo
  {author} {\bibfnamefont {K.}~\bibnamefont {Onomitsu}},\ and\ \bibinfo
  {author} {\bibfnamefont {K.}~\bibnamefont {Muraki}},\ }\bibfield  {title}
  {\bibinfo {title} {Single-edge transport in an {InAs}/{GaSb} quantum spin
  {H}all insulator},\ }\href {https://doi.org/10.1103/physrevb.94.035301}
  {\bibfield  {journal} {\bibinfo  {journal} {Phys. Rev. B}\ }\textbf {\bibinfo
  {volume} {94}},\ \bibinfo {pages} {035301} (\bibinfo {year}
  {2016})}\BibitemShut {NoStop}%
\bibitem [{\citenamefont {Nguyen}\ \emph {et~al.}(2016)\citenamefont {Nguyen},
  \citenamefont {Kiselev}, \citenamefont {Noah}, \citenamefont {Yi},
  \citenamefont {Qu}, \citenamefont {Beukman}, \citenamefont {de~Vries},
  \citenamefont {van Veen}, \citenamefont {Nadj-Perge}, \citenamefont
  {Kouwenhoven}, \citenamefont {Kjaergaard}, \citenamefont {Suominen},
  \citenamefont {Nichele}, \citenamefont {Marcus}, \citenamefont {Manfra},\
  and\ \citenamefont {Sokolich}}]{2016Nguyen}%
  \BibitemOpen
  \bibfield  {author} {\bibinfo {author} {\bibfnamefont {B.-M.}\ \bibnamefont
  {Nguyen}}, \bibinfo {author} {\bibfnamefont {A.~A.}\ \bibnamefont {Kiselev}},
  \bibinfo {author} {\bibfnamefont {R.}~\bibnamefont {Noah}}, \bibinfo {author}
  {\bibfnamefont {W.}~\bibnamefont {Yi}}, \bibinfo {author} {\bibfnamefont
  {F.}~\bibnamefont {Qu}}, \bibinfo {author} {\bibfnamefont {A.~J.~A.}\
  \bibnamefont {Beukman}}, \bibinfo {author} {\bibfnamefont {F.~K.}\
  \bibnamefont {de~Vries}}, \bibinfo {author} {\bibfnamefont {J.}~\bibnamefont
  {van Veen}}, \bibinfo {author} {\bibfnamefont {S.}~\bibnamefont
  {Nadj-Perge}}, \bibinfo {author} {\bibfnamefont {L.~P.}\ \bibnamefont
  {Kouwenhoven}}, \bibinfo {author} {\bibfnamefont {M.}~\bibnamefont
  {Kjaergaard}}, \bibinfo {author} {\bibfnamefont {H.~J.}\ \bibnamefont
  {Suominen}}, \bibinfo {author} {\bibfnamefont {F.}~\bibnamefont {Nichele}},
  \bibinfo {author} {\bibfnamefont {C.~M.}\ \bibnamefont {Marcus}}, \bibinfo
  {author} {\bibfnamefont {M.~J.}\ \bibnamefont {Manfra}},\ and\ \bibinfo
  {author} {\bibfnamefont {M.}~\bibnamefont {Sokolich}},\ }\bibfield  {title}
  {\bibinfo {title} {Decoupling edge versus bulk conductance in the trivial
  regime of an {InAs}/{GaSb} double quantum well using {C}orbino ring
  geometry},\ }\href {https://doi.org/10.1103/physrevlett.117.077701}
  {\bibfield  {journal} {\bibinfo  {journal} {Phys. Rev. Lett.}\ }\textbf
  {\bibinfo {volume} {117}},\ \bibinfo {pages} {077701} (\bibinfo {year}
  {2016})}\BibitemShut {NoStop}%
\bibitem [{\citenamefont {Liu}\ \emph {et~al.}(2008)\citenamefont {Liu},
  \citenamefont {Hughes}, \citenamefont {Qi}, \citenamefont {Wang},\ and\
  \citenamefont {Zhang}}]{2008Liu}%
  \BibitemOpen
  \bibfield  {author} {\bibinfo {author} {\bibfnamefont {C.}~\bibnamefont
  {Liu}}, \bibinfo {author} {\bibfnamefont {T.~L.}\ \bibnamefont {Hughes}},
  \bibinfo {author} {\bibfnamefont {X.-L.}\ \bibnamefont {Qi}}, \bibinfo
  {author} {\bibfnamefont {K.}~\bibnamefont {Wang}},\ and\ \bibinfo {author}
  {\bibfnamefont {S.-C.}\ \bibnamefont {Zhang}},\ }\bibfield  {title} {\bibinfo
  {title} {Quantum spin {H}all effect in inverted type-{II} semiconductors},\
  }\href {https://doi.org/10.1103/physrevlett.100.236601} {\bibfield  {journal}
  {\bibinfo  {journal} {Phys. Rev. Lett.}\ }\textbf {\bibinfo {volume} {100}},\
  \bibinfo {pages} {236601} (\bibinfo {year} {2008})}\BibitemShut {NoStop}%
\bibitem [{\citenamefont {Meyer}\ \emph {et~al.}(2024)\citenamefont {Meyer},
  \citenamefont {F\"ahndrich}, \citenamefont {Schmid}, \citenamefont {Wolf},
  \citenamefont {Krishtopenko}, \citenamefont {Jouault}, \citenamefont
  {Bastard}, \citenamefont {Teppe}, \citenamefont {Hartmann},\ and\
  \citenamefont {H\"{o}fling}}]{2024Meyer}%
  \BibitemOpen
  \bibfield  {author} {\bibinfo {author} {\bibfnamefont {M.}~\bibnamefont
  {Meyer}}, \bibinfo {author} {\bibfnamefont {T.}~\bibnamefont {F\"ahndrich}},
  \bibinfo {author} {\bibfnamefont {S.}~\bibnamefont {Schmid}}, \bibinfo
  {author} {\bibfnamefont {A.}~\bibnamefont {Wolf}}, \bibinfo {author}
  {\bibfnamefont {S.~S.}\ \bibnamefont {Krishtopenko}}, \bibinfo {author}
  {\bibfnamefont {B.}~\bibnamefont {Jouault}}, \bibinfo {author} {\bibfnamefont
  {G.}~\bibnamefont {Bastard}}, \bibinfo {author} {\bibfnamefont
  {F.}~\bibnamefont {Teppe}}, \bibinfo {author} {\bibfnamefont
  {F.}~\bibnamefont {Hartmann}},\ and\ \bibinfo {author} {\bibfnamefont
  {S.}~\bibnamefont {H\"{o}fling}},\ }\bibfield  {title} {\bibinfo {title}
  {Coexistence of topological and normal insulating phases in electro-optically
  tuned {InAs}/{GaSb} bilayer quantum wells},\ }\href
  {https://doi.org/10.1103/physrevb.109.l121303} {\bibfield  {journal}
  {\bibinfo  {journal} {Phys. Rev. B}\ }\textbf {\bibinfo {volume} {109}},\
  \bibinfo {pages} {L121303} (\bibinfo {year} {2024})}\BibitemShut {NoStop}%
\bibitem [{\citenamefont {Du}\ \emph {et~al.}(2017)\citenamefont {Du},
  \citenamefont {Li}, \citenamefont {Lou}, \citenamefont {Wu}, \citenamefont
  {Liu}, \citenamefont {Han}, \citenamefont {Zhang}, \citenamefont {Sullivan},
  \citenamefont {Ikhlassi}, \citenamefont {Chang},\ and\ \citenamefont
  {Du}}]{2017Dua}%
  \BibitemOpen
  \bibfield  {author} {\bibinfo {author} {\bibfnamefont {L.}~\bibnamefont
  {Du}}, \bibinfo {author} {\bibfnamefont {T.}~\bibnamefont {Li}}, \bibinfo
  {author} {\bibfnamefont {W.}~\bibnamefont {Lou}}, \bibinfo {author}
  {\bibfnamefont {X.}~\bibnamefont {Wu}}, \bibinfo {author} {\bibfnamefont
  {X.}~\bibnamefont {Liu}}, \bibinfo {author} {\bibfnamefont {Z.}~\bibnamefont
  {Han}}, \bibinfo {author} {\bibfnamefont {C.}~\bibnamefont {Zhang}}, \bibinfo
  {author} {\bibfnamefont {G.}~\bibnamefont {Sullivan}}, \bibinfo {author}
  {\bibfnamefont {A.}~\bibnamefont {Ikhlassi}}, \bibinfo {author}
  {\bibfnamefont {K.}~\bibnamefont {Chang}},\ and\ \bibinfo {author}
  {\bibfnamefont {R.-R.}\ \bibnamefont {Du}},\ }\bibfield  {title} {\bibinfo
  {title} {Tuning edge states in strained-layer {InAs}/{GaInSb} quantum spin
  {H}all insulators},\ }\href {https://doi.org/10.1103/physrevlett.119.056803}
  {\bibfield  {journal} {\bibinfo  {journal} {Phys. Rev. Lett.}\ }\textbf
  {\bibinfo {volume} {119}},\ \bibinfo {pages} {056803} (\bibinfo {year}
  {2017})}\BibitemShut {NoStop}%
\bibitem [{\citenamefont {Krishtopenko}\ and\ \citenamefont
  {Teppe}(2018{\natexlab{a}})}]{2018Krishtopenkoa}%
  \BibitemOpen
  \bibfield  {author} {\bibinfo {author} {\bibfnamefont {S.~S.}\ \bibnamefont
  {Krishtopenko}}\ and\ \bibinfo {author} {\bibfnamefont {F.}~\bibnamefont
  {Teppe}},\ }\bibfield  {title} {\bibinfo {title} {Quantum spin {H}all
  insulator with a large bandgap, {D}irac fermions, and bilayer graphene
  analog},\ }\href {https://doi.org/10.1126/sciadv.aap7529} {\bibfield
  {journal} {\bibinfo  {journal} {Sci. Adv.}\ }\textbf {\bibinfo {volume}
  {4}},\ \bibinfo {pages} {eaap7529} (\bibinfo {year}
  {2018}{\natexlab{a}})}\BibitemShut {NoStop}%
\bibitem [{\citenamefont {Krishtopenko}\ \emph {et~al.}(2018)\citenamefont
  {Krishtopenko}, \citenamefont {Ruffenach}, \citenamefont {Gonzalez-Posada},
  \citenamefont {Boissier}, \citenamefont {Marcinkiewicz}, \citenamefont
  {Fadeev}, \citenamefont {Kadykov}, \citenamefont {Rumyantsev}, \citenamefont
  {Morozov}, \citenamefont {Gavrilenko}, \citenamefont {Consejo}, \citenamefont
  {Desrat}, \citenamefont {Jouault}, \citenamefont {Knap}, \citenamefont
  {Tourni{\'{e}}},\ and\ \citenamefont {Teppe}}]{2018Krishtopenko}%
  \BibitemOpen
  \bibfield  {author} {\bibinfo {author} {\bibfnamefont {S.~S.}\ \bibnamefont
  {Krishtopenko}}, \bibinfo {author} {\bibfnamefont {S.}~\bibnamefont
  {Ruffenach}}, \bibinfo {author} {\bibfnamefont {F.}~\bibnamefont
  {Gonzalez-Posada}}, \bibinfo {author} {\bibfnamefont {G.}~\bibnamefont
  {Boissier}}, \bibinfo {author} {\bibfnamefont {M.}~\bibnamefont
  {Marcinkiewicz}}, \bibinfo {author} {\bibfnamefont {M.~A.}\ \bibnamefont
  {Fadeev}}, \bibinfo {author} {\bibfnamefont {A.~M.}\ \bibnamefont {Kadykov}},
  \bibinfo {author} {\bibfnamefont {V.~V.}\ \bibnamefont {Rumyantsev}},
  \bibinfo {author} {\bibfnamefont {S.~V.}\ \bibnamefont {Morozov}}, \bibinfo
  {author} {\bibfnamefont {V.~I.}\ \bibnamefont {Gavrilenko}}, \bibinfo
  {author} {\bibfnamefont {C.}~\bibnamefont {Consejo}}, \bibinfo {author}
  {\bibfnamefont {W.}~\bibnamefont {Desrat}}, \bibinfo {author} {\bibfnamefont
  {B.}~\bibnamefont {Jouault}}, \bibinfo {author} {\bibfnamefont
  {W.}~\bibnamefont {Knap}}, \bibinfo {author} {\bibfnamefont {E.}~\bibnamefont
  {Tourni{\'{e}}}},\ and\ \bibinfo {author} {\bibfnamefont {F.}~\bibnamefont
  {Teppe}},\ }\bibfield  {title} {\bibinfo {title} {Temperature-dependent
  terahertz spectroscopy of inverted-band three-layer {InAs}/{GaSb}/{InAs}
  quantum well},\ }\href {https://doi.org/10.1103/physrevb.97.245419}
  {\bibfield  {journal} {\bibinfo  {journal} {Phys. Rev. B}\ }\textbf {\bibinfo
  {volume} {97}},\ \bibinfo {pages} {245419} (\bibinfo {year}
  {2018})}\BibitemShut {NoStop}%
\bibitem [{\citenamefont {Marcinkiewicz}\ \emph {et~al.}(2017)\citenamefont
  {Marcinkiewicz}, \citenamefont {Ruffenach}, \citenamefont {Krishtopenko},
  \citenamefont {Kadykov}, \citenamefont {Consejo}, \citenamefont {But},
  \citenamefont {Desrat}, \citenamefont {Knap}, \citenamefont {Torres},
  \citenamefont {Ikonnikov}, \citenamefont {Spirin}, \citenamefont {Morozov},
  \citenamefont {Gavrilenko}, \citenamefont {Mikhailov}, \citenamefont
  {Dvoretskii},\ and\ \citenamefont {Teppe}}]{2017Marcinkiewicz}%
  \BibitemOpen
  \bibfield  {author} {\bibinfo {author} {\bibfnamefont {M.}~\bibnamefont
  {Marcinkiewicz}}, \bibinfo {author} {\bibfnamefont {S.}~\bibnamefont
  {Ruffenach}}, \bibinfo {author} {\bibfnamefont {S.~S.}\ \bibnamefont
  {Krishtopenko}}, \bibinfo {author} {\bibfnamefont {A.~M.}\ \bibnamefont
  {Kadykov}}, \bibinfo {author} {\bibfnamefont {C.}~\bibnamefont {Consejo}},
  \bibinfo {author} {\bibfnamefont {D.~B.}\ \bibnamefont {But}}, \bibinfo
  {author} {\bibfnamefont {W.}~\bibnamefont {Desrat}}, \bibinfo {author}
  {\bibfnamefont {W.}~\bibnamefont {Knap}}, \bibinfo {author} {\bibfnamefont
  {J.}~\bibnamefont {Torres}}, \bibinfo {author} {\bibfnamefont {A.~V.}\
  \bibnamefont {Ikonnikov}}, \bibinfo {author} {\bibfnamefont {K.~E.}\
  \bibnamefont {Spirin}}, \bibinfo {author} {\bibfnamefont {S.~V.}\
  \bibnamefont {Morozov}}, \bibinfo {author} {\bibfnamefont {V.~I.}\
  \bibnamefont {Gavrilenko}}, \bibinfo {author} {\bibfnamefont {N.~N.}\
  \bibnamefont {Mikhailov}}, \bibinfo {author} {\bibfnamefont {S.~A.}\
  \bibnamefont {Dvoretskii}},\ and\ \bibinfo {author} {\bibfnamefont
  {F.}~\bibnamefont {Teppe}},\ }\bibfield  {title} {\bibinfo {title}
  {Temperature-driven single-valley {D}irac fermions in {HgTe} quantum wells},\
  }\href {https://doi.org/10.1103/physrevb.96.035405} {\bibfield  {journal}
  {\bibinfo  {journal} {Phys. Rev. B}\ }\textbf {\bibinfo {volume} {96}},\
  \bibinfo {pages} {035405} (\bibinfo {year} {2017})}\BibitemShut {NoStop}%
\bibitem [{\citenamefont {Kadykov}\ \emph {et~al.}(2018)\citenamefont
  {Kadykov}, \citenamefont {Krishtopenko}, \citenamefont {Jouault},
  \citenamefont {Desrat}, \citenamefont {Knap}, \citenamefont {Ruffenach},
  \citenamefont {Consejo}, \citenamefont {Torres}, \citenamefont {Morozov},
  \citenamefont {Mikhailov}, \citenamefont {Dvoretskii},\ and\ \citenamefont
  {Teppe}}]{2018Kadykov}%
  \BibitemOpen
  \bibfield  {author} {\bibinfo {author} {\bibfnamefont {A.~M.}\ \bibnamefont
  {Kadykov}}, \bibinfo {author} {\bibfnamefont {S.~S.}\ \bibnamefont
  {Krishtopenko}}, \bibinfo {author} {\bibfnamefont {B.}~\bibnamefont
  {Jouault}}, \bibinfo {author} {\bibfnamefont {W.}~\bibnamefont {Desrat}},
  \bibinfo {author} {\bibfnamefont {W.}~\bibnamefont {Knap}}, \bibinfo {author}
  {\bibfnamefont {S.}~\bibnamefont {Ruffenach}}, \bibinfo {author}
  {\bibfnamefont {C.}~\bibnamefont {Consejo}}, \bibinfo {author} {\bibfnamefont
  {J.}~\bibnamefont {Torres}}, \bibinfo {author} {\bibfnamefont {S.~V.}\
  \bibnamefont {Morozov}}, \bibinfo {author} {\bibfnamefont {N.~N.}\
  \bibnamefont {Mikhailov}}, \bibinfo {author} {\bibfnamefont {S.~A.}\
  \bibnamefont {Dvoretskii}},\ and\ \bibinfo {author} {\bibfnamefont
  {F.}~\bibnamefont {Teppe}},\ }\bibfield  {title} {\bibinfo {title}
  {Temperature-induced topological phase transition in {HgTe} quantum wells},\
  }\href {https://doi.org/10.1103/physrevlett.120.086401} {\bibfield  {journal}
  {\bibinfo  {journal} {Phys. Rev. Lett.}\ }\textbf {\bibinfo {volume} {120}},\
  \bibinfo {pages} {086401} (\bibinfo {year} {2018})}\BibitemShut {NoStop}%
\bibitem [{\citenamefont {Krishtopenko}\ \emph {et~al.}(2017)\citenamefont
  {Krishtopenko}, \citenamefont {Ikonnikov}, \citenamefont {Maremyanin},
  \citenamefont {Bovkun}, \citenamefont {Spirin}, \citenamefont {Kadykov},
  \citenamefont {Marcinkiewicz}, \citenamefont {Ruffenach}, \citenamefont
  {Consejo}, \citenamefont {Teppe}, \citenamefont {Knap}, \citenamefont
  {Semyagin}, \citenamefont {Putyato}, \citenamefont {Emelyanov}, \citenamefont
  {Preobrazhenskii},\ and\ \citenamefont {Gavrilenko}}]{NewRef1}%
  \BibitemOpen
  \bibfield  {author} {\bibinfo {author} {\bibfnamefont {S.~S.}\ \bibnamefont
  {Krishtopenko}}, \bibinfo {author} {\bibfnamefont {A.~V.}\ \bibnamefont
  {Ikonnikov}}, \bibinfo {author} {\bibfnamefont {K.~V.}\ \bibnamefont
  {Maremyanin}}, \bibinfo {author} {\bibfnamefont {L.~S.}\ \bibnamefont
  {Bovkun}}, \bibinfo {author} {\bibfnamefont {K.~E.}\ \bibnamefont {Spirin}},
  \bibinfo {author} {\bibfnamefont {A.~M.}\ \bibnamefont {Kadykov}}, \bibinfo
  {author} {\bibfnamefont {M.}~\bibnamefont {Marcinkiewicz}}, \bibinfo {author}
  {\bibfnamefont {S.}~\bibnamefont {Ruffenach}}, \bibinfo {author}
  {\bibfnamefont {C.}~\bibnamefont {Consejo}}, \bibinfo {author} {\bibfnamefont
  {F.}~\bibnamefont {Teppe}}, \bibinfo {author} {\bibfnamefont
  {W.}~\bibnamefont {Knap}}, \bibinfo {author} {\bibfnamefont {B.~R.}\
  \bibnamefont {Semyagin}}, \bibinfo {author} {\bibfnamefont {M.~A.}\
  \bibnamefont {Putyato}}, \bibinfo {author} {\bibfnamefont {E.~A.}\
  \bibnamefont {Emelyanov}}, \bibinfo {author} {\bibfnamefont {V.~V.}\
  \bibnamefont {Preobrazhenskii}},\ and\ \bibinfo {author} {\bibfnamefont
  {V.~I.}\ \bibnamefont {Gavrilenko}},\ }\bibfield  {title} {\bibinfo {title}
  {Cyclotron resonance of {D}irac fermions in {InAs/GaSb/InAs} quantum wells},\
  }\href {https://doi.org/10.1134/S1063782617010109} {\bibfield  {journal}
  {\bibinfo  {journal} {Semiconductors}\ }\textbf {\bibinfo {volume} {51}},\
  \bibinfo {pages} {38} (\bibinfo {year} {2017})}\BibitemShut {NoStop}%
\bibitem [{\citenamefont {Ruffenach}\ \emph
  {et~al.}(2017{\natexlab{a}})\citenamefont {Ruffenach}, \citenamefont
  {Krishtopenko}, \citenamefont {Bovkun}, \citenamefont {Ikonnikov},
  \citenamefont {Marcinkiewicz}, \citenamefont {Consejo}, \citenamefont
  {Potemski}, \citenamefont {Piot}, \citenamefont {Orlita}, \citenamefont
  {Semyagin}, \citenamefont {Putyato}, \citenamefont {Emelyanov}, \citenamefont
  {Preobrazhenskii}, \citenamefont {Knap}, \citenamefont {Gonzalez-Posada},
  \citenamefont {Boissier}, \citenamefont {Tourni\'{e}}, \citenamefont
  {Teppe},\ and\ \citenamefont {Gavrilenko}}]{NewRef2}%
  \BibitemOpen
  \bibfield  {author} {\bibinfo {author} {\bibfnamefont {S.}~\bibnamefont
  {Ruffenach}}, \bibinfo {author} {\bibfnamefont {S.~S.}\ \bibnamefont
  {Krishtopenko}}, \bibinfo {author} {\bibfnamefont {L.~S.}\ \bibnamefont
  {Bovkun}}, \bibinfo {author} {\bibfnamefont {A.~V.}\ \bibnamefont
  {Ikonnikov}}, \bibinfo {author} {\bibfnamefont {M.}~\bibnamefont
  {Marcinkiewicz}}, \bibinfo {author} {\bibfnamefont {C.}~\bibnamefont
  {Consejo}}, \bibinfo {author} {\bibfnamefont {M.}~\bibnamefont {Potemski}},
  \bibinfo {author} {\bibfnamefont {B.}~\bibnamefont {Piot}}, \bibinfo {author}
  {\bibfnamefont {M.}~\bibnamefont {Orlita}}, \bibinfo {author} {\bibfnamefont
  {B.~R.}\ \bibnamefont {Semyagin}}, \bibinfo {author} {\bibfnamefont {M.~A.}\
  \bibnamefont {Putyato}}, \bibinfo {author} {\bibfnamefont {E.~A.}\
  \bibnamefont {Emelyanov}}, \bibinfo {author} {\bibfnamefont {V.~V.}\
  \bibnamefont {Preobrazhenskii}}, \bibinfo {author} {\bibfnamefont
  {W.}~\bibnamefont {Knap}}, \bibinfo {author} {\bibfnamefont {F.}~\bibnamefont
  {Gonzalez-Posada}}, \bibinfo {author} {\bibfnamefont {G.}~\bibnamefont
  {Boissier}}, \bibinfo {author} {\bibfnamefont {E.}~\bibnamefont
  {Tourni\'{e}}}, \bibinfo {author} {\bibfnamefont {F.}~\bibnamefont {Teppe}},\
  and\ \bibinfo {author} {\bibfnamefont {V.~I.}\ \bibnamefont {Gavrilenko}},\
  }\bibfield  {title} {\bibinfo {title} {Magnetoabsorbtion of {D}irac fermions
  in gapless three-layer {InAs/GaSb/InAs} quantum wells},\ }\href
  {https://doi.org/10.1134/S0021364017230102} {\bibfield  {journal} {\bibinfo
  {journal} {JETP Lett.}\ }\textbf {\bibinfo {volume} {106}},\ \bibinfo {pages}
  {727} (\bibinfo {year} {2017}{\natexlab{a}})}\BibitemShut {NoStop}%
\bibitem [{\citenamefont {Krishtopenko}\ \emph
  {et~al.}(2019{\natexlab{a}})\citenamefont {Krishtopenko}, \citenamefont
  {Desrat}, \citenamefont {Spirin}, \citenamefont {Consejo}, \citenamefont
  {Ruffenach}, \citenamefont {Gonzalez-Posada}, \citenamefont {Jouault},
  \citenamefont {Knap}, \citenamefont {Maremyanin}, \citenamefont {Gavrilenko},
  \citenamefont {Boissier}, \citenamefont {Torres}, \citenamefont {Zaknoune},
  \citenamefont {Tourni\'e},\ and\ \citenamefont {Teppe}}]{2019Krishtopenkoa}%
  \BibitemOpen
  \bibfield  {author} {\bibinfo {author} {\bibfnamefont {S.~S.}\ \bibnamefont
  {Krishtopenko}}, \bibinfo {author} {\bibfnamefont {W.}~\bibnamefont
  {Desrat}}, \bibinfo {author} {\bibfnamefont {K.~E.}\ \bibnamefont {Spirin}},
  \bibinfo {author} {\bibfnamefont {C.}~\bibnamefont {Consejo}}, \bibinfo
  {author} {\bibfnamefont {S.}~\bibnamefont {Ruffenach}}, \bibinfo {author}
  {\bibfnamefont {F.}~\bibnamefont {Gonzalez-Posada}}, \bibinfo {author}
  {\bibfnamefont {B.}~\bibnamefont {Jouault}}, \bibinfo {author} {\bibfnamefont
  {W.}~\bibnamefont {Knap}}, \bibinfo {author} {\bibfnamefont {K.~V.}\
  \bibnamefont {Maremyanin}}, \bibinfo {author} {\bibfnamefont {V.~I.}\
  \bibnamefont {Gavrilenko}}, \bibinfo {author} {\bibfnamefont
  {G.}~\bibnamefont {Boissier}}, \bibinfo {author} {\bibfnamefont
  {J.}~\bibnamefont {Torres}}, \bibinfo {author} {\bibfnamefont
  {M.}~\bibnamefont {Zaknoune}}, \bibinfo {author} {\bibfnamefont
  {E.}~\bibnamefont {Tourni\'e}},\ and\ \bibinfo {author} {\bibfnamefont
  {F.}~\bibnamefont {Teppe}},\ }\bibfield  {title} {\bibinfo {title} {Massless
  {D}irac fermions in {III-V} semiconductor quantum wells},\ }\href
  {https://doi.org/10.1103/physrevb.99.121405} {\bibfield  {journal} {\bibinfo
  {journal} {Phys. Rev. B}\ }\textbf {\bibinfo {volume} {99}},\ \bibinfo
  {pages} {121405(R)} (\bibinfo {year} {2019}{\natexlab{a}})}\BibitemShut
  {NoStop}%
\bibitem [{\citenamefont {Krishtopenko}\ \emph
  {et~al.}(2019{\natexlab{b}})\citenamefont {Krishtopenko}, \citenamefont
  {Ruffenach}, \citenamefont {Gonzalez-Posada}, \citenamefont {Consejo},
  \citenamefont {Desrat}, \citenamefont {Jouault}, \citenamefont {Knap},
  \citenamefont {Fadeev}, \citenamefont {Kadykov}, \citenamefont {Rumyantsev},
  \citenamefont {Morozov}, \citenamefont {Boissier}, \citenamefont
  {Tourni{\'{e}}}, \citenamefont {Gavrilenko},\ and\ \citenamefont
  {Teppe}}]{2019Krishtopenko}%
  \BibitemOpen
  \bibfield  {author} {\bibinfo {author} {\bibfnamefont {S.~S.}\ \bibnamefont
  {Krishtopenko}}, \bibinfo {author} {\bibfnamefont {S.}~\bibnamefont
  {Ruffenach}}, \bibinfo {author} {\bibfnamefont {F.}~\bibnamefont
  {Gonzalez-Posada}}, \bibinfo {author} {\bibfnamefont {C.}~\bibnamefont
  {Consejo}}, \bibinfo {author} {\bibfnamefont {W.}~\bibnamefont {Desrat}},
  \bibinfo {author} {\bibfnamefont {B.}~\bibnamefont {Jouault}}, \bibinfo
  {author} {\bibfnamefont {W.}~\bibnamefont {Knap}}, \bibinfo {author}
  {\bibfnamefont {M.~A.}\ \bibnamefont {Fadeev}}, \bibinfo {author}
  {\bibfnamefont {A.~M.}\ \bibnamefont {Kadykov}}, \bibinfo {author}
  {\bibfnamefont {V.~V.}\ \bibnamefont {Rumyantsev}}, \bibinfo {author}
  {\bibfnamefont {S.~V.}\ \bibnamefont {Morozov}}, \bibinfo {author}
  {\bibfnamefont {G.}~\bibnamefont {Boissier}}, \bibinfo {author}
  {\bibfnamefont {E.}~\bibnamefont {Tourni{\'{e}}}}, \bibinfo {author}
  {\bibfnamefont {V.~I.}\ \bibnamefont {Gavrilenko}},\ and\ \bibinfo {author}
  {\bibfnamefont {F.}~\bibnamefont {Teppe}},\ }\bibfield  {title} {\bibinfo
  {title} {Terahertz spectroscopy of two-dimensional semimetal in three-layer
  {InAs}/{GaSb}/{InAs} quantum well},\ }\href
  {https://doi.org/10.1134/s0021364019020085} {\bibfield  {journal} {\bibinfo
  {journal} {{JETP} Letters}\ }\textbf {\bibinfo {volume} {109}},\ \bibinfo
  {pages} {96} (\bibinfo {year} {2019}{\natexlab{b}})}\BibitemShut {NoStop}%
\bibitem [{\citenamefont {Meyer}\ \emph {et~al.}(2021)\citenamefont {Meyer},
  \citenamefont {Schmid}, \citenamefont {Jabeen}, \citenamefont {Bastard},
  \citenamefont {Hartmann},\ and\ \citenamefont {H{\"o}fling}}]{2021Meyer}%
  \BibitemOpen
  \bibfield  {author} {\bibinfo {author} {\bibfnamefont {M.}~\bibnamefont
  {Meyer}}, \bibinfo {author} {\bibfnamefont {S.}~\bibnamefont {Schmid}},
  \bibinfo {author} {\bibfnamefont {F.}~\bibnamefont {Jabeen}}, \bibinfo
  {author} {\bibfnamefont {G.}~\bibnamefont {Bastard}}, \bibinfo {author}
  {\bibfnamefont {F.}~\bibnamefont {Hartmann}},\ and\ \bibinfo {author}
  {\bibfnamefont {S.}~\bibnamefont {H{\"o}fling}},\ }\bibfield  {title}
  {\bibinfo {title} {Topological band structure in {InAs}/{GaSb}/{InAs} triple
  quantum wells},\ }\href {https://doi.org/10.1103/physrevb.104.085301}
  {\bibfield  {journal} {\bibinfo  {journal} {Phys. Rev. B}\ }\textbf {\bibinfo
  {volume} {104}},\ \bibinfo {pages} {085301} (\bibinfo {year}
  {2021})}\BibitemShut {NoStop}%
\bibitem [{\citenamefont {Schmid}\ \emph {et~al.}(2022)\citenamefont {Schmid},
  \citenamefont {Meyer}, \citenamefont {Jabeen}, \citenamefont {Bastard},
  \citenamefont {Hartmann},\ and\ \citenamefont {H\"{o}fling}}]{2022Schmid}%
  \BibitemOpen
  \bibfield  {author} {\bibinfo {author} {\bibfnamefont {S.}~\bibnamefont
  {Schmid}}, \bibinfo {author} {\bibfnamefont {M.}~\bibnamefont {Meyer}},
  \bibinfo {author} {\bibfnamefont {F.}~\bibnamefont {Jabeen}}, \bibinfo
  {author} {\bibfnamefont {G.}~\bibnamefont {Bastard}}, \bibinfo {author}
  {\bibfnamefont {F.}~\bibnamefont {Hartmann}},\ and\ \bibinfo {author}
  {\bibfnamefont {S.}~\bibnamefont {H\"{o}fling}},\ }\bibfield  {title}
  {\bibinfo {title} {Exploring the phase diagram of {InAs}/{GaSb}/{InAs}
  trilayer quantum wells},\ }\href
  {https://doi.org/10.1103/physrevb.105.155304} {\bibfield  {journal} {\bibinfo
   {journal} {Phys. Rev. B}\ }\textbf {\bibinfo {volume} {105}},\ \bibinfo
  {pages} {155304} (\bibinfo {year} {2022})}\BibitemShut {NoStop}%
\bibitem [{\citenamefont {Meyer}\ \emph {et~al.}(2023)\citenamefont {Meyer},
  \citenamefont {Schmid}, \citenamefont {Jabeen}, \citenamefont {Bastard},
  \citenamefont {Hartmann},\ and\ \citenamefont {Höfling}}]{2023Meyer}%
  \BibitemOpen
  \bibfield  {author} {\bibinfo {author} {\bibfnamefont {M.}~\bibnamefont
  {Meyer}}, \bibinfo {author} {\bibfnamefont {S.}~\bibnamefont {Schmid}},
  \bibinfo {author} {\bibfnamefont {F.}~\bibnamefont {Jabeen}}, \bibinfo
  {author} {\bibfnamefont {G.}~\bibnamefont {Bastard}}, \bibinfo {author}
  {\bibfnamefont {F.}~\bibnamefont {Hartmann}},\ and\ \bibinfo {author}
  {\bibfnamefont {S.}~\bibnamefont {Höfling}},\ }\bibfield  {title} {\bibinfo
  {title} {Voltage control of the quantum scattering time in
  {InAs}/{GaSb}/{InAs} trilayer quantum wells},\ }\href
  {https://doi.org/10.1088/1367-2630/acbab7} {\bibfield  {journal} {\bibinfo
  {journal} {New J. Phys.}\ }\textbf {\bibinfo {volume} {25}},\ \bibinfo
  {pages} {023035} (\bibinfo {year} {2023})}\BibitemShut {NoStop}%
\bibitem [{\citenamefont {Avogadri}\ \emph {et~al.}(2022)\citenamefont
  {Avogadri}, \citenamefont {Gebert}, \citenamefont {Krishtopenko},
  \citenamefont {Castillo}, \citenamefont {Consejo}, \citenamefont {Ruffenach},
  \citenamefont {Roblin}, \citenamefont {Bray}, \citenamefont {Krupko},
  \citenamefont {Juillaguet}, \citenamefont {Contreras}, \citenamefont {Wolf},
  \citenamefont {Hartmann}, \citenamefont {H\"{o}fling}, \citenamefont
  {Boissier}, \citenamefont {Rodriguez}, \citenamefont {Nanot}, \citenamefont
  {Tourni{\'{e}}}, \citenamefont {Teppe},\ and\ \citenamefont
  {Jouault}}]{2022Avogadri}%
  \BibitemOpen
  \bibfield  {author} {\bibinfo {author} {\bibfnamefont {C.}~\bibnamefont
  {Avogadri}}, \bibinfo {author} {\bibfnamefont {S.}~\bibnamefont {Gebert}},
  \bibinfo {author} {\bibfnamefont {S.~S.}\ \bibnamefont {Krishtopenko}},
  \bibinfo {author} {\bibfnamefont {I.}~\bibnamefont {Castillo}}, \bibinfo
  {author} {\bibfnamefont {C.}~\bibnamefont {Consejo}}, \bibinfo {author}
  {\bibfnamefont {S.}~\bibnamefont {Ruffenach}}, \bibinfo {author}
  {\bibfnamefont {C.}~\bibnamefont {Roblin}}, \bibinfo {author} {\bibfnamefont
  {C.}~\bibnamefont {Bray}}, \bibinfo {author} {\bibfnamefont {Y.}~\bibnamefont
  {Krupko}}, \bibinfo {author} {\bibfnamefont {S.}~\bibnamefont {Juillaguet}},
  \bibinfo {author} {\bibfnamefont {S.}~\bibnamefont {Contreras}}, \bibinfo
  {author} {\bibfnamefont {A.}~\bibnamefont {Wolf}}, \bibinfo {author}
  {\bibfnamefont {F.}~\bibnamefont {Hartmann}}, \bibinfo {author}
  {\bibfnamefont {S.}~\bibnamefont {H\"{o}fling}}, \bibinfo {author}
  {\bibfnamefont {G.}~\bibnamefont {Boissier}}, \bibinfo {author}
  {\bibfnamefont {J.-B.}\ \bibnamefont {Rodriguez}}, \bibinfo {author}
  {\bibfnamefont {S.}~\bibnamefont {Nanot}}, \bibinfo {author} {\bibfnamefont
  {E.}~\bibnamefont {Tourni{\'{e}}}}, \bibinfo {author} {\bibfnamefont
  {F.}~\bibnamefont {Teppe}},\ and\ \bibinfo {author} {\bibfnamefont
  {B.}~\bibnamefont {Jouault}},\ }\bibfield  {title} {\bibinfo {title} {Large
  inverted band gap in strained three-layer {InAs}/{GaInSb} quantum wells},\
  }\href {https://doi.org/10.1103/physrevresearch.4.l042042} {\bibfield
  {journal} {\bibinfo  {journal} {Phys. Rev. Research}\ }\textbf {\bibinfo
  {volume} {4}},\ \bibinfo {pages} {L042042} (\bibinfo {year}
  {2022})}\BibitemShut {NoStop}%
\bibitem [{\citenamefont {Schmidt}\ \emph {et~al.}(2012)\citenamefont
  {Schmidt}, \citenamefont {Rachel}, \citenamefont {von Oppen},\ and\
  \citenamefont {Glazman}}]{NewRef3}%
  \BibitemOpen
  \bibfield  {author} {\bibinfo {author} {\bibfnamefont {T.~L.}\ \bibnamefont
  {Schmidt}}, \bibinfo {author} {\bibfnamefont {S.}~\bibnamefont {Rachel}},
  \bibinfo {author} {\bibfnamefont {F.}~\bibnamefont {von Oppen}},\ and\
  \bibinfo {author} {\bibfnamefont {L.~I.}\ \bibnamefont {Glazman}},\
  }\bibfield  {title} {\bibinfo {title} {Inelastic electron backscattering in a
  generic helical edge channel},\ }\href
  {https://doi.org/10.1103/PhysRevLett.108.156402} {\bibfield  {journal}
  {\bibinfo  {journal} {Phys. Rev. Lett.}\ }\textbf {\bibinfo {volume} {108}},\
  \bibinfo {pages} {156402} (\bibinfo {year} {2012})}\BibitemShut {NoStop}%
\bibitem [{\citenamefont {Kainaris}\ \emph {et~al.}(2014)\citenamefont
  {Kainaris}, \citenamefont {Gornyi}, \citenamefont {Carr},\ and\ \citenamefont
  {Mirlin}}]{NewRef4}%
  \BibitemOpen
  \bibfield  {author} {\bibinfo {author} {\bibfnamefont {N.}~\bibnamefont
  {Kainaris}}, \bibinfo {author} {\bibfnamefont {I.~V.}\ \bibnamefont
  {Gornyi}}, \bibinfo {author} {\bibfnamefont {S.~T.}\ \bibnamefont {Carr}},\
  and\ \bibinfo {author} {\bibfnamefont {A.~D.}\ \bibnamefont {Mirlin}},\
  }\bibfield  {title} {\bibinfo {title} {Conductivity of a generic helical
  liquid},\ }\href {https://doi.org/10.1103/PhysRevB.90.075118} {\bibfield
  {journal} {\bibinfo  {journal} {Phys. Rev. B}\ }\textbf {\bibinfo {volume}
  {90}},\ \bibinfo {pages} {075118} (\bibinfo {year} {2014})}\BibitemShut
  {NoStop}%
\bibitem [{\citenamefont {V\"ayrynen}\ \emph {et~al.}(2013)\citenamefont
  {V\"ayrynen}, \citenamefont {Goldstein},\ and\ \citenamefont
  {Glazman}}]{2013Vaeyrynen}%
  \BibitemOpen
  \bibfield  {author} {\bibinfo {author} {\bibfnamefont {J.~I.}\ \bibnamefont
  {V\"ayrynen}}, \bibinfo {author} {\bibfnamefont {M.}~\bibnamefont
  {Goldstein}},\ and\ \bibinfo {author} {\bibfnamefont {L.~I.}\ \bibnamefont
  {Glazman}},\ }\bibfield  {title} {\bibinfo {title} {Helical edge resistance
  introduced by charge puddles},\ }\href
  {https://doi.org/10.1103/physrevlett.110.216402} {\bibfield  {journal}
  {\bibinfo  {journal} {Phys. Rev. Lett.}\ }\textbf {\bibinfo {volume} {110}},\
  \bibinfo {pages} {216402} (\bibinfo {year} {2013})}\BibitemShut {NoStop}%
\bibitem [{\citenamefont {V\"ayrynen}\ \emph {et~al.}(2014)\citenamefont
  {V\"ayrynen}, \citenamefont {Goldstein}, \citenamefont {Gefen},\ and\
  \citenamefont {Glazman}}]{2014Vaeyrynen}%
  \BibitemOpen
  \bibfield  {author} {\bibinfo {author} {\bibfnamefont {J.~I.}\ \bibnamefont
  {V\"ayrynen}}, \bibinfo {author} {\bibfnamefont {M.}~\bibnamefont
  {Goldstein}}, \bibinfo {author} {\bibfnamefont {Y.}~\bibnamefont {Gefen}},\
  and\ \bibinfo {author} {\bibfnamefont {L.~I.}\ \bibnamefont {Glazman}},\
  }\bibfield  {title} {\bibinfo {title} {Resistance of helical edges formed in
  a semiconductor heterostructure},\ }\href
  {https://doi.org/10.1103/physrevb.90.115309} {\bibfield  {journal} {\bibinfo
  {journal} {Phys. Rev. B}\ }\textbf {\bibinfo {volume} {90}},\ \bibinfo
  {pages} {115309} (\bibinfo {year} {2014})}\BibitemShut {NoStop}%
\bibitem [{\citenamefont {V\"ayrynen}\ \emph {et~al.}(2016)\citenamefont
  {V\"ayrynen}, \citenamefont {Geissler},\ and\ \citenamefont
  {Glazman}}]{2016Vaeyrynen}%
  \BibitemOpen
  \bibfield  {author} {\bibinfo {author} {\bibfnamefont {J.~I.}\ \bibnamefont
  {V\"ayrynen}}, \bibinfo {author} {\bibfnamefont {F.}~\bibnamefont
  {Geissler}},\ and\ \bibinfo {author} {\bibfnamefont {L.~I.}\ \bibnamefont
  {Glazman}},\ }\bibfield  {title} {\bibinfo {title} {Magnetic moments in a
  helical edge can make weak correlations seem strong},\ }\href
  {https://doi.org/10.1103/PhysRevB.93.241301} {\bibfield  {journal} {\bibinfo
  {journal} {Phys. Rev. B}\ }\textbf {\bibinfo {volume} {93}},\ \bibinfo
  {pages} {241301(R)} (\bibinfo {year} {2016})}\BibitemShut {NoStop}%
\bibitem [{\citenamefont {Hsu}\ \emph {et~al.}(2021)\citenamefont {Hsu},
  \citenamefont {Stano}, \citenamefont {Klinovaja},\ and\ \citenamefont
  {Loss}}]{2021Hsu}%
  \BibitemOpen
  \bibfield  {author} {\bibinfo {author} {\bibfnamefont {C.-H.}\ \bibnamefont
  {Hsu}}, \bibinfo {author} {\bibfnamefont {P.}~\bibnamefont {Stano}}, \bibinfo
  {author} {\bibfnamefont {J.}~\bibnamefont {Klinovaja}},\ and\ \bibinfo
  {author} {\bibfnamefont {D.}~\bibnamefont {Loss}},\ }\bibfield  {title}
  {\bibinfo {title} {Helical liquids in semiconductors},\ }\href
  {https://doi.org/10.1088/1361-6641/ac2c27} {\bibfield  {journal} {\bibinfo
  {journal} {Semicond Sci. Technol.}\ }\textbf {\bibinfo {volume} {36}},\
  \bibinfo {pages} {123003} (\bibinfo {year} {2021})}\BibitemShut {NoStop}%
\bibitem [{\citenamefont {Knez}\ \emph {et~al.}(2014)\citenamefont {Knez},
  \citenamefont {Rettner}, \citenamefont {Yang}, \citenamefont {Parkin},
  \citenamefont {Du}, \citenamefont {Du},\ and\ \citenamefont
  {Sullivan}}]{2014Knez}%
  \BibitemOpen
  \bibfield  {author} {\bibinfo {author} {\bibfnamefont {I.}~\bibnamefont
  {Knez}}, \bibinfo {author} {\bibfnamefont {C.~T.}\ \bibnamefont {Rettner}},
  \bibinfo {author} {\bibfnamefont {S.-H.}\ \bibnamefont {Yang}}, \bibinfo
  {author} {\bibfnamefont {S.~S.~P.}\ \bibnamefont {Parkin}}, \bibinfo {author}
  {\bibfnamefont {L.}~\bibnamefont {Du}}, \bibinfo {author} {\bibfnamefont
  {R.-R.}\ \bibnamefont {Du}},\ and\ \bibinfo {author} {\bibfnamefont
  {G.}~\bibnamefont {Sullivan}},\ }\bibfield  {title} {\bibinfo {title}
  {Observation of edge transport in the disordered regime of topologically
  insulating {InAs}/{GaSb} quantum wells},\ }\href
  {https://doi.org/10.1103/physrevlett.112.026602} {\bibfield  {journal}
  {\bibinfo  {journal} {Phys. Rev. Lett.}\ }\textbf {\bibinfo {volume} {112}},\
  \bibinfo {pages} {026602} (\bibinfo {year} {2014})}\BibitemShut {NoStop}%
\bibitem [{\citenamefont {Li}\ \emph {et~al.}(2015)\citenamefont {Li},
  \citenamefont {Wang}, \citenamefont {Fu}, \citenamefont {Du}, \citenamefont
  {Schreiber}, \citenamefont {Mu}, \citenamefont {Liu}, \citenamefont
  {Sullivan}, \citenamefont {Cs{\'{a}}thy}, \citenamefont {Lin},\ and\
  \citenamefont {Du}}]{2015Li}%
  \BibitemOpen
  \bibfield  {author} {\bibinfo {author} {\bibfnamefont {T.}~\bibnamefont
  {Li}}, \bibinfo {author} {\bibfnamefont {P.}~\bibnamefont {Wang}}, \bibinfo
  {author} {\bibfnamefont {H.}~\bibnamefont {Fu}}, \bibinfo {author}
  {\bibfnamefont {L.}~\bibnamefont {Du}}, \bibinfo {author} {\bibfnamefont
  {K.~A.}\ \bibnamefont {Schreiber}}, \bibinfo {author} {\bibfnamefont
  {X.}~\bibnamefont {Mu}}, \bibinfo {author} {\bibfnamefont {X.}~\bibnamefont
  {Liu}}, \bibinfo {author} {\bibfnamefont {G.}~\bibnamefont {Sullivan}},
  \bibinfo {author} {\bibfnamefont {G.~A.}\ \bibnamefont {Cs{\'{a}}thy}},
  \bibinfo {author} {\bibfnamefont {X.}~\bibnamefont {Lin}},\ and\ \bibinfo
  {author} {\bibfnamefont {R.-R.}\ \bibnamefont {Du}},\ }\bibfield  {title}
  {\bibinfo {title} {Observation of a helical luttinger liquid in {InAs}/{GaSb}
  quantum spin {H}all edges},\ }\href
  {https://doi.org/10.1103/physrevlett.115.136804} {\bibfield  {journal}
  {\bibinfo  {journal} {Phys. Rev. Lett.}\ }\textbf {\bibinfo {volume} {115}},\
  \bibinfo {pages} {136804} (\bibinfo {year} {2015})}\BibitemShut {NoStop}%
\bibitem [{\citenamefont {Spanton}\ \emph {et~al.}(2014)\citenamefont
  {Spanton}, \citenamefont {Nowack}, \citenamefont {Du}, \citenamefont
  {Sullivan}, \citenamefont {Du},\ and\ \citenamefont {Moler}}]{2014Spanton}%
  \BibitemOpen
  \bibfield  {author} {\bibinfo {author} {\bibfnamefont {E.~M.}\ \bibnamefont
  {Spanton}}, \bibinfo {author} {\bibfnamefont {K.~C.}\ \bibnamefont {Nowack}},
  \bibinfo {author} {\bibfnamefont {L.}~\bibnamefont {Du}}, \bibinfo {author}
  {\bibfnamefont {G.}~\bibnamefont {Sullivan}}, \bibinfo {author}
  {\bibfnamefont {R.-R.}\ \bibnamefont {Du}},\ and\ \bibinfo {author}
  {\bibfnamefont {K.~A.}\ \bibnamefont {Moler}},\ }\bibfield  {title} {\bibinfo
  {title} {Images of edge current in {InAs}/{GaSb} quantum wells},\ }\href
  {https://doi.org/10.1103/physrevlett.113.026804} {\bibfield  {journal}
  {\bibinfo  {journal} {Phys. Rev. Lett.}\ }\textbf {\bibinfo {volume} {113}},\
  \bibinfo {pages} {026804} (\bibinfo {year} {2014})}\BibitemShut {NoStop}%
\bibitem [{\citenamefont {Li}\ \emph {et~al.}(2017)\citenamefont {Li},
  \citenamefont {Wang}, \citenamefont {Sullivan}, \citenamefont {Lin},\ and\
  \citenamefont {Du}}]{NewRef7}%
  \BibitemOpen
  \bibfield  {author} {\bibinfo {author} {\bibfnamefont {T.}~\bibnamefont
  {Li}}, \bibinfo {author} {\bibfnamefont {P.}~\bibnamefont {Wang}}, \bibinfo
  {author} {\bibfnamefont {G.}~\bibnamefont {Sullivan}}, \bibinfo {author}
  {\bibfnamefont {X.}~\bibnamefont {Lin}},\ and\ \bibinfo {author}
  {\bibfnamefont {R.-R.}\ \bibnamefont {Du}},\ }\bibfield  {title} {\bibinfo
  {title} {Low-temperature conductivity of weakly interacting quantum spin
  {H}all edges in strained-layer {InAs/GaInSb}},\ }\href
  {https://doi.org/10.1103/PhysRevB.96.241406} {\bibfield  {journal} {\bibinfo
  {journal} {Phys. Rev. B}\ }\textbf {\bibinfo {volume} {96}},\ \bibinfo
  {pages} {241406(R)} (\bibinfo {year} {2017})}\BibitemShut {NoStop}%
\bibitem [{\citenamefont {Bandurin}\ \emph {et~al.}(2016)\citenamefont
  {Bandurin}, \citenamefont {Torre}, \citenamefont {Kumar}, \citenamefont
  {Ben~Shalom}, \citenamefont {Tomadin}, \citenamefont {Principi},
  \citenamefont {Auton}, \citenamefont {Khestanova}, \citenamefont {Novoselov},
  \citenamefont {Grigorieva}, \citenamefont {Ponomarenko}, \citenamefont
  {Geim},\ and\ \citenamefont {Polini}}]{2016Bandurin}%
  \BibitemOpen
  \bibfield  {author} {\bibinfo {author} {\bibfnamefont {D.~A.}\ \bibnamefont
  {Bandurin}}, \bibinfo {author} {\bibfnamefont {I.}~\bibnamefont {Torre}},
  \bibinfo {author} {\bibfnamefont {R.~K.}\ \bibnamefont {Kumar}}, \bibinfo
  {author} {\bibfnamefont {M.}~\bibnamefont {Ben~Shalom}}, \bibinfo {author}
  {\bibfnamefont {A.}~\bibnamefont {Tomadin}}, \bibinfo {author} {\bibfnamefont
  {A.}~\bibnamefont {Principi}}, \bibinfo {author} {\bibfnamefont {G.~H.}\
  \bibnamefont {Auton}}, \bibinfo {author} {\bibfnamefont {E.}~\bibnamefont
  {Khestanova}}, \bibinfo {author} {\bibfnamefont {K.~S.}\ \bibnamefont
  {Novoselov}}, \bibinfo {author} {\bibfnamefont {I.~V.}\ \bibnamefont
  {Grigorieva}}, \bibinfo {author} {\bibfnamefont {L.~A.}\ \bibnamefont
  {Ponomarenko}}, \bibinfo {author} {\bibfnamefont {A.~K.}\ \bibnamefont
  {Geim}},\ and\ \bibinfo {author} {\bibfnamefont {M.}~\bibnamefont {Polini}},\
  }\bibfield  {title} {\bibinfo {title} {Negative local resistance caused by
  viscous electron backflow in graphene},\ }\href
  {https://doi.org/10.1126/science.aad0201} {\bibfield  {journal} {\bibinfo
  {journal} {Science}\ }\textbf {\bibinfo {volume} {351}},\ \bibinfo {pages}
  {1055} (\bibinfo {year} {2016})}\BibitemShut {NoStop}%
\bibitem [{\citenamefont {Abanin}\ \emph {et~al.}(2007)\citenamefont {Abanin},
  \citenamefont {Novoselov}, \citenamefont {Zeitler}, \citenamefont {Lee},
  \citenamefont {Geim},\ and\ \citenamefont {Levitov}}]{2007Abanin}%
  \BibitemOpen
  \bibfield  {author} {\bibinfo {author} {\bibfnamefont {D.~A.}\ \bibnamefont
  {Abanin}}, \bibinfo {author} {\bibfnamefont {K.~S.}\ \bibnamefont
  {Novoselov}}, \bibinfo {author} {\bibfnamefont {U.}~\bibnamefont {Zeitler}},
  \bibinfo {author} {\bibfnamefont {P.~A.}\ \bibnamefont {Lee}}, \bibinfo
  {author} {\bibfnamefont {A.~K.}\ \bibnamefont {Geim}},\ and\ \bibinfo
  {author} {\bibfnamefont {L.~S.}\ \bibnamefont {Levitov}},\ }\bibfield
  {title} {\bibinfo {title} {Dissipative quantum {H}all effect in graphene near
  the {D}irac point},\ }\href {https://doi.org/10.1103/physrevlett.98.196806}
  {\bibfield  {journal} {\bibinfo  {journal} {Phys. Rev. Lett.}\ }\textbf
  {\bibinfo {volume} {98}},\ \bibinfo {pages} {196806} (\bibinfo {year}
  {2007})}\BibitemShut {NoStop}%
\bibitem [{\citenamefont {Driscoll}(2021)}]{2021Driscoll}%
  \BibitemOpen
  \bibfield  {author} {\bibinfo {author} {\bibfnamefont {T.~A.}\ \bibnamefont
  {Driscoll}},\ }\href {https://doi.org/10.5281/ZENODO.5245134} {\bibinfo
  {title} {Schwarz–{C}hristoffel toolbox for conformal mapping in {MATLAB}}}
  (\bibinfo {year} {2021})\BibitemShut {NoStop}%
\bibitem [{\citenamefont {Durand}\ \emph {et~al.}(2016)\citenamefont {Durand},
  \citenamefont {Zhang}, \citenamefont {Hus}, \citenamefont {Ma}, \citenamefont
  {McGuire}, \citenamefont {Xu}, \citenamefont {Cao}, \citenamefont
  {Miotkowski}, \citenamefont {Chen},\ and\ \citenamefont {Li}}]{2016Durand}%
  \BibitemOpen
  \bibfield  {author} {\bibinfo {author} {\bibfnamefont {C.}~\bibnamefont
  {Durand}}, \bibinfo {author} {\bibfnamefont {X.-G.}\ \bibnamefont {Zhang}},
  \bibinfo {author} {\bibfnamefont {S.~M.}\ \bibnamefont {Hus}}, \bibinfo
  {author} {\bibfnamefont {C.}~\bibnamefont {Ma}}, \bibinfo {author}
  {\bibfnamefont {M.~A.}\ \bibnamefont {McGuire}}, \bibinfo {author}
  {\bibfnamefont {Y.}~\bibnamefont {Xu}}, \bibinfo {author} {\bibfnamefont
  {H.}~\bibnamefont {Cao}}, \bibinfo {author} {\bibfnamefont {I.}~\bibnamefont
  {Miotkowski}}, \bibinfo {author} {\bibfnamefont {Y.~P.}\ \bibnamefont
  {Chen}},\ and\ \bibinfo {author} {\bibfnamefont {A.-P.}\ \bibnamefont {Li}},\
  }\bibfield  {title} {\bibinfo {title} {Differentiation of surface and bulk
  conductivities in topological insulators via four-probe spectroscopy},\
  }\href {https://doi.org/10.1021/acs.nanolett.5b04425} {\bibfield  {journal}
  {\bibinfo  {journal} {Nano Lett.}\ }\textbf {\bibinfo {volume} {16}},\
  \bibinfo {pages} {2213} (\bibinfo {year} {2016})}\BibitemShut {NoStop}%
\bibitem [{\citenamefont {Krishtopenko}\ \emph {et~al.}(2016)\citenamefont
  {Krishtopenko}, \citenamefont {Yahniuk}, \citenamefont {But}, \citenamefont
  {Gavrilenko}, \citenamefont {Knap},\ and\ \citenamefont
  {Teppe}}]{2016Krishtopenko}%
  \BibitemOpen
  \bibfield  {author} {\bibinfo {author} {\bibfnamefont {S.~S.}\ \bibnamefont
  {Krishtopenko}}, \bibinfo {author} {\bibfnamefont {I.}~\bibnamefont
  {Yahniuk}}, \bibinfo {author} {\bibfnamefont {D.~B.}\ \bibnamefont {But}},
  \bibinfo {author} {\bibfnamefont {V.~I.}\ \bibnamefont {Gavrilenko}},
  \bibinfo {author} {\bibfnamefont {W.}~\bibnamefont {Knap}},\ and\ \bibinfo
  {author} {\bibfnamefont {F.}~\bibnamefont {Teppe}},\ }\bibfield  {title}
  {\bibinfo {title} {Pressure- and temperature-driven phase transitions in
  {HgTe} quantum wells},\ }\href {https://doi.org/10.1103/physrevb.94.245402}
  {\bibfield  {journal} {\bibinfo  {journal} {Phys. Rev. B}\ }\textbf {\bibinfo
  {volume} {94}},\ \bibinfo {pages} {245402} (\bibinfo {year}
  {2016})}\BibitemShut {NoStop}%
\bibitem [{\citenamefont {Vurgaftman}\ \emph {et~al.}(2001)\citenamefont
  {Vurgaftman}, \citenamefont {Meyer},\ and\ \citenamefont
  {Ram-Mohan}}]{2001Vurgaftman}%
  \BibitemOpen
  \bibfield  {author} {\bibinfo {author} {\bibfnamefont {I.}~\bibnamefont
  {Vurgaftman}}, \bibinfo {author} {\bibfnamefont {J.~R.}\ \bibnamefont
  {Meyer}},\ and\ \bibinfo {author} {\bibfnamefont {L.~R.}\ \bibnamefont
  {Ram-Mohan}},\ }\bibfield  {title} {\bibinfo {title} {Band parameters for
  {III–V} compound semiconductors and their alloys},\ }\href
  {https://doi.org/10.1063/1.1368156} {\bibfield  {journal} {\bibinfo
  {journal} {J. Appl. Phys.}\ }\textbf {\bibinfo {volume} {89}},\ \bibinfo
  {pages} {5815} (\bibinfo {year} {2001})}\BibitemShut {NoStop}%
\bibitem [{\citenamefont {Bernevig}\ \emph {et~al.}(2006)\citenamefont
  {Bernevig}, \citenamefont {Hughes},\ and\ \citenamefont
  {Zhang}}]{2006Bernevig}%
  \BibitemOpen
  \bibfield  {author} {\bibinfo {author} {\bibfnamefont {B.~A.}\ \bibnamefont
  {Bernevig}}, \bibinfo {author} {\bibfnamefont {T.~L.}\ \bibnamefont
  {Hughes}},\ and\ \bibinfo {author} {\bibfnamefont {S.-C.}\ \bibnamefont
  {Zhang}},\ }\bibfield  {title} {\bibinfo {title} {Quantum spin {H}all effect
  and topological phase transition in {HgTe} quantum wells},\ }\href
  {https://doi.org/10.1126/science.1133734} {\bibfield  {journal} {\bibinfo
  {journal} {Science}\ }\textbf {\bibinfo {volume} {314}},\ \bibinfo {pages}
  {1757} (\bibinfo {year} {2006})}\BibitemShut {NoStop}%
\bibitem [{\citenamefont {Krishtopenko}\ and\ \citenamefont
  {Teppe}(2018{\natexlab{b}})}]{2018Krishtopenkob}%
  \BibitemOpen
  \bibfield  {author} {\bibinfo {author} {\bibfnamefont {S.~S.}\ \bibnamefont
  {Krishtopenko}}\ and\ \bibinfo {author} {\bibfnamefont {F.}~\bibnamefont
  {Teppe}},\ }\bibfield  {title} {\bibinfo {title} {Realistic picture of
  helical edge states in {HgTe} quantum wells},\ }\href
  {https://doi.org/10.1103/physrevb.97.165408} {\bibfield  {journal} {\bibinfo
  {journal} {Phys. Rev. B}\ }\textbf {\bibinfo {volume} {97}},\ \bibinfo
  {pages} {165408} (\bibinfo {year} {2018}{\natexlab{b}})}\BibitemShut
  {NoStop}%
\bibitem [{\citenamefont {Maciejko}\ \emph {et~al.}(2009)\citenamefont
  {Maciejko}, \citenamefont {Liu}, \citenamefont {Oreg}, \citenamefont {Qi},
  \citenamefont {Wu},\ and\ \citenamefont {Zhang}}]{NewRef8}%
  \BibitemOpen
  \bibfield  {author} {\bibinfo {author} {\bibfnamefont {J.}~\bibnamefont
  {Maciejko}}, \bibinfo {author} {\bibfnamefont {C.}~\bibnamefont {Liu}},
  \bibinfo {author} {\bibfnamefont {Y.}~\bibnamefont {Oreg}}, \bibinfo {author}
  {\bibfnamefont {X.-L.}\ \bibnamefont {Qi}}, \bibinfo {author} {\bibfnamefont
  {C.}~\bibnamefont {Wu}},\ and\ \bibinfo {author} {\bibfnamefont {S.-C.}\
  \bibnamefont {Zhang}},\ }\bibfield  {title} {\bibinfo {title} {Kondo effect
  in the helical edge liquid of the quantum spin {H}all state},\ }\href
  {https://doi.org/10.1103/PhysRevLett.102.256803} {\bibfield  {journal}
  {\bibinfo  {journal} {Phys. Rev. Lett.}\ }\textbf {\bibinfo {volume} {102}},\
  \bibinfo {pages} {256803} (\bibinfo {year} {2009})}\BibitemShut {NoStop}%
\bibitem [{\citenamefont {Tanaka}\ \emph {et~al.}(2011)\citenamefont {Tanaka},
  \citenamefont {Furusaki},\ and\ \citenamefont {Matveev}}]{NewRef9}%
  \BibitemOpen
  \bibfield  {author} {\bibinfo {author} {\bibfnamefont {Y.}~\bibnamefont
  {Tanaka}}, \bibinfo {author} {\bibfnamefont {A.}~\bibnamefont {Furusaki}},\
  and\ \bibinfo {author} {\bibfnamefont {K.~A.}\ \bibnamefont {Matveev}},\
  }\bibfield  {title} {\bibinfo {title} {Conductance of a helical edge liquid
  coupled to a magnetic impurity},\ }\href
  {https://doi.org/10.1103/PhysRevLett.106.236402} {\bibfield  {journal}
  {\bibinfo  {journal} {Phys. Rev. Lett.}\ }\textbf {\bibinfo {volume} {106}},\
  \bibinfo {pages} {236402} (\bibinfo {year} {2011})}\BibitemShut {NoStop}%
\bibitem [{\citenamefont {Bendias}\ \emph {et~al.}(2018)\citenamefont
  {Bendias}, \citenamefont {Shamim}, \citenamefont {Herrmann}, \citenamefont
  {Budewitz}, \citenamefont {Shekhar}, \citenamefont {Leubner}, \citenamefont
  {Kleinlein}, \citenamefont {E.~Bocquillon},\ and\ \citenamefont
  {Molenkamp}}]{NewRef11}%
  \BibitemOpen
  \bibfield  {author} {\bibinfo {author} {\bibfnamefont {K.}~\bibnamefont
  {Bendias}}, \bibinfo {author} {\bibfnamefont {S.}~\bibnamefont {Shamim}},
  \bibinfo {author} {\bibfnamefont {O.}~\bibnamefont {Herrmann}}, \bibinfo
  {author} {\bibfnamefont {A.}~\bibnamefont {Budewitz}}, \bibinfo {author}
  {\bibfnamefont {P.}~\bibnamefont {Shekhar}}, \bibinfo {author} {\bibfnamefont
  {P.}~\bibnamefont {Leubner}}, \bibinfo {author} {\bibfnamefont
  {J.}~\bibnamefont {Kleinlein}}, \bibinfo {author} {\bibfnamefont {H.~B.}\
  \bibnamefont {E.~Bocquillon}},\ and\ \bibinfo {author} {\bibfnamefont
  {L.~W.}\ \bibnamefont {Molenkamp}},\ }\bibfield  {title} {\bibinfo {title}
  {High mobility {HgTe} microstructures for quantum spin {H}all studies},\
  }\href {https://doi.org/10.1021/acs.nanolett.8b01405} {\bibfield  {journal}
  {\bibinfo  {journal} {Nano Lett.}\ }\textbf {\bibinfo {volume} {18}},\
  \bibinfo {pages} {4831} (\bibinfo {year} {2018})}\BibitemShut {NoStop}%
\bibitem [{\citenamefont {Lunczer}\ \emph {et~al.}(2019)\citenamefont
  {Lunczer}, \citenamefont {Leubner}, \citenamefont {Endres}, \citenamefont
  {M\"uller}, \citenamefont {Br\"une}, \citenamefont {Buhmann},\ and\
  \citenamefont {Molenkamp}}]{NewRef12}%
  \BibitemOpen
  \bibfield  {author} {\bibinfo {author} {\bibfnamefont {L.}~\bibnamefont
  {Lunczer}}, \bibinfo {author} {\bibfnamefont {P.}~\bibnamefont {Leubner}},
  \bibinfo {author} {\bibfnamefont {M.}~\bibnamefont {Endres}}, \bibinfo
  {author} {\bibfnamefont {V.~L.}\ \bibnamefont {M\"uller}}, \bibinfo {author}
  {\bibfnamefont {C.}~\bibnamefont {Br\"une}}, \bibinfo {author} {\bibfnamefont
  {H.}~\bibnamefont {Buhmann}},\ and\ \bibinfo {author} {\bibfnamefont {L.~W.}\
  \bibnamefont {Molenkamp}},\ }\bibfield  {title} {\bibinfo {title}
  {Approaching quantization in macroscopic quantum spin {H}all devices through
  gate training},\ }\href {https://doi.org/10.1103/PhysRevLett.123.047701}
  {\bibfield  {journal} {\bibinfo  {journal} {Phys. Rev. Lett.}\ }\textbf
  {\bibinfo {volume} {123}},\ \bibinfo {pages} {047701} (\bibinfo {year}
  {2019})}\BibitemShut {NoStop}%
\bibitem [{\citenamefont {Fuchs}\ \emph {et~al.}(2023)\citenamefont {Fuchs},
  \citenamefont {Shamim}, \citenamefont {Shekhar}, \citenamefont {F\"urst},
  \citenamefont {Kleinlein}, \citenamefont {V\"ayrynen}, \citenamefont
  {Buhmann},\ and\ \citenamefont {Molenkamp}}]{NewRef13}%
  \BibitemOpen
  \bibfield  {author} {\bibinfo {author} {\bibfnamefont {C.}~\bibnamefont
  {Fuchs}}, \bibinfo {author} {\bibfnamefont {S.}~\bibnamefont {Shamim}},
  \bibinfo {author} {\bibfnamefont {P.}~\bibnamefont {Shekhar}}, \bibinfo
  {author} {\bibfnamefont {L.}~\bibnamefont {F\"urst}}, \bibinfo {author}
  {\bibfnamefont {J.}~\bibnamefont {Kleinlein}}, \bibinfo {author}
  {\bibfnamefont {J.~I.}\ \bibnamefont {V\"ayrynen}}, \bibinfo {author}
  {\bibfnamefont {H.}~\bibnamefont {Buhmann}},\ and\ \bibinfo {author}
  {\bibfnamefont {L.~W.}\ \bibnamefont {Molenkamp}},\ }\bibfield  {title}
  {\bibinfo {title} {Kondo interaction of quantum spin {H}all edge channels
  with charge puddles},\ }\href {https://doi.org/10.1103/PhysRevB.108.205302}
  {\bibfield  {journal} {\bibinfo  {journal} {Phys. Rev. B}\ }\textbf {\bibinfo
  {volume} {108}},\ \bibinfo {pages} {205302} (\bibinfo {year}
  {2023})}\BibitemShut {NoStop}%
\bibitem [{\citenamefont {Dietl}(2023{\natexlab{a}})}]{NewRef5}%
  \BibitemOpen
  \bibfield  {author} {\bibinfo {author} {\bibfnamefont {T.}~\bibnamefont
  {Dietl}},\ }\bibfield  {title} {\bibinfo {title} {Effects of charge dopants
  in quantum spin {H}all materials},\ }\href
  {https://doi.org/10.1103/PhysRevLett.130.086202} {\bibfield  {journal}
  {\bibinfo  {journal} {Phys. Rev. Lett.}\ }\textbf {\bibinfo {volume} {130}},\
  \bibinfo {pages} {086202} (\bibinfo {year} {2023}{\natexlab{a}})}\BibitemShut
  {NoStop}%
\bibitem [{\citenamefont {Dietl}(2023{\natexlab{b}})}]{NewRef6}%
  \BibitemOpen
  \bibfield  {author} {\bibinfo {author} {\bibfnamefont {T.}~\bibnamefont
  {Dietl}},\ }\bibfield  {title} {\bibinfo {title} {Quantitative theory of
  backscattering in topological {HgTe} and {(Hg,Mn)Te} quantum wells: acceptor
  states, {K}ondo effect, precessional dephasing, and bound magnetic polaron},\
  }\href {https://doi.org/10.1103/PhysRevB.107.085421} {\bibfield  {journal}
  {\bibinfo  {journal} {Phys. Rev. B}\ }\textbf {\bibinfo {volume} {107}},\
  \bibinfo {pages} {085421} (\bibinfo {year} {2023}{\natexlab{b}})}\BibitemShut
  {NoStop}%
\bibitem [{\citenamefont {Nikolaev}\ \emph {et~al.}(2020)\citenamefont
  {Nikolaev}, \citenamefont {Uaman-Svetikova}, \citenamefont {Rumyantsev},
  \citenamefont {Zholudev}, \citenamefont {Kozlov}, \citenamefont {Morozov},
  \citenamefont {Dvoretsky}, \citenamefont {Mikhailov}, \citenamefont
  {Gavrilenko},\ and\ \citenamefont {Ikonnikov}}]{NewRef14}%
  \BibitemOpen
  \bibfield  {author} {\bibinfo {author} {\bibfnamefont {I.~D.}\ \bibnamefont
  {Nikolaev}}, \bibinfo {author} {\bibfnamefont {T.~A.}\ \bibnamefont
  {Uaman-Svetikova}}, \bibinfo {author} {\bibfnamefont {V.~V.}\ \bibnamefont
  {Rumyantsev}}, \bibinfo {author} {\bibfnamefont {M.~S.}\ \bibnamefont
  {Zholudev}}, \bibinfo {author} {\bibfnamefont {D.~V.}\ \bibnamefont
  {Kozlov}}, \bibinfo {author} {\bibfnamefont {S.~V.}\ \bibnamefont {Morozov}},
  \bibinfo {author} {\bibfnamefont {S.~A.}\ \bibnamefont {Dvoretsky}}, \bibinfo
  {author} {\bibfnamefont {N.~N.}\ \bibnamefont {Mikhailov}}, \bibinfo {author}
  {\bibfnamefont {V.~I.}\ \bibnamefont {Gavrilenko}},\ and\ \bibinfo {author}
  {\bibfnamefont {A.~V.}\ \bibnamefont {Ikonnikov}},\ }\bibfield  {title}
  {\bibinfo {title} {Probing states of a double acceptor in {CdHgTe}
  heterostructures via optical gating},\ }\href
  {https://doi.org/10.1134/S0021364020100124} {\bibfield  {journal} {\bibinfo
  {journal} {JETP Lett.}\ }\textbf {\bibinfo {volume} {111}},\ \bibinfo {pages}
  {575} (\bibinfo {year} {2020})}\BibitemShut {NoStop}%
\bibitem [{\citenamefont {Ruffenach}\ \emph
  {et~al.}(2017{\natexlab{b}})\citenamefont {Ruffenach}, \citenamefont
  {Kadykov}, \citenamefont {Rumyantsev}, \citenamefont {Torres}, \citenamefont
  {Coquillat}, \citenamefont {But}, \citenamefont {Krishtopenko}, \citenamefont
  {Consejo}, \citenamefont {Knap}, \citenamefont {Winnerl}, \citenamefont
  {Helm}, \citenamefont {Fadeev}, \citenamefont {Mikhailov}, \citenamefont
  {Dvoretskii}, \citenamefont {Gavrilenko}, \citenamefont {Morozov},\ and\
  \citenamefont {Teppe}}]{NewRef15}%
  \BibitemOpen
  \bibfield  {author} {\bibinfo {author} {\bibfnamefont {S.}~\bibnamefont
  {Ruffenach}}, \bibinfo {author} {\bibfnamefont {A.}~\bibnamefont {Kadykov}},
  \bibinfo {author} {\bibfnamefont {V.~V.}\ \bibnamefont {Rumyantsev}},
  \bibinfo {author} {\bibfnamefont {J.}~\bibnamefont {Torres}}, \bibinfo
  {author} {\bibfnamefont {D.}~\bibnamefont {Coquillat}}, \bibinfo {author}
  {\bibfnamefont {D.}~\bibnamefont {But}}, \bibinfo {author} {\bibfnamefont
  {S.~S.}\ \bibnamefont {Krishtopenko}}, \bibinfo {author} {\bibfnamefont
  {C.}~\bibnamefont {Consejo}}, \bibinfo {author} {\bibfnamefont
  {W.}~\bibnamefont {Knap}}, \bibinfo {author} {\bibfnamefont {S.}~\bibnamefont
  {Winnerl}}, \bibinfo {author} {\bibfnamefont {M.}~\bibnamefont {Helm}},
  \bibinfo {author} {\bibfnamefont {M.~A.}\ \bibnamefont {Fadeev}}, \bibinfo
  {author} {\bibfnamefont {N.~N.}\ \bibnamefont {Mikhailov}}, \bibinfo {author}
  {\bibfnamefont {S.~A.}\ \bibnamefont {Dvoretskii}}, \bibinfo {author}
  {\bibfnamefont {V.~I.}\ \bibnamefont {Gavrilenko}}, \bibinfo {author}
  {\bibfnamefont {S.~V.}\ \bibnamefont {Morozov}},\ and\ \bibinfo {author}
  {\bibfnamefont {F.}~\bibnamefont {Teppe}},\ }\bibfield  {title} {\bibinfo
  {title} {{HgCdTe}-based heterostructures for terahertz photonics},\ }\href
  {https://doi.org/10.1063/1.4977781} {\bibfield  {journal} {\bibinfo
  {journal} {APL Mater.}\ }\textbf {\bibinfo {volume} {5}},\ \bibinfo {pages}
  {035503} (\bibinfo {year} {2017}{\natexlab{b}})}\BibitemShut {NoStop}%
\bibitem [{\citenamefont {Rumyantsev}\ \emph {et~al.}(2024)\citenamefont
  {Rumyantsev}, \citenamefont {Mazhukina}, \citenamefont {Utochkin},
  \citenamefont {Kudryavtsev}, \citenamefont {Dubinov}, \citenamefont
  {Aleshkin}, \citenamefont {Razova}, \citenamefont {Kuritsin}, \citenamefont
  {Fadeev}, \citenamefont {Antonov}, \citenamefont {Mikhailov}, \citenamefont
  {Dvoretsky}, \citenamefont {Gavrilenko}, \citenamefont {Teppe},\ and\
  \citenamefont {Morozov}}]{NewRef16}%
  \BibitemOpen
  \bibfield  {author} {\bibinfo {author} {\bibfnamefont {V.~V.}\ \bibnamefont
  {Rumyantsev}}, \bibinfo {author} {\bibfnamefont {K.~A.}\ \bibnamefont
  {Mazhukina}}, \bibinfo {author} {\bibfnamefont {V.~V.}\ \bibnamefont
  {Utochkin}}, \bibinfo {author} {\bibfnamefont {K.~E.}\ \bibnamefont
  {Kudryavtsev}}, \bibinfo {author} {\bibfnamefont {A.~A.}\ \bibnamefont
  {Dubinov}}, \bibinfo {author} {\bibfnamefont {V.~Y.}\ \bibnamefont
  {Aleshkin}}, \bibinfo {author} {\bibfnamefont {A.~A.}\ \bibnamefont
  {Razova}}, \bibinfo {author} {\bibfnamefont {D.~I.}\ \bibnamefont
  {Kuritsin}}, \bibinfo {author} {\bibfnamefont {M.~A.}\ \bibnamefont
  {Fadeev}}, \bibinfo {author} {\bibfnamefont {A.~V.}\ \bibnamefont {Antonov}},
  \bibinfo {author} {\bibfnamefont {N.~N.}\ \bibnamefont {Mikhailov}}, \bibinfo
  {author} {\bibfnamefont {S.~A.}\ \bibnamefont {Dvoretsky}}, \bibinfo {author}
  {\bibfnamefont {V.~I.}\ \bibnamefont {Gavrilenko}}, \bibinfo {author}
  {\bibfnamefont {F.}~\bibnamefont {Teppe}},\ and\ \bibinfo {author}
  {\bibfnamefont {S.~V.}\ \bibnamefont {Morozov}},\ }\bibfield  {title}
  {\bibinfo {title} {Optically pumped stimulated emission in {HgCdTe}-based
  quantum wells: Toward continuous wave lasing in very long-wavelength infrared
  range},\ }\href {https://doi.org/10.1063/5.0186292} {\bibfield  {journal}
  {\bibinfo  {journal} {Appl. Phys. Lett.}\ }\textbf {\bibinfo {volume}
  {124}},\ \bibinfo {pages} {161111} (\bibinfo {year} {2024})}\BibitemShut
  {NoStop}%
\bibitem [{\citenamefont {Virkkala}\ \emph {et~al.}(2012)\citenamefont
  {Virkkala}, \citenamefont {Havu}, \citenamefont {Tuomisto},\ and\
  \citenamefont {Puska}}]{NewRef17}%
  \BibitemOpen
  \bibfield  {author} {\bibinfo {author} {\bibfnamefont {V.}~\bibnamefont
  {Virkkala}}, \bibinfo {author} {\bibfnamefont {V.}~\bibnamefont {Havu}},
  \bibinfo {author} {\bibfnamefont {F.}~\bibnamefont {Tuomisto}},\ and\
  \bibinfo {author} {\bibfnamefont {M.~J.}\ \bibnamefont {Puska}},\ }\bibfield
  {title} {\bibinfo {title} {Native point defect energetics in {GaSb}: Enabling
  $p$-type conductivity of undoped {GaSb}},\ }\href
  {https://doi.org/10.1103/PhysRevB.86.144101} {\bibfield  {journal} {\bibinfo
  {journal} {Phys. Rev. B}\ }\textbf {\bibinfo {volume} {86}},\ \bibinfo
  {pages} {144101} (\bibinfo {year} {2012})}\BibitemShut {NoStop}%
\bibitem [{\citenamefont {Kujala}\ \emph {et~al.}(2014)\citenamefont {Kujala},
  \citenamefont {Segercrantz}, \citenamefont {Tuomisto},\ and\ \citenamefont
  {Slotte}}]{NewRef18}%
  \BibitemOpen
  \bibfield  {author} {\bibinfo {author} {\bibfnamefont {J.}~\bibnamefont
  {Kujala}}, \bibinfo {author} {\bibfnamefont {N.}~\bibnamefont {Segercrantz}},
  \bibinfo {author} {\bibfnamefont {F.}~\bibnamefont {Tuomisto}},\ and\
  \bibinfo {author} {\bibfnamefont {J.}~\bibnamefont {Slotte}},\ }\bibfield
  {title} {\bibinfo {title} {Native point defects in {GaSb}},\ }\href
  {https://doi.org/10.1063/1.4898082} {\bibfield  {journal} {\bibinfo
  {journal} {J. Appl. Phys.}\ }\textbf {\bibinfo {volume} {116}},\ \bibinfo
  {pages} {143508} (\bibinfo {year} {2014})}\BibitemShut {NoStop}%
\bibitem [{\citenamefont {Segercrantz}\ \emph {et~al.}(2014)\citenamefont
  {Segercrantz}, \citenamefont {Slotte}, \citenamefont {Makkonen},
  \citenamefont {Kujala}, \citenamefont {Tuomisto}, \citenamefont {Song},\ and\
  \citenamefont {Wang}}]{NewRef19}%
  \BibitemOpen
  \bibfield  {author} {\bibinfo {author} {\bibfnamefont {N.}~\bibnamefont
  {Segercrantz}}, \bibinfo {author} {\bibfnamefont {J.}~\bibnamefont {Slotte}},
  \bibinfo {author} {\bibfnamefont {I.}~\bibnamefont {Makkonen}}, \bibinfo
  {author} {\bibfnamefont {J.}~\bibnamefont {Kujala}}, \bibinfo {author}
  {\bibfnamefont {F.}~\bibnamefont {Tuomisto}}, \bibinfo {author}
  {\bibfnamefont {Y.}~\bibnamefont {Song}},\ and\ \bibinfo {author}
  {\bibfnamefont {S.}~\bibnamefont {Wang}},\ }\bibfield  {title} {\bibinfo
  {title} {Point defect balance in epitaxial {GaSb}},\ }\href
  {https://doi.org/10.1063/1.4894473} {\bibfield  {journal} {\bibinfo
  {journal} {Appl. Phys. Lett.}\ }\textbf {\bibinfo {volume} {105}},\ \bibinfo
  {pages} {082113} (\bibinfo {year} {2014})}\BibitemShut {NoStop}%
\bibitem [{\citenamefont {Segercrantz}\ \emph {et~al.}(2017)\citenamefont
  {Segercrantz}, \citenamefont {Slotte}, \citenamefont {Tuomisto},
  \citenamefont {Mizohata},\ and\ \citenamefont
  {R\"{a}is\"{a}nen}}]{2017Segercrantz}%
  \BibitemOpen
  \bibfield  {author} {\bibinfo {author} {\bibfnamefont {N.}~\bibnamefont
  {Segercrantz}}, \bibinfo {author} {\bibfnamefont {J.}~\bibnamefont {Slotte}},
  \bibinfo {author} {\bibfnamefont {F.}~\bibnamefont {Tuomisto}}, \bibinfo
  {author} {\bibfnamefont {K.}~\bibnamefont {Mizohata}},\ and\ \bibinfo
  {author} {\bibfnamefont {J.}~\bibnamefont {R\"{a}is\"{a}nen}},\ }\bibfield
  {title} {\bibinfo {title} {Instability of the {Sb} vacancy in {GaSb}},\
  }\href {https://doi.org/10.1103/physrevb.95.184103} {\bibfield  {journal}
  {\bibinfo  {journal} {Phys. Rev. B}\ }\textbf {\bibinfo {volume} {95}},\
  \bibinfo {pages} {184103} (\bibinfo {year} {2017})}\BibitemShut {NoStop}%
\bibitem [{\citenamefont {Nowack}\ \emph {et~al.}(2013)\citenamefont {Nowack},
  \citenamefont {Spanton}, \citenamefont {Baenninger}, \citenamefont {Konig},
  \citenamefont {Kirtley}, \citenamefont {Kalisky}, \citenamefont {Ames},
  \citenamefont {Leubner}, \citenamefont {Brune}, \citenamefont {Buhmann},
  \citenamefont {Molenkamp}, \citenamefont {Goldhaber-Gordon},\ and\
  \citenamefont {Moler}}]{NewRef20}%
  \BibitemOpen
  \bibfield  {author} {\bibinfo {author} {\bibfnamefont {K.~C.}\ \bibnamefont
  {Nowack}}, \bibinfo {author} {\bibfnamefont {E.~M.}\ \bibnamefont {Spanton}},
  \bibinfo {author} {\bibfnamefont {M.}~\bibnamefont {Baenninger}}, \bibinfo
  {author} {\bibfnamefont {M.}~\bibnamefont {Konig}}, \bibinfo {author}
  {\bibfnamefont {J.~R.}\ \bibnamefont {Kirtley}}, \bibinfo {author}
  {\bibfnamefont {B.}~\bibnamefont {Kalisky}}, \bibinfo {author} {\bibfnamefont
  {C.}~\bibnamefont {Ames}}, \bibinfo {author} {\bibfnamefont {P.}~\bibnamefont
  {Leubner}}, \bibinfo {author} {\bibfnamefont {C.}~\bibnamefont {Brune}},
  \bibinfo {author} {\bibfnamefont {H.}~\bibnamefont {Buhmann}}, \bibinfo
  {author} {\bibfnamefont {L.~W.}\ \bibnamefont {Molenkamp}}, \bibinfo {author}
  {\bibfnamefont {D.}~\bibnamefont {Goldhaber-Gordon}},\ and\ \bibinfo {author}
  {\bibfnamefont {K.~A.}\ \bibnamefont {Moler}},\ }\bibfield  {title} {\bibinfo
  {title} {Imaging currents in {HgTe} quantum wells in the quantum spin {H}all
  regime},\ }\href {https://doi.org/10.1038/nmat3682} {\bibfield  {journal}
  {\bibinfo  {journal} {Nat. Mater.}\ }\textbf {\bibinfo {volume} {12}},\
  \bibinfo {pages} {787} (\bibinfo {year} {2013})}\BibitemShut {NoStop}%
\bibitem [{\citenamefont {Gusev}\ \emph {et~al.}(2019)\citenamefont {Gusev},
  \citenamefont {Kvon}, \citenamefont {Olshanetsky},\ and\ \citenamefont
  {Mikhailov}}]{2019Gusev}%
  \BibitemOpen
  \bibfield  {author} {\bibinfo {author} {\bibfnamefont {G.}~\bibnamefont
  {Gusev}}, \bibinfo {author} {\bibfnamefont {Z.}~\bibnamefont {Kvon}},
  \bibinfo {author} {\bibfnamefont {E.}~\bibnamefont {Olshanetsky}},\ and\
  \bibinfo {author} {\bibfnamefont {N.}~\bibnamefont {Mikhailov}},\ }\bibfield
  {title} {\bibinfo {title} {Mesoscopic transport in two-dimensional
  topological insulators},\ }\href {https://doi.org/10.1016/j.ssc.2019.113701}
  {\bibfield  {journal} {\bibinfo  {journal} {Solid State Commun.}\ }\textbf
  {\bibinfo {volume} {302}},\ \bibinfo {pages} {113701} (\bibinfo {year}
  {2019})}\BibitemShut {NoStop}%
\bibitem [{\citenamefont {Tagliacozzo}\ \emph {et~al.}(2019)\citenamefont
  {Tagliacozzo}, \citenamefont {Campagnano}, \citenamefont {Giuliano},
  \citenamefont {Lucignano},\ and\ \citenamefont {Jouault}}]{2019Tagliacozzo}%
  \BibitemOpen
  \bibfield  {author} {\bibinfo {author} {\bibfnamefont {A.}~\bibnamefont
  {Tagliacozzo}}, \bibinfo {author} {\bibfnamefont {G.}~\bibnamefont
  {Campagnano}}, \bibinfo {author} {\bibfnamefont {D.}~\bibnamefont
  {Giuliano}}, \bibinfo {author} {\bibfnamefont {P.}~\bibnamefont
  {Lucignano}},\ and\ \bibinfo {author} {\bibfnamefont {B.}~\bibnamefont
  {Jouault}},\ }\bibfield  {title} {\bibinfo {title} {Thermal transport driven
  by charge imbalance in graphene in a magnetic field close to the charge
  neutrality point at low temperature: Nonlocal resistance},\ }\href
  {https://doi.org/10.1103/physrevb.99.155417} {\bibfield  {journal} {\bibinfo
  {journal} {Phys. Rev. B}\ }\textbf {\bibinfo {volume} {99}},\ \bibinfo
  {pages} {155417} (\bibinfo {year} {2019})}\BibitemShut {NoStop}%
\end{thebibliography}
\end{document}